\documentclass[aps,prd,twocolumn,showpacs,superscriptaddress,groupedaddress,amsmath,amssymb]{revtex4}

\usepackage{graphicx}
\usepackage{graphics}
\usepackage{dcolumn}
\usepackage{bm}
\usepackage{amssymb} 
\usepackage{multirow} 

\newcommand {\Bd} {\ensuremath{B^0_d}}
\newcommand {\Bs} {\ensuremath{B^0_s}}
\newcommand {\Bq} {\ensuremath{B^0_q}}
\newcommand {\Ds} {\ensuremath{D_s}}

\newcommand {\barBq} {\ensuremath{\bar{B}^0_q}}
\newcommand {\asld} {\ensuremath{a^d_{\mathrm{sl}}}}
\newcommand {\asls} {\ensuremath{a^s_{\mathrm{sl}}}}
\newcommand {\aslq} {\ensuremath{a^q_{\mathrm{sl}}}}
\newcommand {\aslb} {\ensuremath{A^b_{\mathrm{sl}}}}
\newcommand {\ktomu} {\ensuremath{K \to \mu}}
\newcommand {\pitomu} {\ensuremath{\pi \to \mu}}
\newcommand {\ptomu} {\ensuremath{p \to \mu}}
\newcommand {\ks} {\ensuremath{K_S}}
\newcommand {\kstp} {\ensuremath{K^{*+}}}
\newcommand {\kstpm} {\ensuremath{K^{*\pm}}}
\newcommand {\kstneu} {\ensuremath{K^{*0}}}
\newcommand {\phin} {\ensuremath{\phi(1020)}}

\begin{document}

\hspace{5.2in} \mbox{Fermilab-Pub-10/114-E}

\title{
Evidence for an anomalous like-sign dimuon charge asymmetry
}

%
\affiliation{Universidad de Buenos Aires, Buenos Aires, Argentina}
\affiliation{LAFEX, Centro Brasileiro de Pesquisas F{\'\i}sicas, Rio de Janeiro, Brazil}
\affiliation{Universidade do Estado do Rio de Janeiro, Rio de Janeiro, Brazil}
\affiliation{Universidade Federal do ABC, Santo Andr\'e, Brazil}
\affiliation{Instituto de F\'{\i}sica Te\'orica, Universidade Estadual Paulista, S\~ao Paulo, Brazil}
\affiliation{Simon Fraser University, Vancouver, British Columbia, and York University, Toronto, Ontario, Canada}
\affiliation{University of Science and Technology of China, Hefei, People's Republic of China}
\affiliation{Universidad de los Andes, Bogot\'{a}, Colombia}
\affiliation{Charles University, Faculty of Mathematics and Physics, Center for Particle Physics, Prague, Czech Republic}
\affiliation{Czech Technical University in Prague, Prague, Czech Republic}
\affiliation{Center for Particle Physics, Institute of Physics, Academy of Sciences of the Czech Republic, Prague, Czech Republic}
\affiliation{Universidad San Francisco de Quito, Quito, Ecuador}
\affiliation{LPC, Universit\'e Blaise Pascal, CNRS/IN2P3, Clermont, France}
\affiliation{LPSC, Universit\'e Joseph Fourier Grenoble 1, CNRS/IN2P3, Institut National Polytechnique de Grenoble, Grenoble, France}
\affiliation{CPPM, Aix-Marseille Universit\'e, CNRS/IN2P3, Marseille, France}
\affiliation{LAL, Universit\'e Paris-Sud, CNRS/IN2P3, Orsay, France}
\affiliation{LPNHE, Universit\'es Paris VI and VII, CNRS/IN2P3, Paris, France}
\affiliation{CEA, Irfu, SPP, Saclay, France}
\affiliation{IPHC, Universit\'e de Strasbourg, CNRS/IN2P3, Strasbourg, France}
\affiliation{IPNL, Universit\'e Lyon 1, CNRS/IN2P3, Villeurbanne, France and Universit\'e de Lyon, Lyon, France}
\affiliation{III. Physikalisches Institut A, RWTH Aachen University, Aachen, Germany}
\affiliation{Physikalisches Institut, Universit{\"a}t Freiburg, Freiburg, Germany}
\affiliation{II. Physikalisches Institut, Georg-August-Universit{\"a}t G\"ottingen, G\"ottingen, Germany}
\affiliation{Institut f{\"u}r Physik, Universit{\"a}t Mainz, Mainz, Germany}
\affiliation{Ludwig-Maximilians-Universit{\"a}t M{\"u}nchen, M{\"u}nchen, Germany}
\affiliation{Fachbereich Physik, Bergische  Universit{\"a}t Wuppertal, Wuppertal, Germany}
\affiliation{Panjab University, Chandigarh, India}
\affiliation{Delhi University, Delhi, India}
\affiliation{Tata Institute of Fundamental Research, Mumbai, India}
\affiliation{University College Dublin, Dublin, Ireland}
\affiliation{Korea Detector Laboratory, Korea University, Seoul, Korea}
\affiliation{SungKyunKwan University, Suwon, Korea}
\affiliation{CINVESTAV, Mexico City, Mexico}
\affiliation{FOM-Institute NIKHEF and University of Amsterdam/NIKHEF, Amsterdam, The Netherlands}
\affiliation{Radboud University Nijmegen/NIKHEF, Nijmegen, The Netherlands}
\affiliation{Joint Institute for Nuclear Research, Dubna, Russia}
\affiliation{Institute for Theoretical and Experimental Physics, Moscow, Russia}
\affiliation{Moscow State University, Moscow, Russia}
\affiliation{Institute for High Energy Physics, Protvino, Russia}
\affiliation{Petersburg Nuclear Physics Institute, St. Petersburg, Russia}
\affiliation{Stockholm University, Stockholm and Uppsala University, Uppsala, Sweden }
\affiliation{Lancaster University, Lancaster LA1 4YB, United Kingdom}
\affiliation{Imperial College London, London SW7 2AZ, United Kingdom}
\affiliation{The University of Manchester, Manchester M13 9PL, United Kingdom}
\affiliation{University of Arizona, Tucson, Arizona 85721, USA}
\affiliation{University of California Riverside, Riverside, California 92521, USA}
\affiliation{Florida State University, Tallahassee, Florida 32306, USA}
\affiliation{Fermi National Accelerator Laboratory, Batavia, Illinois 60510, USA}
\affiliation{University of Illinois at Chicago, Chicago, Illinois 60607, USA}
\affiliation{Northern Illinois University, DeKalb, Illinois 60115, USA}
\affiliation{Northwestern University, Evanston, Illinois 60208, USA}
\affiliation{Indiana University, Bloomington, Indiana 47405, USA}
\affiliation{Purdue University Calumet, Hammond, Indiana 46323, USA}
\affiliation{University of Notre Dame, Notre Dame, Indiana 46556, USA}
\affiliation{Iowa State University, Ames, Iowa 50011, USA}
\affiliation{University of Kansas, Lawrence, Kansas 66045, USA}
\affiliation{Kansas State University, Manhattan, Kansas 66506, USA}
\affiliation{Louisiana Tech University, Ruston, Louisiana 71272, USA}
\affiliation{University of Maryland, College Park, Maryland 20742, USA}
\affiliation{Boston University, Boston, Massachusetts 02215, USA}
\affiliation{Northeastern University, Boston, Massachusetts 02115, USA}
\affiliation{University of Michigan, Ann Arbor, Michigan 48109, USA}
\affiliation{Michigan State University, East Lansing, Michigan 48824, USA}
\affiliation{University of Mississippi, University, Mississippi 38677, USA}
\affiliation{University of Nebraska, Lincoln, Nebraska 68588, USA}
\affiliation{Rutgers University, Piscataway, New Jersey 08855, USA}
\affiliation{Princeton University, Princeton, New Jersey 08544, USA}
\affiliation{State University of New York, Buffalo, New York 14260, USA}
\affiliation{Columbia University, New York, New York 10027, USA}
\affiliation{University of Rochester, Rochester, New York 14627, USA}
\affiliation{State University of New York, Stony Brook, New York 11794, USA}
\affiliation{Brookhaven National Laboratory, Upton, New York 11973, USA}
\affiliation{Langston University, Langston, Oklahoma 73050, USA}
\affiliation{University of Oklahoma, Norman, Oklahoma 73019, USA}
\affiliation{Oklahoma State University, Stillwater, Oklahoma 74078, USA}
\affiliation{Brown University, Providence, Rhode Island 02912, USA}
\affiliation{University of Texas, Arlington, Texas 76019, USA}
\affiliation{Southern Methodist University, Dallas, Texas 75275, USA}
\affiliation{Rice University, Houston, Texas 77005, USA}
\affiliation{University of Virginia, Charlottesville, Virginia 22901, USA}
\affiliation{University of Washington, Seattle, Washington 98195, USA}
\author{V.M.~Abazov} \affiliation{Joint Institute for Nuclear Research, Dubna, Russia}
\author{B.~Abbott} \affiliation{University of Oklahoma, Norman, Oklahoma 73019, USA}
\author{M.~Abolins} \affiliation{Michigan State University, East Lansing, Michigan 48824, USA}
\author{B.S.~Acharya} \affiliation{Tata Institute of Fundamental Research, Mumbai, India}
\author{M.~Adams} \affiliation{University of Illinois at Chicago, Chicago, Illinois 60607, USA}
\author{T.~Adams} \affiliation{Florida State University, Tallahassee, Florida 32306, USA}
\author{E.~Aguilo} \affiliation{Simon Fraser University, Vancouver, British Columbia, and York University, Toronto, Ontario, Canada}
\author{G.D.~Alexeev} \affiliation{Joint Institute for Nuclear Research, Dubna, Russia}
\author{G.~Alkhazov} \affiliation{Petersburg Nuclear Physics Institute, St. Petersburg, Russia}
\author{A.~Alton$^{a}$} \affiliation{University of Michigan, Ann Arbor, Michigan 48109, USA}
% ...
%
%
\collaboration{The D0 Collaboration\footnote{with visitors from
$^{a}$  
Augustana College, Sioux Falls, SD, USA,
$^{b}$  
The University of Liverpool, Liverpool, UK,
$^{c}$  
SLAC, Menlo Park, CA, USA,
$^{d}$  
ICREA/IFAE, Barcelona, Spain,
$^{e}$  
Centro de Investigacion en Computacion - IPN,
  Mexico City, Mexico,
$^{f}$  
ECFM, Universidad Autonoma de Sinaloa, Culiac\'an, Mexico,
$^{g}$  
and Universit{\"a}t Bern, Bern, Switzerland.
}} \noaffiliation
\vskip 0.25cm

\date{May 16, 2010}

\begin{abstract}

\noindent
We measure the charge asymmetry $A$ of like-sign dimuon events
in 6.1 fb$^{-1}$ of $p\overline{p}$ collisions
recorded with the D0 detector at a center-of-mass energy $\sqrt{s} = 1.96$ TeV
at the Fermilab Tevatron collider.
From $A$, we extract the like-sign dimuon charge asymmetry
in semileptonic $b$-hadron decays:
$\aslb = -0.00957 \pm 0.00251~({\rm stat}) \pm 0.00146~({\rm syst})$.
This result differs by 3.2 standard deviations from the standard model prediction
$\aslb(SM) = (-2.3^{+0.5}_{-0.6}) \times 10^{-4}$ and provides
first evidence of anomalous CP-violation in the mixing
of neutral $B$ mesons.


\end{abstract}


\maketitle


\section{Introduction}

Studies of particle production and decay under the reversal of
discrete symmetries (charge, parity and time reversal) have yielded
considerable insight on the structure of the theories that
describe high energy phenomena. Of particular interest is the
observation of CP violation, a phenomenon well established
in the $K^0$ and $B^0_d$ systems, but not yet observed for
the $B^0_s$ system, where all $CP$ violation effects are expected
to be small in the standard model (SM)~\cite{Nierste} (See~\cite{pdg} and references
therein for a review of the experimental results and of the theoretical
framework for describing $CP$ violation in neutral mesons decays).
The violation of $CP$ symmetry is a necessary condition for
baryogenesis, the process thought to be responsible for the
matter-antimatter asymmetry of the universe~\cite{sakharov}.
However, the observed $CP$ violation in the $K^0$ and $B^0_d$ systems, consistent with 
the standard model expectation,
is not sufficient to explain this
asymmetry, suggesting the presence of additional sources of
$CP$ violation, beyond the standard model.

The D0 experiment at the Fermilab Tevatron proton-antiproton
($p\bar{p}$) collider, operating at a center-of-mass energy
$\sqrt{s}=1.96$~TeV, is in a unique position to study possible
effects of $CP$ violation, in particular through the study of
charge asymmetries in generic final states, given that the
initial state is $CP$-symmetric. The high center-of-mass
energy provides access to mass states beyond the reach
of the B-factories. The periodic reversal of the D0 solenoid and toroid polarities
results in a cancellation at the first order of most detector-related
asymmetries. In this paper we present a measurement of
the like-sign dimuon charge asymmetry $A$, defined as
\begin{equation}
A \equiv \frac{N^{++} - N^{--}}{N^{++} + N^{--}},
\label{o_defA}
\end{equation}
where $N^{++}$ and $N^{--}$ represent, respectively,
the number of events in which the two muons of
highest transverse momentum satisfying the kinematic selections
have the same positive or negative charge. After removing the
contributions from backgrounds and from residual detector effects, we
observe a net asymmetry that is significantly different from zero.

We interpret this result assuming that the only source of this
asymmetry is the mixing of neutral $B$ mesons that decay semileptonically,
and obtain a measurement of the asymmetry $\aslb$ defined as
\begin{equation}
\aslb \equiv \frac{N^{++}_{b} - N^{--}_{b}}{N^{++}_{b} + N^{--}_{b}},
\end{equation}
where $N^{++}_{b}$ and $N^{--}_{b}$ represent the number of events
containing two $b$ hadrons decaying semileptonically and producing two positive or
two negative muons, respectively.  As shown in Appendix~\ref{theory} each neutral
$\Bq$ meson $(q=d,s)$ contributes a term to this asymmetry given by:
\begin{equation}
\aslq = \frac{\Delta \Gamma_q}{\Delta M_q} \tan \phi_q,
\label{i_phiq}
\end{equation}
where $\phi_q$ is the $CP$-violating phase, and $\Delta M_q$ and $\Delta \Gamma_q$ are
the mass and width differences between the eigenstates of the mass matrices of
the neutral $\Bq$ mesons. The SM predicts the values $\phi_s = 0.0042 \pm 0.0014$ and
$\phi_d = -0.096^{+0.026}_{-0.038}$ \cite{Nierste}. These values set the scale for the
expected asymmetries in the semileptonic decays of $\Bq$ mesons that are negligible
compared to the present experimental sensitivity~\cite{Nierste}. In the standard model $\aslb$ is
\begin{equation}
\aslb({\rm SM}) = (-2.3^{+0.5}_{-0.6}) \times 10^{-4},
\label{in_aslbsm}
\end{equation}
where the uncertainty is mainly due to experimental measurement of the fraction
of $\Bq$ mesons produced in $p \bar p$ collisions at the Tevatron,
and of the parameters controlling the mixing of neutral $B$ mesons.
The $\Bd$ semileptonic charge asymmetry, which constrains the
phase $\phi_d$, has been measured at $e^+ e^-$ colliders~\cite{pdg},
and the most precise results reported by the BaBar and Belle
Collaborations, given in Refs.~\cite{babar,belle}, are in agreement
with the SM prediction. Extensions of the
SM could produce additional contributions to the Feynman box diagrams responsible for
$\Bq$ mixing and other corrections that can provide larger values of
$\phi_q$~\cite{Randall,Hewett,Hou,Soni}. Measurements of $\aslb$ or $\phi_q$ that differ
significantly from the SM expectations would indicate the presence of new physics.

The asymmetry $\aslb$ is also equal to the charge
asymmetry $a^b_{\rm sl}$ of semileptonic decays of $b$ hadrons to muons
of ``wrong charge" (i.e. a muon charge opposite to the charge of the original $b$ quark)
induced through $\Bq\barBq$ oscillations~\cite{Grossman}:
\begin{equation}
a^b_{\rm sl} \equiv \frac{\Gamma(\bar B \to \mu^+ X) - \Gamma(B \to \mu^- X)}
             {\Gamma(\bar B \to \mu^+ X) + \Gamma(B \to \mu^- X)} = \aslb.
\label{asl-w}
\end{equation}

We extract $\aslb$ from two observables. The first is
the like-sign dimuon charge asymmetry $A$ of Eq.~(\ref{o_defA}), and
the second observable is the inclusive muon charge asymmetry $a$ defined as
\begin{equation}
a \equiv \frac{n^+ - n^-}{n^+ + n^-},
\label{asym_a}
\end{equation}
where $n^+$ and $n^-$ correspond to the number of detected positive and negative muons,
respectively.

At the Fermilab Tevatron collider, $b$ quarks are produced mainly in $b \bar{b}$ pairs.
The signal for the asymmetry $A$ is composed of like-sign dimuon events, with one muon
arising from direct semileptonic $b$-hadron decay $b \rightarrow \mu^- X$~\cite{charge},
and the other muon resulting from $\Bq \barBq$ oscillation,
followed by the direct semileptonic $\barBq$ meson decay
$B^0_q \rightarrow \bar{B}^0_q \rightarrow \mu^- X$.
Consequently the second muon has the ``wrong sign"
due to $\Bq\barBq$  mixing.
For the asymmetry $a$, the signal comes from mixing, followed
by the semileptonic decay $B^0_q \rightarrow \bar{B}^0_q \rightarrow \mu^- X$.
The main backgrounds for these measurements arise from events with
at least one muon from kaon or pion decay, or
from the sequential decay of $b$ quarks $b \rightarrow c \rightarrow \mu^+ X$.
For the asymmetry $a$, there is an additional background from direct production
of $c$-quarks followed by their semileptonic decays.


The data used in this analysis
were recorded with the D0 detector~\cite{run2muon, run2det,layer0}
at the Fermilab Tevatron proton-antiproton collider
between April 2002 and June 2009 and correspond to
an integrated luminosity of $6.1 \pm 0.4$~fb$^{-1}$.
The result presented in this Article
supersedes our previous measurement~\cite{D01}
based on the initial data set corresponding
to 1 fb$^{-1}$ of integrated luminosity.
In addition to the larger data set, the main difference between these two analyses
is that almost all quantities in the present measurement
are obtained directly from data, with minimal input from simulation.
To avoid any bias, the central value of the asymmetry was extracted from the full data
set only after all other aspects of the analysis and all systematic
uncertainties had been finalized.

The outline of the paper is as follows.
In Sec.~\ref{strategy},
we present the strategy of the measurement.
The detector and data selections are discussed in Sec.~\ref{selection}, and
in Sec.~\ref{sec_mc} we describe the Monte Carlo simulations
used in this analysis.
Sections~\ref{sec_fk}-\ref{sec_Ab} provide further details.
Section~\ref{sec_ah} presents the results,
Sec.~\ref{sec_consistency} describes consistency checks,
Sec.~\ref{sec_comp} compares the obtained result with other existing measurements,
and, finally, Sec.~\ref{conclusions} gives the conclusions.
Appendices~\ref{theory}--\ref{sec_alt} provide additional technical details on
aspects of the analysis.

\section{Measurement method}
\label{strategy}

We measure the dimuon charge asymmetry $A$ defined in Eq.~(\ref{o_defA})
and the inclusive muon charge asymmetry $a$ of Eq.~(\ref{asym_a}),
starting from a dimuon data sample and an inclusive muon sample
respectively. Background processes and detector asymmetries contribute
to these asymmetries. These contributions are measured directly
in data and used to correct the asymmetries. After applying these corrections,
the only expected source of residual asymmetry in both the inclusive
muon and dimuon samples is from the asymmetry $\aslb$.
Simulations are used to relate the residual asymmetries to the asymmetry \aslb,
and to obtain two independent measurements of \aslb.
These measurements are combined to take advantage of the
correlated contributions from backgrounds, and to reduce the
total uncertainties in the determination of $\aslb$.

The source of the asymmetry $a$ has its nominal origin in
the semileptonic charge asymmetry of neutral $B$ mesons, defined in Eq.~(\ref{asl-w}).
However, various detector and material-related processes also contribute to $n^\pm$.
We classify all muons into two categories according to their origin.
The first category, ``short", denoted in the following as ``$S$",
includes muons from weak decays of $b$ and $c$ quarks and $\tau$ leptons,
and from electromagnetic decays of the short-lived mesons ($\phi$, $\omega$, $\eta$, $\rho^0$).
The muons in the second, ``long", category denoted as ``$L$",
come from decays of charged kaons and pions 
and from other processes: charged kaons, pions, and protons not fully absorbed by the calorimeter and reaching the muon detectors (``punch-through"),
and false matches of central tracks produced by
kaons, pions or protons to a track segment reconstructed in the muon detector. Thus,
the $L$ sample contains only the contribution
from long-lived particles. The total number of muons in the inclusive muon sample is
\begin{equation}
n = n^+ + n^- = n_S + n_L,
\label{n}
\end{equation}
where $n_S$ is the number of $S$ muons, and $n_L$
is the number of $L$ muons.

The initial number of observed
$\mu^+$ (upper signs)
or $\mu^-$ (lower signs) is
\begin{eqnarray}
n^\pm & \propto & f_S (1 \pm a_S) (1 \pm \delta) + f_K (1 \pm a_K) \nonumber \\
& & + f_\pi (1 \pm a_\pi) + f_p (1 \pm a_p) .
\label{inclusive_mu}
\end{eqnarray}
In this expression, the quantity $\delta$ is the charge asymmetry
related to muon detection and identification, $f_K$ is
the fraction of muons from charged kaon decay, punch-through, or
false association with a kaon track, and $a_K$ is their charge asymmetry.
This asymmetry is measured directly in data as described in Sec.~\ref{sec_ak},
and therefore, by definition, includes the contribution from $\delta$.
The analogous quantities $f_\pi$ and $f_p$ represent the fraction of
muons from charged pion decay, punch-through or false muon association
with a pion track, and proton punch-through or false muon association
with a proton track, respectively, while $a_\pi$ and $a_p$ represent the
corresponding charge asymmetries. The fraction $f_p$ also includes a
contribution from the association of falsely identified tracks with muons.
The quantity $f_{\rm bkg} \equiv n_L/(n_S + n_L) = f_K + f_\pi + f_p$
is the $L$ background fraction, $f_S \equiv n_S/(n_S + n_L) = 1 - f_{\rm bkg}$
is the fraction of $S$ muons, and $a_S$ is related to the semileptonic
charge asymmetry $\aslb$, as discussed in Sec.~\ref{sec_Ab}. The charge
asymmetry $a$ can be expressed in terms of these quantities as
\begin{equation}
a = f_S (a_S + \delta) + f_K a_K + f_\pi a_\pi + f_p a_p,
\label{inclusive_mu_a}
\end {equation}
where, because of the small values of $\delta$ and $a_S$,
only terms that depend linearly on the asymmetries are considered.

The most important background term is $f_K a_K$, which measures
the contribution from kaon decay and punch-through. The asymmetry
$a_K$ reflects the fact that the inelastic interaction length of the $K^+$ meson
is greater than that of the $K^-$ meson~\cite{pdg}. This difference arises
from additional hyperon production channels in $K^-$-nucleon reactions,
which are absent for their $K^+$-nucleon analogs. Since the interaction
probability of $K^+$ mesons is smaller, they travel further than $K^-$ in the
detector material, and have a greater chance of decaying to muons, and a
larger probability to punch-through the absorber material thereby
mimicking a muon signal. As a result, the asymmetry $a_K$ is positive.
Since all other asymmetries are at least a factor of  ten smaller than $a_K$,
neglecting the quadratic terms in Eq.~(\ref{inclusive_mu_a}) produces
an impact of $<1$\% on the final result.

In analogy with Eq.~(\ref{n}), the number of like-sign dimuon events
can be written as
\begin{equation}
N = N^{++} + N^{--} = N_{SS} + N_{SL} + N_{LL},
\label{N}
\end{equation}
where $N_{SS}$ ($N_{LL}$) is the number of like-sign dimuon events
with two $S$ ($L$) muons, and, similarly,
$N_{SL}$ is the number of events with one $S$ and one $L$ muon.
A particle producing an $L$ muon can be a kaon, pion or proton, and,
correspondingly, we define
the numbers $N_{SL}^x$ with $x=K, \pi$ and $p$.
In a similar way, we define $N_{LL}^{xy}$ with $x,y = K, \pi, p$.
The corresponding fractions, defined per like-sign dimuon event, are
$F_{SL}^x  \equiv  N_{SL}^x / N$ and $F_{LL}^{xy}  \equiv  N_{LL}^{xy} / N$.
We also define $F_{SS} \equiv N_{SS}/N$, $F_{SL} \equiv N_{SL}/N$, and $F_{LL} \equiv N_{LL}/N$.

The number of observed like-sign dimuon events
$\mu^+ \mu^+$ (upper signs) or $\mu^- \mu^-$ (lower signs) is
\begin{eqnarray}
N^{\pm\pm} & \propto & F_{SS} (1 \pm A_S) (1 \pm \Delta)^2 \nonumber \\
&+ & \sum_{x=K,\pi,p} F_{SL}^x (1 \pm A_x) (1 \pm a_S) (1 \pm \Delta) \nonumber \\
&+ & \sum_{x,y=K,\pi,} \sum_{p;~y \geq x} F_{LL}^{xy} (1 \pm A_x) (1 \pm A_y).
\label{l_s_dimu}
\end{eqnarray}
The charge asymmetry of $N_{SS}$ events contains the
contribution from the expected asymmetry $A_S$ that
we want to measure, and the charge asymmetry $\Delta$
related to the detection and identification of muons.
The asymmetry of the $N_{SL}$ events contains the
contribution of background asymmetries $A_x$ ($x=K,\pi, p$)
for one muon, and the asymmetry  $(1 \pm a_S)(1 \pm \Delta)$
for the other muon. The asymmetry of $N_{LL}$ events contains
the contribution from background asymmetries $A_x$
for both muons. By definition, the detection asymmetry $\Delta$
is included in the values of $A_K$, $A_\pi$, and $A_p$.

Keeping only the terms linear in asymmetries,
the uncorrected dimuon charge asymmetry defined
in Eq.~(\ref{o_defA}) can be expressed as
\begin{eqnarray}
A & \equiv & \frac{N^{++} - N^{--}}{N^{++} + N^{--}} \nonumber \\
  & & = F_{SS} A_S + F_{SL} a_S \nonumber \\
  & & + (2 - F_{\rm bkg}) \Delta + F_K A_K + F_\pi A_\pi + F_p A_p,
\label{dimuon_A}
\end{eqnarray}
where $F_K = F_{SL}^K + F_{LL}^{K \pi} + F_{LL}^{Kp} + 2F_{LL}^{KK}$
is the total number of muons from charged kaon
decay or punch-through per like-sign dimuon event,
and the quantities $F_\pi$ and $F_p$ are defined similarly for charged pions and
protons. The background fraction $F_{\rm bkg}$ is
\begin{equation}
F_{\rm bkg} \equiv F_K + F_\pi + F_p = F_{SL} + 2 F_{LL}.
\label{F_bkg}
\end{equation}
From Eqs.~(\ref{N}) and~(\ref{F_bkg}), it follows that
\begin{equation}
F_{SS} + F_{\rm bkg} - F_{LL} = 1.
\label{sumF}
\end{equation}

As in Eq.~(\ref{inclusive_mu_a}), the largest background contribution in Eq.~(\ref{dimuon_A})
is from the term $F_K A_K$, and all other terms are found to be at least a factor of ten smaller.
The estimated contribution from the neglected quadratic terms in Eq.~(\ref{dimuon_A})
is $\approx 2 \times 10^{-5}$, which corresponds to $\approx 4\%$ of the statistical
uncertainty on $A$.

In the following sections, we determine from data all the
parameters in Eqs.~(\ref{inclusive_mu_a}) and~(\ref{dimuon_A})
used to relate the measured uncorrected asymmetries $a$ and $A$ to
the asymmetries $a_S$ and $A_S$. The detection charge asymmetry
$\Delta$ can differ from $\delta$ due to differences in the muon transverse
momentum $p_T$ and pseudorapidity $\eta$~\cite{rapidity} distributions
of the like-sign dimuon and inclusive muon data samples.
For the same reason, we expect the fractions $f_x$  in Eq.~(\ref{inclusive_mu_a})
and $F_x$ in Eq.~(\ref{dimuon_A}) for $x = K, \pi$ and $p$ to
differ. On physics grounds we expect the asymmetries
$a_x$ and $A_x$ to be identical for any particle of given $p_T$ and
$\eta$. 

All measurements are performed as a function of the muon $p_T$ measured
in the central tracker. The range of $p_T$ values between 1.5 and 25
GeV is divided into five bins, as shown in Table~\ref{tab7}.
The term $f_K a_K$ is obtained by the weighted average of the measured values of
$f_K^i a_K^i$, $i=0,1,2,3,4$, with weights given by the fraction of muons in a given $p_T$
interval, $f_\mu^i$, in the inclusive muon sample:
\begin{equation}
f_K a_K = \sum_{i=0}^4 f_\mu^i f_K^i a_K^i.
\label{fkak}
\end{equation}
Similarly, the term $F_K A_K$ is computed as
\begin{equation}
F_K A_K = \sum_{i=0}^4 F_\mu^i F_K^i a_K^i,
\label{FkAk}
\end{equation}
where $F_\mu^i$ is the fraction of muons in a given $p_T$ interval in the like-sign dimuon
sample. 
Since the kaon asymmetry is determined by the properties of
the particle and not those of the event, we use the same asymmetry $a_K^i$ for a given
$p_T$ interval in both the inclusive muon and the like-sign dimuon sample.
We verify in Sec.~\ref{sec_consistency} that the final result does not depend
significantly on muon $\eta$, nor upon kinematic properties of events,
luminosity or the mass of the $\mu \mu$ system.
The definition of the muon $p_T$ intervals and the values of $f_\mu^i$ and $F_\mu^i$ are given in Table~\ref{tab7}.
The same procedure is applied to all other terms in Eqs.~(\ref{inclusive_mu_a})
and~(\ref{dimuon_A}), e.g.,
\begin{equation}
 (2 - F_{\rm bkg}) \Delta = \sum_{i=0}^4 F_\mu^i (2 - F^i_{\rm bkg}) \delta_i.
\end{equation}

\begin{table}
\caption{\label{tab7}
Fractions of muon candidates in the inclusive muon
($f_\mu^i$) and in the like-sign dimuon ($F_\mu^i$, with two
entries per event) samples.
}
\begin{ruledtabular}
\newcolumntype{A}{D{A}{\pm}{-1}}
\newcolumntype{B}{D{B}{-}{-1}}
\begin{tabular}{cBcc}
Bin & \multicolumn{1}{c}{Muon $p_T$ range (GeV)}       &  $f_\mu^i$ & $F_\mu^i$ \\
\hline
0  & 1.5\ B \ 2.5 & 0.0055 & 0.0442 \\
1  & 2.5\ B \ 4.2 & 0.1636 & 0.2734 \\
2  & 4.2\ B \ 7.0 & 0.6587 & 0.5017 \\
3  & 7.0\ B \ 10.0 & 0.1175 & 0.1238 \\
4  & 10.0\ B \ 25.0 & 0.0547 & 0.0569 \\
\end{tabular}
\end{ruledtabular}
\end{table}

As in the case of $a_S$, the source of the asymmetry $A_S$
is the charge asymmetry in semileptonic $B$-meson decays.
Thus, two independent measurements of $\aslb$ can be performed using the inclusive
muon and like-sign dimuon data samples. The asymmetry $a_S$ is dominated by detector effects,
mostly due to the asymmetry arising from the different interaction lengths of charged kaons.
However, $A_S$ is far more sensitive to the asymmetry $\aslb$ because of the
definition of $A$ in Eq.~(\ref{o_defA}), which has the number of like-sign dimuon events,
rather than all dimuon events in the denominator. Although a weighted average of these
$\aslb$ measurements can be made, we take advantage of correlations among  backgrounds and
asymmetries to further improve the precision of $\aslb$ through a linear combination of
$A$ and $a$. In this combination, which is discussed in Sec.~\ref{sec_ah}, the detector
effects and related systematic uncertainties cancel to a large degree, resulting
in an improved measurement of $\aslb$.

\section{Detector and data selection}
\label{selection}

The D0 detector is described
in Refs.~\cite{run2muon, run2det,layer0}.
It consists of a magnetic central-tracking system that
comprises a silicon microstrip tracker (SMT) and a central fiber
tracker (CFT), both located within a $1.9$~T superconducting solenoidal
magnet~\cite{run2det}. The SMT has $\approx$ 800,000 individual strips,
with a typical pitch of $50-80$ $\mu$m, and a design optimized for
tracking and vertexing for $|\eta|<2.5$.
The system has a six-barrel longitudinal structure, each with a set
of four layers arranged axially around the beam pipe, and interspersed
with 16 radial disks. In the spring of 2006, a ``Layer 0" barrel detector with
$12288$ additional strips was installed~\cite{layer0},
and two radial disks were removed.
The sensors of Layer 0 are located at a radius of $17$ mm from the colliding beams.
The CFT has eight thin coaxial barrels, each
supporting two doublets of overlapping scintillating fibers of 0.835~mm
diameter, one doublet parallel to the collision axis, and the
other alternating by $\pm 3^{\circ}$ relative to the axis. Light signals
are transferred via clear fibers to visual light photon counters (VLPCs)
that have $\approx 80$\% quantum efficiency.

The muon system~\cite{run2muon} is located beyond the liquid Argon-Uranium
calorimeters that surround the central tracking system, and consists of
a layer A of tracking detectors and scintillation trigger counters
before 1.8~T iron toroids, followed by two similar layers B and C after
the toroids. Tracking for $|\eta|<1$ relies on 10-cm wide drift
tubes, while 1-cm minidrift tubes are used for
$1<|\eta|<2$.

The trigger and data acquisition systems are designed to handle
the high instantaneous luminosities. Based on information from
tracking, calorimetry, and muon systems, the output of the first level
of the trigger is used to limit the rate for accepted events to
$< 2$~kHz. At the next trigger stage, with more refined
information, the rate is reduced further to $<1$~kHz. These
first two levels of triggering rely mainly on hardware and firmware.
The third and final level of the trigger, with access to full event
information, uses software algorithms and a computing farm, and reduces
the output rate to $<200$~Hz, which is written to tape.

The single muon and dimuon triggers used in this analysis are based on the information provided
by the muon detectors, combined with the tracks reconstructed by the
tracking system.
The single muon triggers with the lowest $p_T$ threshold are prescaled at high instantaneous luminosity,
have a higher average $p_T$ threshold than the dimuon triggers and
cover a smaller range of pseudorapidity than the dimuon triggers.

In this analysis we select events with one or two muons.
We therefore first apply track selections, and then require
either one or two muons.

\textit{Track selection:} we select tracks with $p_T$
in the range $ 1.5 < p_T < 25$~GeV and $|\eta| < 2.2$.
The upper limit on the transverse momentum is applied to suppress the contribution
of muons from $W$ and $Z$ boson decays. To ensure that the muon candidate can pass through
the detector, including all three layers of the muon system,
we require either $p_T > 4.2$~GeV or a longitudinal momentum component $|p_z| > 6.4$~GeV.
The selected tracks have to satisfy the following quality requirements:
at least 2 axial and 1 stereo hits in the SMT,
and at least 3 axial and 3 stereo hits in the CFT.
The primary interaction vertex closest to this track must contain at least
five charged particles. This vertex is determined for each event using all reconstructed tracks.
The average position of the collision point in the plane transverse to the beam is measured
for each run and is used as a constraint. The precision of the primary vertex reconstruction for
each event is on average $\approx 20$ $\mu$m in the transverse plane
and  $\approx 40$ $\mu$m along the beam direction.
The transverse impact parameter of the selected track relative
to the closest primary vertex must be $<0.3$ cm, with the longitudinal distance from
the point of closest approach to this vertex $<0.5$ cm.

\textit{Single muon selection:} the selected track must have a matching
track segment reconstructed in the muon system, with at least two hits in the layer A
chambers, at least two hits in the layer
B or C chambers, and at least one scintillator hit associated with the track.
The $\chi^2$ for the difference between the track parameters
measured in the central tracker and in the muon system must be less than 40
(with 5 d.o.f.);
the measured time in at least one
of the scintillators associated with the muon candidate must be
within 5 ns of the
expected time. 
The muon is assigned the charge of the track reconstructed
in the central tracker. For muon $p_T < 25$~GeV, the fraction of muons
with mismeasured charge and their contribution to the asymmetries are
found to be negligible. The scintillator
timing and the track impact parameter requirements reduce the
background from cosmic rays and from beam halo to a negligible level.

\textit{Dimuon selection:} The two highest transverse momentum
muons in the event must pass all the
selections described above, and be associated to the same interaction
vertex, applying the same requirements
on the transverse impact parameter and on the distance of closest
approach to the primary vertex along the beam axis used in the single muon
selection. To remove events in which the two muons
originate from the decay of the same $b$ hadron, we require that
the invariant mass of the two muons be  $>2.8$~GeV.

These requirements define the \textit{reference selections},
which are changed while performing consistency checks of the analysis. Unless
stated otherwise all figures, tables and results in this article refer
to these \textit{reference selections}.

This analysis uses two data samples. The \textit{inclusive muon sample}
contains all events with at least one muon candidate
passing the muon selection and at least one single muon trigger.
If an event contains more than one muon,
each muon is included in the inclusive muon sample.
Such events constitute about 0.5\% of the total inclusive muon sample.
The \textit{like-sign dimuon sample} contains all events
with at least two muon candidates of the same charge
that pass the reference dimuon selection and at least one dimuon trigger.
If more than two muons pass the single muon selection, the two muons with the highest $p_T$
are selected for inclusion in the dimuon sample.
Such events comprise $\approx 0.7$\% of the total like-sign dimuon sample.

The polarities of the toroidal and solenoidal magnetic fields are reversed
on average every two weeks so that the four solenoid-toroid polarity
combinations are exposed to approximately the same
integrated luminosity. This allows for a cancellation of first order
effects related with the instrumental asymmetry~\cite{D01}. To ensure such cancellation,
the events are weighted according to the integrated luminosity for each
dataset corresponding to a different configuration of the magnets' polarities.
These weights are given in Table~\ref{tab00}.

\begin{figure}
\begin{center}
\includegraphics[width=0.50\textwidth]{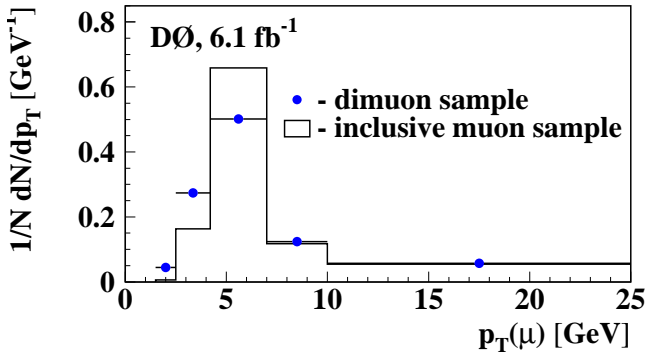}
\caption{ The normalized muon $p_T$ distribution.
The points correspond to
the like-sign dimuon sample and the histogram correspond to the inclusive
muon sample. The distribution for the like-sign
dimuon sample contains two entries per event.
}
\label{fig_distr}
\end{center}
\end{figure}

The normalized $p_T$ distributions of muons in the selected
data samples are shown in Fig.~\ref{fig_distr}. Differences in these distributions
are caused by the trigger requirements.

\begin{table}
\caption{\label{tab00}
Weights assigned to the events with different solenoid and toroid polarities
in the inclusive muon and like-sign dimuon samples.
}
\begin{ruledtabular}
\newcolumntype{A}{D{A}{\pm}{-1}}
\newcolumntype{B}{D{B}{-}{-1}}
\begin{tabular}{cccc}
Solenoid & Toroid   & Weight & Weight \\
polarity & polarity & inclusive muon & like-sign dimuon \\
\hline
$-1$ & $-1$ & 0.895 & 0.879 \\
$-1$ & +1 & 1.000 & 1.000 \\
+1 & $-1$ & 0.954 & 0.961 \\
+1 & +1 & 0.939 & 0.955
\end{tabular}
\end{ruledtabular}
\end{table}

\section{Monte Carlo simulation}
\label{sec_mc}

Since almost all quantities are extracted from data, the MC simulations
are used in only a limited way. The simulations of QCD processes used in this analysis are:
\begin{itemize}
\item Inclusive $p \bar p$ collisions containing a minimum transverse
energy $E^{\rm min}_T > 10$ or $20$~GeV at the generator level.
\item Inclusive $p \bar p \to b \bar{b} X$ and $p \bar p \to c \bar{c} X $
final states containing a muon,
with an additional requirement that the $b$ or $c$ quark has transverse
momentum $p_T > 3$ GeV, and that the produced muon has $p_T > 1.5$ GeV and $|\eta| < 2.1$.
\end{itemize}
The samples with different $E^{\rm min}_T$ are used to study the impact
of the kinematics of generated events on the parameters extracted from the simulation.

In all cases we use the {\sc pythia v6.409}~\cite{pythia}
event generator,
interfaced to the {\sc evtgen} decay package~\cite{evtgen} and
the {\sc CTEQ6L1}~\cite{pdf} parton distribution functions.
The generated events are propagated through the D0 detector
using a {\sc geant}~\cite{geant}
based program with full detector simulation.
The response in the detector is digitized, and
the effects of multiple interactions at high luminosity are modeled by overlaying
hits from randomly triggered $p\bar{p}$ collisions on the digitized hits from MC.
The complete events are reconstructed with the same program as used for data, and,
finally, analyzed using the same selection criteria described above for data.

\section{Measurement of $f_K$, $F_K$}
\label{sec_fk}

A kaon, pion, or proton can be misidentified as a muon and thus contribute to the inclusive muon
and the like-sign dimuon samples.
This can happen because of pion and kaon decays in flight,
punch-through, or muon misidentification. 
We do not distinguish these individual processes, but rather measure the total fraction
of such particles using data.
In the following, the notation \ktomu~ stands for the phrase ``kaon misidentified as a muon,"
and the notations \pitomu~ and \ptomu~ have corresponding meanings for pions and protons.
In this Section we discuss the measurement of $f_K$ and $F_K$. The measurement
of the corresponding factors for pions and protons and of the asymmetries are
discussed in the following Sections.

The fraction $f_K$ in the inclusive muon sample is measured using
$K^{*0} \to K^+ \pi^-$ decays~\cite{charge} with \ktomu. The fraction $f_{K^{*0}}$
of these decays is related to the fraction $f_K$ by
\begin{equation}
f_{K^{*0}} =  \varepsilon_0 f_K R(K^{*0}) ,
\label{kstf}
\end{equation}
where $R(K^{*0})$ is the fraction of all kaons that result from $K^{*0} \to K^+ \pi^-$ decays, and
$\varepsilon_0$ is the efficiency to reconstruct the pion from the $K^{*0} \to K^+ \pi^-$
decay, provided that the \ktomu~ track is reconstructed.

We also select $K_S$ mesons and reconstruct $K^{*+} \to K_S \pi^+$ decays. The number of
these decays is
\begin{equation}
N(K^{*+} \to K_S \pi^+) =  \varepsilon_c N(K_S) R(K^{*+}) ,
\end{equation}
where $R(K^{*+})$ is the fraction of $K_S$ that result from $K^{*+} \to K_S \pi^+$ decays, and
$\varepsilon_c$ is the efficiency to reconstruct the additional pion in the
$K^{*+} \to K_S \pi^+$ decay, provided that the $K_S$ meson is reconstructed.
We use isospin invariance to set
\begin{equation}
R(K^{*0}) = R(K^{*+}).
\label{assume1}
\end{equation}
This relation is also confirmed by data as
discussed in Sec.~\ref{sec_syst}. We apply the same kinematic selection criteria to
the charged kaon and $K_S$ candidates, and use exactly the same criteria to select an additional
pion and reconstruct the $K^{*0} \to K^+ \pi^-$ and $K^{*+} \to K_S \pi^+$ decays.
Therefore we set
\begin{equation}
\varepsilon_0 = \varepsilon_c.
\label{assume2}
\end{equation}
This relation is confirmed by simulation. We assign a systematic uncertainty related
to this relation, as discussed in Sec.~\ref{sec_syst}.
From Eqs.~(\ref{kstf})--(\ref{assume2}), we obtain
\begin{equation}
f_K = \frac{N(K_S)}{N(K^{*+} \to K_S \pi^+)} f_{K^{*0}}.
\label{fkst}
\end{equation}

We use a similar relation to obtain the quantity $F_K$ of \ktomu~ tracks in the like-sign
dimuon sample:
\begin{equation}
F_K = \frac{N(K_S)}{N(K^{*+} \to K_S \pi^+)} F_{K^{*0}},
\label{Fkst}
\end{equation}
where $F_{K^{*0}}$ is the fraction of $\kstneu \to K^+ \pi^-$ decays with
\ktomu~ in the like-sign dimuon sample. The numbers $N(K_S)$ and $N(K^{*+} \to K_S \pi^+)$
are obtained from the inclusive muon sample.

Since the kaon track parameters must be known to reconstruct the $K^{*0}$ meson,
these measurements of $f_K$ and $F_K$ require the kaons to decay after being
reconstructed in the central tracking system. A small number of kaon decays occur
close to the interaction point, so that the muon track is reconstructed by the
tracker. These muons are counted in the inclusive
muon and the like-sign dimuon samples, but do not contribute to the measurement
of the \ktomu~ fraction, because their parameters differ significantly from the parameters of the
original kaon, and they do not produce  a narrow $K^{*0}$ meson peak.
The fractions $F_K$ and $f_K$ measured in exclusive decays are therefore
divided by a factor $C$ that corresponds to the fraction of correctly
reconstructed kaons among all \ktomu~ tracks. This factor is calculated
from simulation as
\begin{equation}
C = 0.938 \pm 0.006.
\label{coeffC}
\end{equation}
Since the mean decay length of kaons in the laboratory frame is much longer than the size
of the D0 detector, the value of $C$ is determined mainly by the detector geometry, and its
value is similar for both $K \to \mu$ and $\pi \to \mu$ tracks.
Therefore, we use the same coefficient $C$ for the computation of the fraction of
$\pi \to \mu$ described in Sec.~\ref{sec_fpi}.
The difference in this coefficient for kaon and pion tracks observed in simulation is taken
as the uncertainty on its value.
The uncertainties from the event generation and reconstruction
produce a smaller impact on this coefficient.

\begin{figure}
\begin{center}
\includegraphics[width=0.50\textwidth]{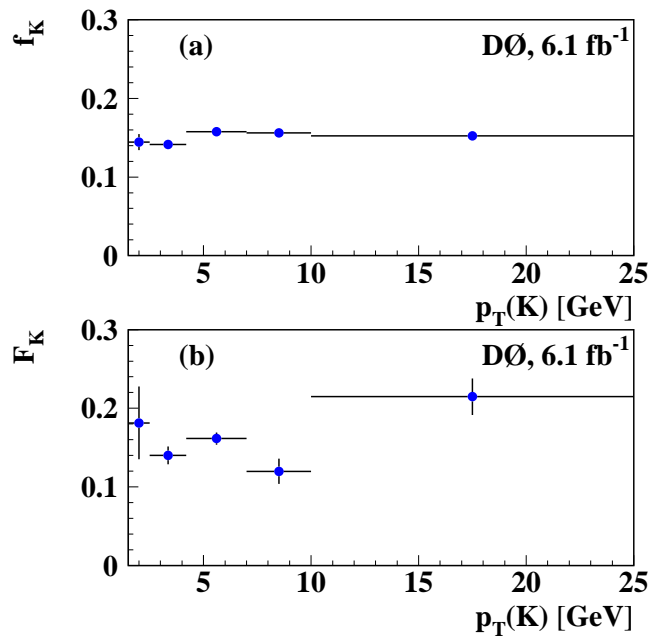}
\caption{The fraction of \ktomu~ tracks in the inclusive muon sample (a)
and the like-sign dimuon sample (b), both as a function of the $p_T$ of the kaon.}
\label{fig_fk1}
\end{center}
\end{figure}

Details of $K_S \to \pi^+ \pi^-$, $K^{*0} \to K^+ \pi^-$, and $K^{*+} \to K_S \pi^+$ selections
and the fitting procedure to measure the number of these decays are given
in Appendix~\ref{sec_fits}.
All quantities in Eqs.~(\ref{fkst}) and~(\ref{Fkst}) are obtained as a
function of the measured transverse momentum of the kaon.
The measured number of $K^{*0} \to K^+ \pi^-$ decays with $\ktomu$
in a given $p_T$ range is normalized by the total number of muons in that interval.
The fraction $F_{\kstneu}$ includes a multiplicative factor of two,
because there are two muons in a like-sign dimuon event, and
by definition it is normalized to the number
of like-sign dimuon events.
Figure~\ref{fig_fk1} and Table~\ref{tab3} give the resulting fractions $f_K$ and $F_K$ for different
$p_T$ bins. Only statistical uncertainties are given; systematic
uncertainties are discussed in Sec.~\ref{sec_syst}.

\begin{table}
\caption{\label{tab3}
Fractions $f_K$ and $F_K$ for different muon $p_T$ bins.
The correspondence between the bin number and the $p_T$ range is given in Table~\ref{tab7}.
The last line shows the weighted average of these quantities
obtained with weights given by the fraction of muons in a given $p_T$
interval $f_\mu^i$ ($F_\mu^i$) in the inclusive muon (dimuon) sample.
Only the statistical uncertainties are given.
}
\begin{ruledtabular}
\newcolumntype{A}{D{A}{\pm}{-1}}
\newcolumntype{B}{D{B}{-}{-1}}
\begin{tabular}{cAA}
Bin &  \multicolumn{1}{c}{$f_K \times 10^2$} & \multicolumn{1}{c}{$F_K \times 10^2$} \\
\hline
0     & 14.45\ A \ 1.02 & 18.13\ A \ 4.62  \\
1     & 14.14\ A \ 0.26 & 14.00\ A \ 1.14  \\
2     & 15.78\ A \ 0.20 & 16.14\ A \ 0.77  \\
3     & 15.63\ A \ 0.35 & 11.97\ A \ 1.60  \\
4     & 15.26\ A \ 0.56 & 21.47\ A \ 2.31  \\ \hline
All   & 15.46\ A \ 0.14 & 15.38\ A \ 0.57
\end{tabular}
\end{ruledtabular}
\end{table}

\section{Measurement of $P(\pitomu)/P(\ktomu)$ and $P(\ptomu)/P(\ktomu)$}
\label{sec_prob}

The probability $P(\ktomu)$ for a kaon to be misidentified as a muon
is measured using $\phi \to K^+ K^-$ decays. Similarly, we use
$K_S \to \pi^+ \pi^-$ and $\Lambda \to p \pi^-$ decays~\cite{charge} to
measure the probabilities $P(\pitomu)$ and $P(\ptomu)$, respectively.
In all cases we measure the number $N_\mu$
of decays in which the candidate particle
satisfies the muon selection criteria defined in Sec.~\ref{selection},
and the number of decays $N_{\rm tr}$ in which the tested particle satisfies
the track selection criteria. When both kaons (pions) from $\phi \to K^+ K^-$
($K_S \to \pi^+ \pi^-$) satisfy the selection criteria, they contribute twice.
The details of the event selections and of the fitting procedure used to extract
the number of $\phi$, $K_S$, and $\Lambda$ decays are given in Appendix~\ref{sec_fits}.
The ratio of $N_\mu(\phi)$ to $N_{\rm tr}(\phi)$ defines $P(\ktomu) \varepsilon(\mu)$,
where $\varepsilon(\mu)$ is the efficiency of muon identification. In the same way,
the ratio of $N_\mu(K_S)$ to $N_{\rm tr}(K_S)$ yields $P(\pitomu) \varepsilon(\mu)$,
and the ratio of $N_\mu(\Lambda)$ to $N_{\rm tr}(\Lambda)$ gives the quantity
$P(\ptomu) \varepsilon(\mu)$. The ratios
$P(\pitomu)/P(\ktomu)$ and $P(\ptomu)/P(\ktomu)$ are obtained from
\begin{eqnarray}
\frac{P(\pitomu)}{P(\ktomu)} & = & \frac{N_\mu(K_S)/N_{\rm tr}(K_S)}
{N_\mu(\phi)/N_{\rm tr}(\phi)}, \nonumber \\
\frac{P(\ptomu)}{P(\ktomu)} & = & \frac{N_\mu(\Lambda)/N_{\rm tr}(\Lambda)}
{N_\mu(\phi)/N_{\rm tr}(\phi)}.
\label{prob}
\end{eqnarray}

Since the initial selection for this measurement requires at least
one identified muon, we determine all these quantities
in the sub-sample of single muon triggers that contain
at least one muon not associated with
the \ktomu, \pitomu~, or \ptomu~ transitions.

\begin{figure}
\begin{center}
\includegraphics[width=0.50\textwidth]{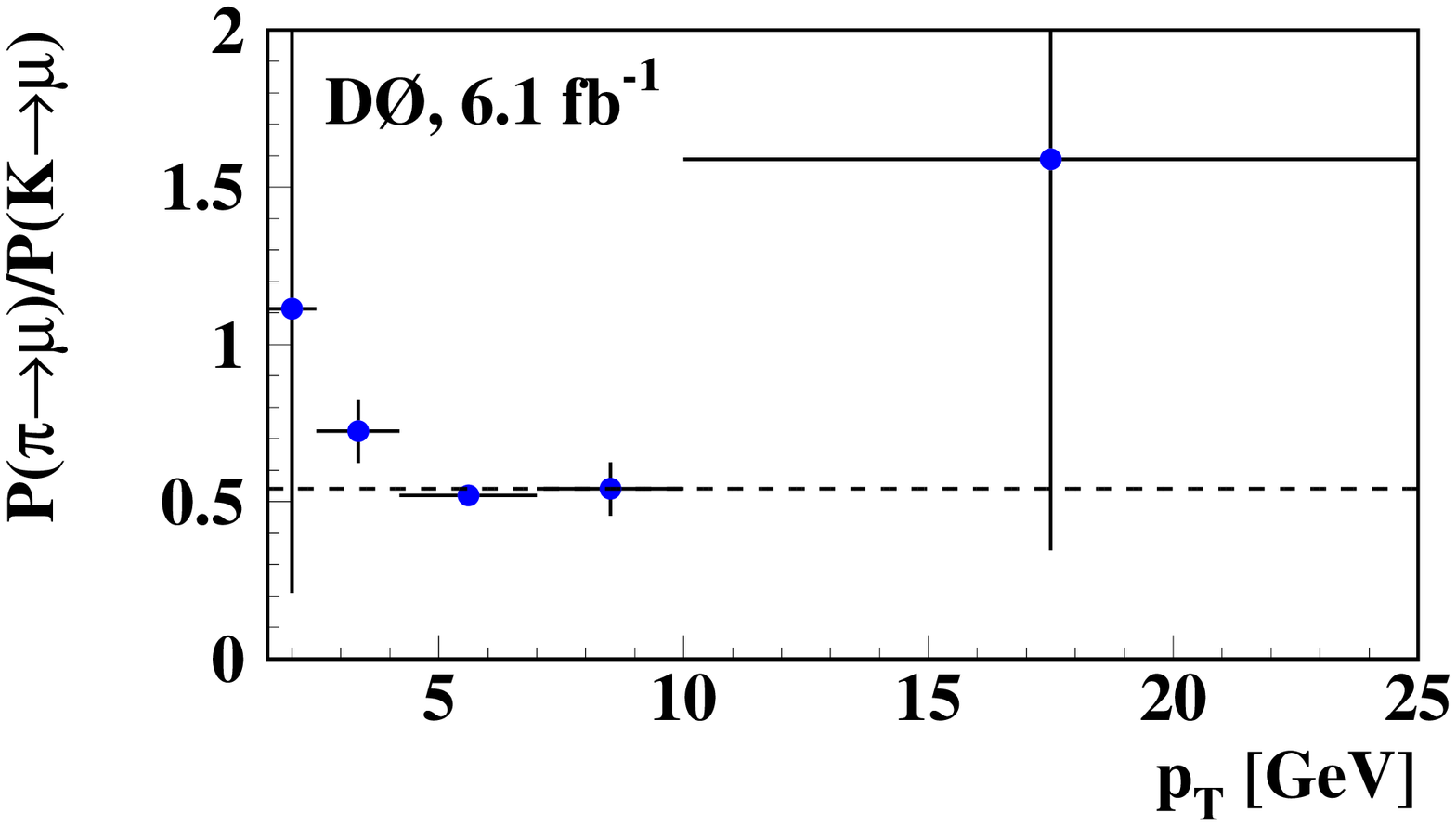}
\caption{The ratio $P(\pitomu)/P(\ktomu)$  as a function of the hadron transverse momentum.
The horizontal dashed line shows the mean value of this ratio.
}
\label{fig_ppimu}
\end{center}
\end{figure}

\begin{figure}
\begin{center}
\includegraphics[width=0.50\textwidth]{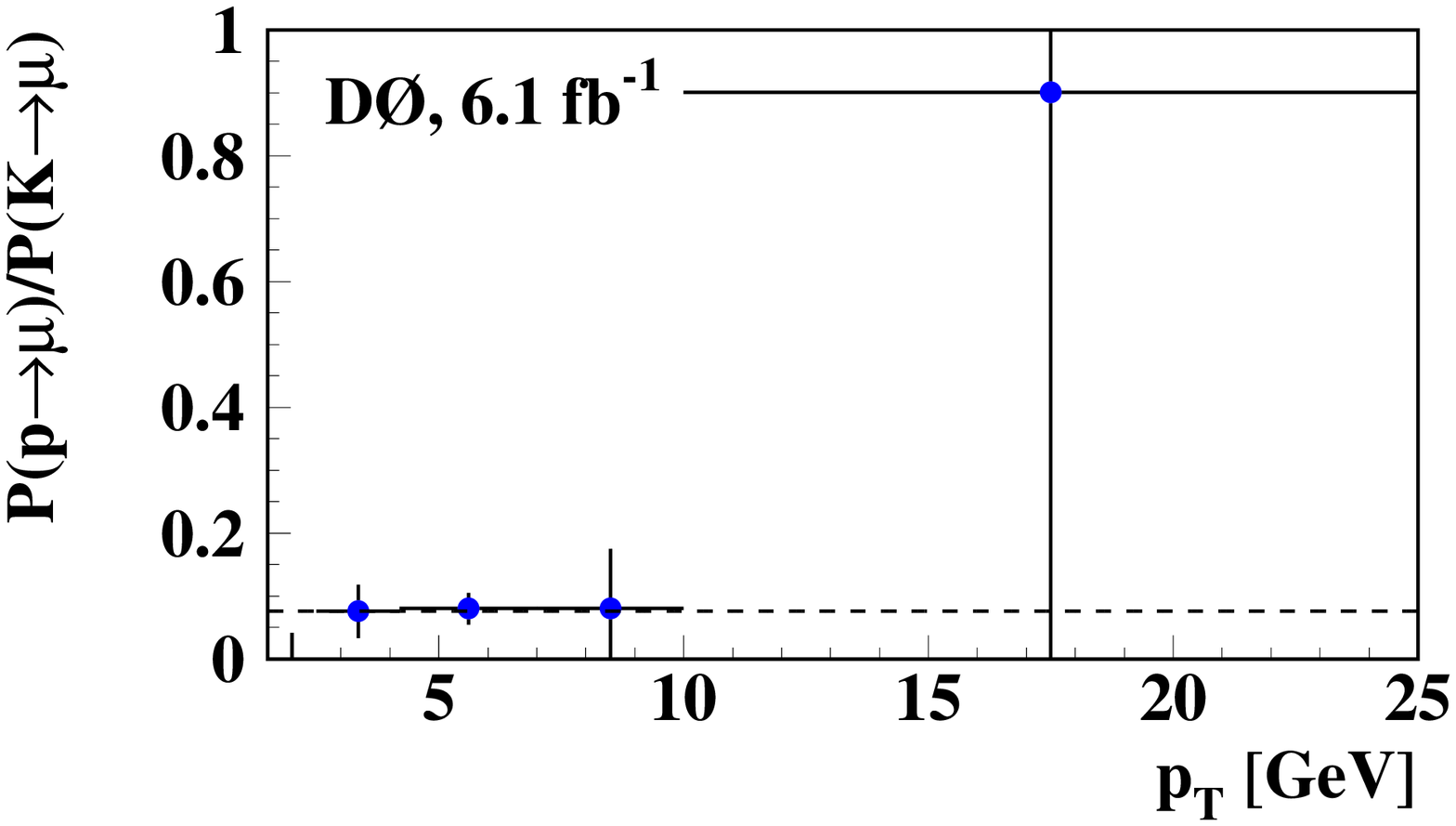}
\caption{The ratio $P(\ptomu)/P(\ktomu)$  as a function of the particle transverse momentum.
The horizontal dashed line shows the mean value of this ratio.
}
\label{fig_ppmu}
\end{center}
\end{figure}

We measure all these parameters as a function of the original particle's transverse momentum.
Figures~\ref{fig_ppimu} and~\ref{fig_ppmu} show the ratio $P(\pitomu)/P(\ktomu)$ and
$P(\ptomu)/P(\ktomu)$ respectively, with the mean values averaged
over $p_T$ determined to be
\begin{eqnarray}
P(\pitomu)/P(\ktomu) & = 0.540 \pm 0.029, \label{rpitomu} \\
P(\ptomu)/P(\ktomu) & = 0.076 \pm 0.021.\label{rptomu}
\end{eqnarray}
The dominant uncertainty in Eqs.~(\ref{rpitomu}) and~(\ref{rptomu}) stems from the limited
statistics of data, and the contribution of all other uncertainties is much smaller.
The probability of a pion to be misidentified as a muon is much larger than that of proton because
the dominant contribution to this probability comes from the $\pi^- \to \mu^- \bar{\nu}$ decay.
The measured ratios~(\ref{rpitomu}) and~(\ref{rptomu}) agree well with the results obtained
from MC, where we obtain $P(\pitomu)/P(\ktomu)({\rm MC}) = 0.530 \pm 0.011$ and
$P(\ptomu)/P(\ktomu)({\rm MC}) = 0.050 \pm 0.003$.

\section{Measurement of $f_\pi$, $f_p$, $F_\pi$, $F_p$}
\label{sec_fpi}

The fraction $f_\pi$ of \pitomu~ tracks in the inclusive muon sample can be expressed as
\begin{equation}
f_\pi = f_K \frac{P(\pitomu)}{P(\ktomu)} \frac{n_\pi}{n_K},
\end{equation}
where the measurement of the fraction $f_K$ is described in Sec.~\ref{sec_fk}, that of the ratio
$P(\pitomu)/P(\ktomu)$ in Sec.~\ref{sec_prob}, and the
quantities $n_\pi$ and $n_K$ are the mean multiplicities of pions and kaons in $p \bar{p}$ interactions.
In a similar way, the fraction $f_p$ of \ptomu~ tracks is determined from
\begin{equation}
f_p = f_K \frac{P(\ptomu)}{P(\ktomu)} \frac{n_p + n_{\rm f}}{n_K},
\end{equation}
where $n_p$ is the average number of protons produced
  in $p \bar{p}$ interactions.  We include in the fraction $f_p$
  the contribution from the number $n_f$ of false tracks, reconstructed
  from random combinations of hits.
The impact of false tracks on the final result is found to be small
and is taken into account in the systematic uncertainty.

\begin{figure}
\begin{center}
\includegraphics[width=0.50\textwidth]{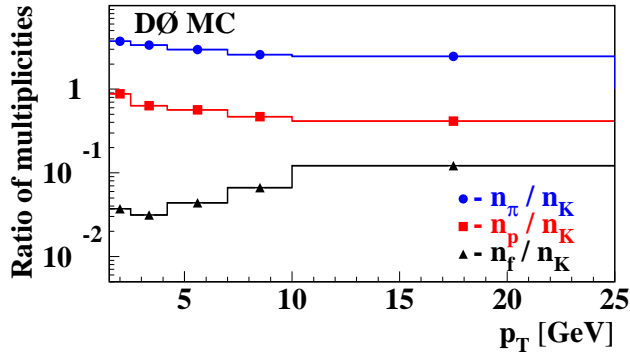}
\caption{The ratio of multiplicities $n_\pi/n_K$, $n_p/n_K$ and $n_{\rm f}/n_K$
as a function of the transverse momentum obtained from {\sc pythia}.
}
\label{fig_mult}
\end{center}
\end{figure}

The values of  $n_K$, $n_\pi$, $n_p$, and $n_{\rm f}$ are taken from the {\sc pythia} simulation
of inclusive hadronic interactions. We count the number of particles satisfying
the track selection criteria in the simulated interactions, and obtain the dependence
of the ratios $n_\pi/n_K$, $n_p/n_K$ and $n_{\rm f}/n_K$ on the particle
$p_T$ shown in Fig.~\ref{fig_mult}.

Both $f_\pi$ and $f_p$ are measured as a function of the particle $p_T$.
However, they are poorly defined in the first and last bins due to low statistics.
Therefore, we combine these quantities for bins 0 and 1 and for bins 3 and 4.

\begin{figure}
\begin{center}
\includegraphics[width=0.50\textwidth]{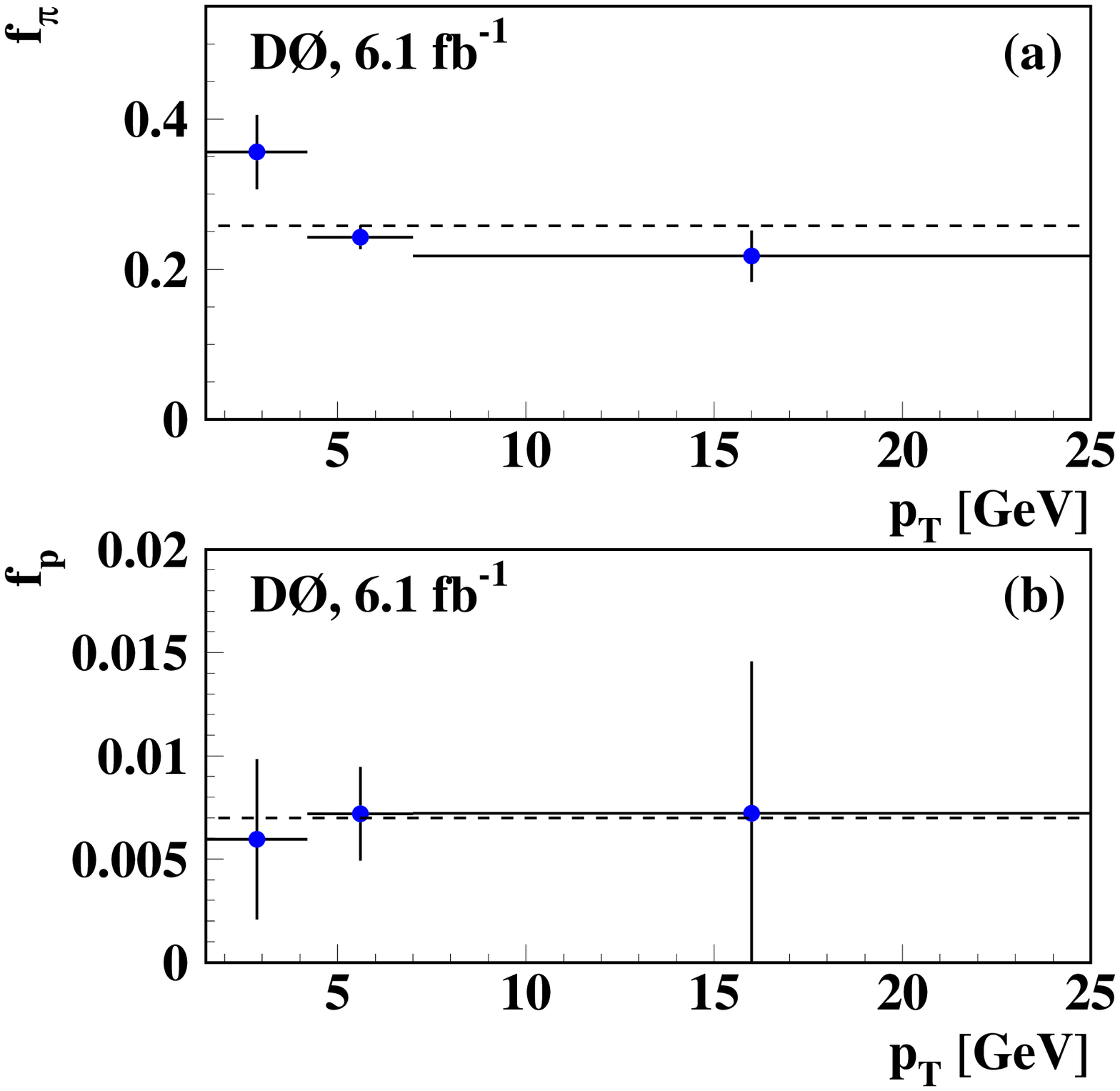}
\caption{The fraction of (a) \pitomu~ tracks
and (b) \ptomu~ tracks in the inclusive muon sample
as a function of the track transverse momentum. The horizontal dashed
lines show the mean values of these fractions.}
\label{fig_fpi1}
\end{center}
\end{figure}

Figure~\ref{fig_fpi1} and Table~\ref{tab4} provide the measured fractions $f_\pi$ and $f_p$
for different $p_T$ bins. Only statistical uncertainties are given.
The systematic uncertainties related to these quantities are discussed
in Sec.~\ref{sec_syst}.

\begin{table}
\caption{\label{tab4}
Fractions $f_\pi$, $f_p$, $F_\pi$, and $F_p$ for different $p_T$ bins.
The correspondence between the bin number and the momentum range is given in
Table~\ref{tab7}.
The last line shows the weighted averages obtained with
weights given by the fraction of muons in a given $p_T$
interval $f_\mu^i$ in the inclusive muon sample. Only the statistical uncertainties are given.
}
\begin{ruledtabular}
\newcolumntype{A}{D{A}{\pm}{-1}}
\newcolumntype{B}{D{B}{^+}{-1}}
\begin{tabular}{cAAAA}
Bin &  \multicolumn{1}{c}{$f_\pi \times 10^2$} & \multicolumn{1}{c}{$f_p \times 10^2$} &
\multicolumn{1}{c}{$F_\pi \times 10^2$} & \multicolumn{1}{c}{$F_p \times 10^2$} \\
\hline
0--1  & 35.6\ A \ 4.9 &  0.6\ A \ 0.4 &  32.2\ A \ 5.1  &  0.5\ A \ 0.4  \\
2     & 24.3\ A \ 1.6 &  0.7\ A \ 0.2 &  22.4\ A \ 1.8  &  0.7\ A \ 0.2  \\
3--4  & 21.7\ A \ 3.4 &  0.7\ A \ 0.7 &  19.0\ A \ 3.4  &  0.6\ A \ 0.6  \\ \hline
All   & 25.8\ A \ 1.4 &  0.7\ A \ 0.2 &  24.9\ A \ 1.5  &  0.6\ A \ 0.2
\end{tabular}
\end{ruledtabular}
\end{table}

\begin{figure}
\begin{center}
\includegraphics[width=0.50\textwidth]{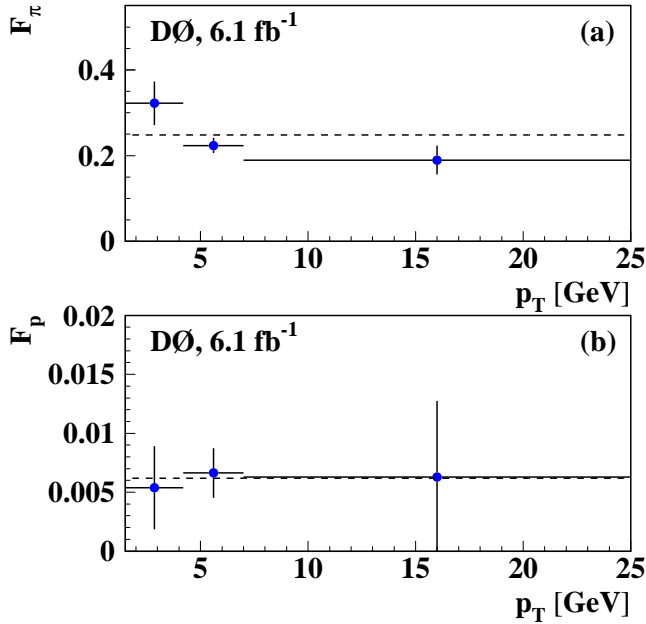}
\caption{The fraction of (a) \pitomu~ tracks
and (b) \ptomu~ tracks in the like-sign dimuon sample as a function of
the track transverse momentum. The horizontal dashed lines show the mean
values of these fractions.}
\label{fig_fpi2}
\end{center}
\end{figure}

The fractions $F_\pi$ and $F_p$ in the like-sign dimuon sample
are determined in a similar way:
\begin{eqnarray}
F_\pi & = & F_K \frac{P(\pitomu)}{P(\ktomu)} \frac{N_\pi}{N_K}, \nonumber \\
F_p & = & F_K \frac{P(\ptomu)}{P(\ktomu)} \frac{N_p + N_{\rm f}}{N_K},
\end{eqnarray}
where the quantities $N_K$, $N_\pi$, $N_p$, and $N_{\rm f}$ represent the average numbers of
kaons, pions, protons and false tracks 
for events with two identified muons with the same charge. The simulation
shows that the ratio $N_\pi / N_K$ can be approximated as
\begin{equation}
\frac{N_\pi}{N_K} = (0.90 \pm 0.05) \frac{n_\pi}{n_K}.
\label{npi}
\end{equation}
The main uncertainty in this value is due to the simulation of pion and
kaon multiplicities in $p \bar p$ interactions and is discussed in Sec.~\ref{sec_syst}.
The ratio $N_p/N_K$ is also consistent with the factor given in Eq.~\ref{npi}.
The value of $N_\pi/N_K$ is smaller than that of $n_\pi/n_K$ because the main contribution
in the sample with one identified muon comes from semileptonic decays of $b$
and $c$ quarks, which usually also contain at least one kaon.
Since the number of simulated events with one identified muon is small,
we obtain the ratios $N_\pi/N_K$ and $(N_p+N_{\rm f})/N_K$ using the approximation
of Eq.~(\ref{npi}), i.e., multiplying the quantities
$n_\pi/n_K$ and $(n_p+n_{\rm f})/n_K$ by the factor $0.90 \pm 0.05$.
Figure~\ref{fig_fpi2} and Table~\ref{tab4} give the fractions $F_\pi$ and $F_p$
for different $p_T$ bins. Only the statistical uncertainties of the
simulation are given. The systematic uncertainties related to these
quantities are discussed in Sec.~\ref{sec_syst}. As in the case of $f_\pi$ and $f_p$,
the mean value of these quantities are used in bins 0 and 1 and for bins 3 and 4.

\section{Systematic uncertainties of background fractions}
\label{sec_syst}

We use Eqs.~(\ref{assume1}) and~(\ref{assume2}) to derive the fractions $f_K$
and $F_K$, and verify the validity of Eq.\ (\ref{assume2}) in simulation, and find
that
\begin{equation}
(\varepsilon_c / \varepsilon_0)_{\rm MC} = 0.986 \pm 0.029,
\end{equation}
where the uncertainty reflects only the statistics of the Monte Carlo.
The ratio $R(K^{*+})/R(K^{*0})$ measured in simulation is
\begin{equation}
R(K^{*+})/R(K^{*0}) = 0.959 \pm 0.035.
\end{equation}
The validity of Eq.~(\ref{assume1}) in simulation relies
mainly on the assumptions used in the fragmentation and hadronization processes
in the event generator. To confirm the validity of Eq.~(\ref{assume1}),
we use the existing experimental data on $K^+$, $K_S$, $K^{*0}$, and $K^{*+}$
multiplicities in jets, which were obtained at $e^+ e^-$ colliders at different
center-of-mass energies~\cite{pdg}. From these data we obtain:
\begin{equation}
R(K^{*+})/R(K^{*0}) = 1.039 \pm 0.075.
\end{equation}
The simulation and data are consistent, and we assign a relative
uncertainty of 7.5\% to both $f_K$ and $F_K$ from the assumption
of Eq.~(\ref{assume1}).
We also assign an uncertainty of 4\% due to the fitting procedure used to extract
the numbers of $K^{*+}$ and $K^{*0}$ events. This uncertainty is obtained by varying the background
parametrization and the fitting range. Since the same background model is used to obtain the number
of $K^{*0}$ events, both in the inclusive muon and the like-sign dimuon samples, this uncertainty
is taken to be the same for $f_K$ and $F_K$.
Adding all contributions in quadrature, and including the uncertainty in Eq.~(\ref{coeffC}), we
find a relative systematic uncertainty of 9.0\% in $f_K$ and $F_K$, with
a 100\% correlation between the two.

We assign an additional uncertainty of 2.0\% on $F_K$ due to 
the description of the background in the
inclusive muon and like-sign dimuon events.
This uncertainty is estimated by varying the background parametrization and range used for fitting, and by
comparing with the results of the alternative fitting method presented in Appendix~\ref{sec_alt}.

We use the ratio $N(K_S)/N(K^{*+} \to K_S \pi^+)$ to measure both
$f_K$ and $F_K$, which is equivalent to the statement that the ratios $F_{K^{*0}}/F_K$
and $f_{K^{*0}}/f_K$ are identical, i.e., that the fraction of kaons originating from $K^{*0}$
is the same in the inclusive muon and in the like-sign dimuon samples.
This is validated in simulation with an uncertainty of 3\% due to the statistics
of the simulation, which we assign as an additional systematic uncertainty to the fraction $F_K$.

The uncertainty on the background fractions $f_\pi$, $F_\pi$, $f_p$, and $F_p$ have an additional
contribution from
the ratios of multiplicities $n_\pi / n_K$ and $n_p / n_K$ extracted from the simulation. To test
the validity of the simulation, we measure the multiplicity of kaons in the inclusive muon sample.
We select events with one reconstructed muon and at least one additional charged particle that
satisfies the track selection criteria. We determine the fraction of kaons among these tracks using
the same method as in Sec.~\ref{sec_fk}, i.e., we find the fraction of tracks
from the $K^{*0} \to K^+ \pi^-$ decay
and convert this into the fraction of kaons. We compare the kaon multiplicity in data using this
method with that measured in the simulation, and we find that they agree within 10\%.
Since part of this difference can be attributed to the uncertainties from
the assumptions of Eqs.~(\ref{assume1}) and~(\ref{assume2}),
and part is due to the fitting procedure described above,
we find that the uncertainty of the kaon multiplicity in the simulation does
not exceed 4\%, and we assign this uncertainty to both quantities $n_\pi / n_K$ and $n_p / n_K$.
We also assign this 4\% uncertainty to the Eq.\ (\ref{npi}) used to
derive the values of $N_\pi / N_K$ and $N_p/N_K$.

Any falsely reconstructed track identified as a muon is treated in the analysis
in the same way as a proton. We check the impact of this approach by completely
removing the contribution of false tracks, or by increasing their contribution by
a factor of ten, and the final value of $\aslb$ changes by less than 0.00016.
We include this difference as the systematic uncertainty on the contribution from
false tracks.

\section{Measurement of $f_S$, $F_{SS}$}
\label{sec_fmu}

We use the measurements of the fractions of background muons in
the ``long" category, obtained in the previous sections to evaluate
the fraction of muons in the ``short" category. In the inclusive
muon sample the fraction $f_S$ is determined as
\begin{equation}
f_S = 1 - f_K - f_\pi - f_p.
\end{equation}
We check through simulation that the contribution from all other
sources to the inclusive muon sample, such as $K_L \to \pi \mu \nu$
decays or the semileptonic decays of hyperons, is negligible.
The muons from $\tau \to \mu \bar{\nu}_\mu \nu_\tau$ are included
by definition in the $f_S$. The fraction $f_S$ is measured separately
in each muon $p_T$ bin and then a weighted average is calculated
with weights given by the fraction of muons in a given $p_T$ interval
$f_\mu^i$ in the inclusive muon samples. From data, we obtain
\begin{equation}
f_S = 0.581 \pm 0.014~({\rm stat}) \pm 0.039~({\rm syst}),
\label{f_h}
\end{equation}
where the systematic uncertainty comes from the uncertainty on the
background fractions described in Sec.~\ref{sec_syst}.

To check the procedure for determining the background fractions,
the composition of the inclusive muon sample in data is compared
to that from simulation in Table~\ref{tab_sim}, where only
statistical uncertainties for both data and simulation are shown.
The agreement between data and simulation is very good, and the
remaining differences are within the assigned systematic uncertainties.
Although the values given in Table~\ref{tab_sim} for data and for
simulation are not independent, some, such as $f_K$ and
$P(\pi \to \mu)/ P(K \to \mu)$ used to derive $f_\pi$,
are measured directly in data. As a consequence, this result
can be used as an additional confirmation of the validity of our method.

\begin{table}
\caption{\label{tab_sim}
Fractions $f_S$, $f_K$, $f_\pi$, and $f_p+f_{\rm f}$, measured in data and in
simulation (MC). Only the statistical uncertainties on these measurements are shown.
}
\begin{ruledtabular}
\newcolumntype{A}{D{A}{\pm}{-1}}
\newcolumntype{B}{D{B}{-}{-1}}
\begin{tabular}{cAAAA}
 &  \multicolumn{1}{c}{$f_S \times 10^2$} & \multicolumn{1}{c}{$f_K \times 10^2$} &
\multicolumn{1}{c}{$f_\pi \times 10^2$} & \multicolumn{1}{c}{$(f_p+f_{\rm f}) \times 10^2$} \\
\hline
Data   & 58.1\ A \ 1.4  &  15.5\ A \ 0.2 &  25.9\ A \ 1.4 & 0.7\ A \ 0.2  \\
MC     & 59.0\ A \ 0.3  &  14.5\ A \ 0.2 &  25.7\ A \ 0.3 & 0.8\ A \ 0.1  \\
\end{tabular}
\end{ruledtabular}
\end{table}

The background fractions $F_K$, $F_\pi$, and $F_p$ are obtained from the same
weighted average used for Eq.~(\ref{f_h}) of the quantities measured in each
$p_T$ interval $i$, starting from the values given in Secs.~\ref{sec_fk} and~\ref{sec_fpi}
($F_\mu^i$ is used as weight instead of $f_\mu^i$).
Using Eq.~(\ref{F_bkg}), we obtain
\begin{equation}
F_{\rm bkg} = 0.409 \pm 0.019~({\rm stat}) \pm 0.040~({\rm syst}).
\label{Fbkg}
\end{equation}

To evaluate the fraction $F_{SS}$ in the like-sign dimuon sample, we take into
account that, in some events, both muons belong to the $L$ category.
The fraction of these events in all events with at least one $L$ muon is measured
in simulation and found to be
\begin{equation}
\frac{F_{LL}}{F_{SL}+F_{LL}} = 0.220 \pm 0.012.
\label{Fl}
\end{equation}
The uncertainty in Eq.~(\ref{Fl}) includes the 4\% systematic uncertainty related to
the multiplicity of different particles in the simulation, as discussed in
Sec.~\ref{sec_syst}. Using Eqs.~(\ref{F_bkg}), (\ref{Fbkg}), and~(\ref{Fl}), we obtain
\begin{equation}
F_{LL} = 0.074 \pm 0.003~({\rm stat}) \pm 0.008~({\rm syst}),
\label{Fl1}
\end{equation}
and, finally, from Eqs.~(\ref{sumF}), (\ref{Fbkg}), and~(\ref{Fl1}) we obtain
\begin{equation}
F_{SS} = 0.665 \pm 0.016~({\rm stat}) \pm 0.033~({\rm syst}).
\label{F_h}
\end{equation}

\section{Measurement of $\delta$}
\label{sec_delta}

Table~III of Ref.~\cite{D01} gives a complete list of the contributions to the
dimuon charge asymmetry that are caused by detector effects. The largest
of these effects is $\approx3$\%. The reversal of magnet polarities is a
characteristic of the D0 experiment that allows the cancellation at first order
of these detector effects, reducing any charge asymmetry introduced by the track
reconstruction considerably~\cite{D01}.

Higher-order effects result in a small residual reconstruction asymmetry at
the 10$^{-3}$ level.
This asymmetry is measured using $J/\psi \to \mu^+ \mu^-$ decays.
We select events that pass at least one dimuon trigger and have at least one identified
muon and one additional particle of opposite charge that satisfies the track
selection criteria of Sec.~\ref{selection}. We verify in Appendices~\ref{sec_tras}
and~\ref{sec_trig} that the track reconstruction and trigger selection do not
introduce an additional charge asymmetry. The residual asymmetry is measured as
a function of the muon transverse momentum.
The probability to identify a muon with charge $Q = \pm 1$ and
$p_T$ corresponding to bin $i$ is denoted by $P_i (1 + Q \delta_i)$,
where $P_i$ is the mean probability for positive and negative muons,
and $\delta_i$ is the muon detection asymmetry we want to measure.
The probability of identifying the second muon with $p_T$ in bin $j$,
provided that the first muon has $p_T$ in bin $i$, is denoted by
$P^i_j (1 + Q \delta_j)$.

The number of events $N_{ij}$ with a positive muon in bin $i$ and negative
muon in bin $j$ is
\begin{equation}
N_{ij} = N P_i (1 + \delta_i) P^i_j (1 - \delta_j),
\label{nij}
\end{equation}
where $N$ is the total number of selected $J/\psi \to \mu^+ \mu^-$ decays.
The number of events with only one selected muon of charge $Q$ is
\begin{equation}
N_{iQ} = N P_i (1 + Q \delta_i) \left ( 1 - \sum_{j=0}^4 P^i_j (1 - Q \delta_j) \right ),
\end{equation}
where the sum extends over the five transverse momentum intervals.

The probabilities $P_i$ and $P^i_j$ are not independent.
From Eq.~(\ref{nij}), we have the following normalization condition
\begin{equation}
P_i P^i_j = P_j P^j_i.
\label{pij}
\end{equation}
In addition, since the total probability to identify the muon is
$P_{\rm tot} = \sum_{i=0}^4 P_i$, we get the following normalization condition
\begin{equation}
(P_{\rm tot})^2 = \sum_{i,j=0}^4 P_i P^i_j.
\label{pnorm}
\end{equation}

Experimentally we measure the quantities $N_{ii}$, $\Sigma_{ij}$, $\Delta_{ij}$ $(i < j)$,
$\Sigma_i$, and $\Delta_i$, which can be expressed as
\begin{eqnarray}
N_{ii} & = & N P_i P^i_i,
\label{array}
\\
\Sigma_{ij}  \equiv  N_{ij} + N_{ji} & = & 2 N P_i P^i_j, \nonumber \\
\Delta_{ij}  \equiv  N_{ij} - N_{ji} & = & 2 N P_i P^i_j (\delta_i - \delta_j), \nonumber \\
\Sigma_i  \equiv  N_{i+} + N_{i-} & = & 2 N P_i (1 - P_{\rm sum}^i), \nonumber \\
\Delta_i  \equiv  N_{i+} - N_{i-} & = & 2 N P_i (1 - P_{\rm sum}^i)
(\delta_i + \delta_{\rm sum}^i), \nonumber
\end{eqnarray}
where $P_{\rm sum}^i = \sum_{j=0}^4 P^i_j$ and
$\delta_{\rm sum}^i = (\sum_{j=0}^4 P^i_j \delta_j)/(1-P_{\rm sum}^i)$.

\begin{figure}
\begin{center}
\includegraphics[width=0.50\textwidth]{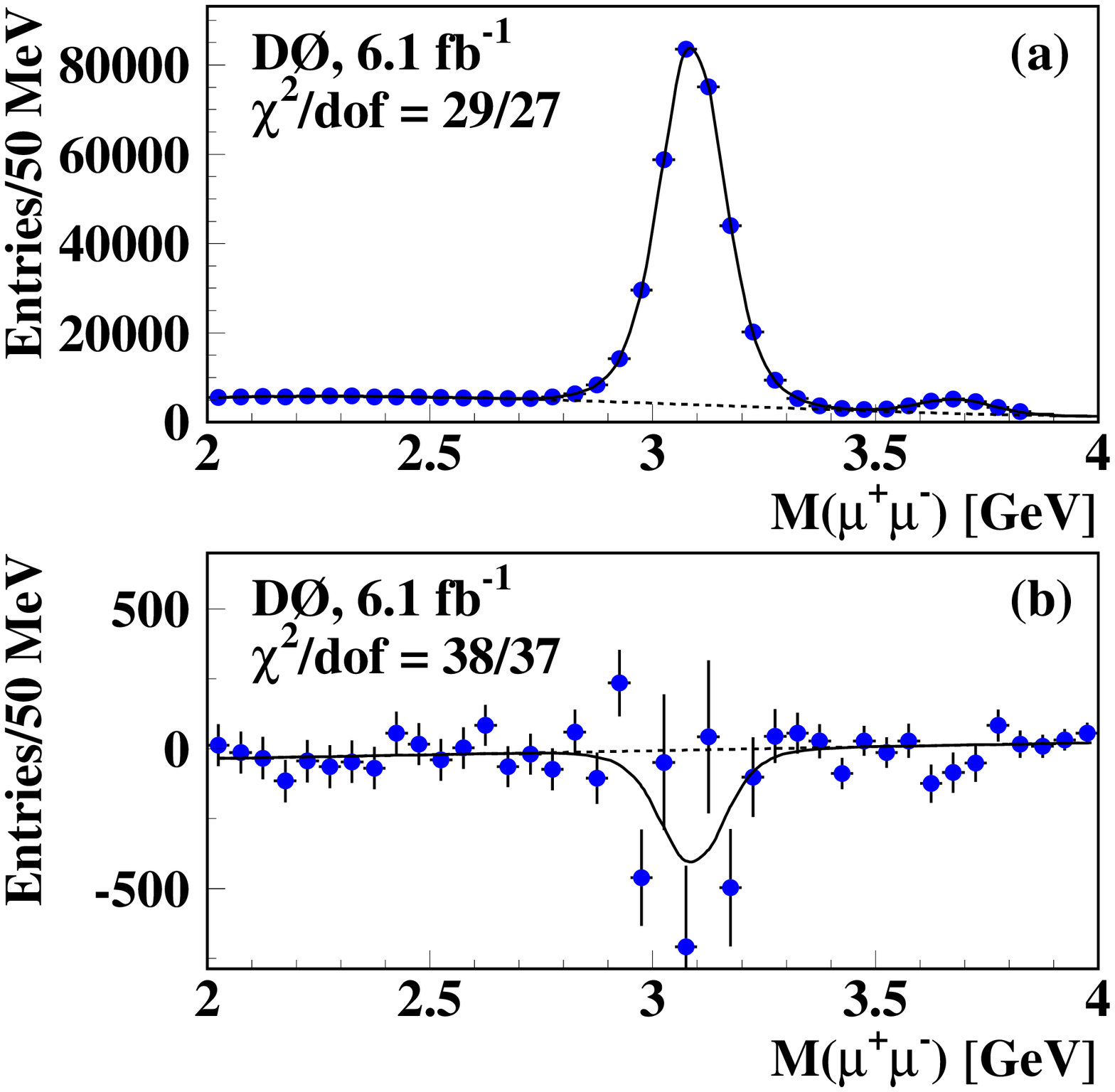}
\caption{The $\mu^+ \mu^-$ invariant mass distributions used to obtain
(a) $\Sigma_{23} = N_{23}+N_{32}$ and (b) $\Delta_{23} = N_{23} - N_{32}$.
The solid line presents the result of the fit; the dashed line shows the non-resonant
background contribution.}
\label{fig_jpsi_s23}
\end{center}
\end{figure}

\begin{figure}
\begin{center}
\includegraphics[width=0.50\textwidth]{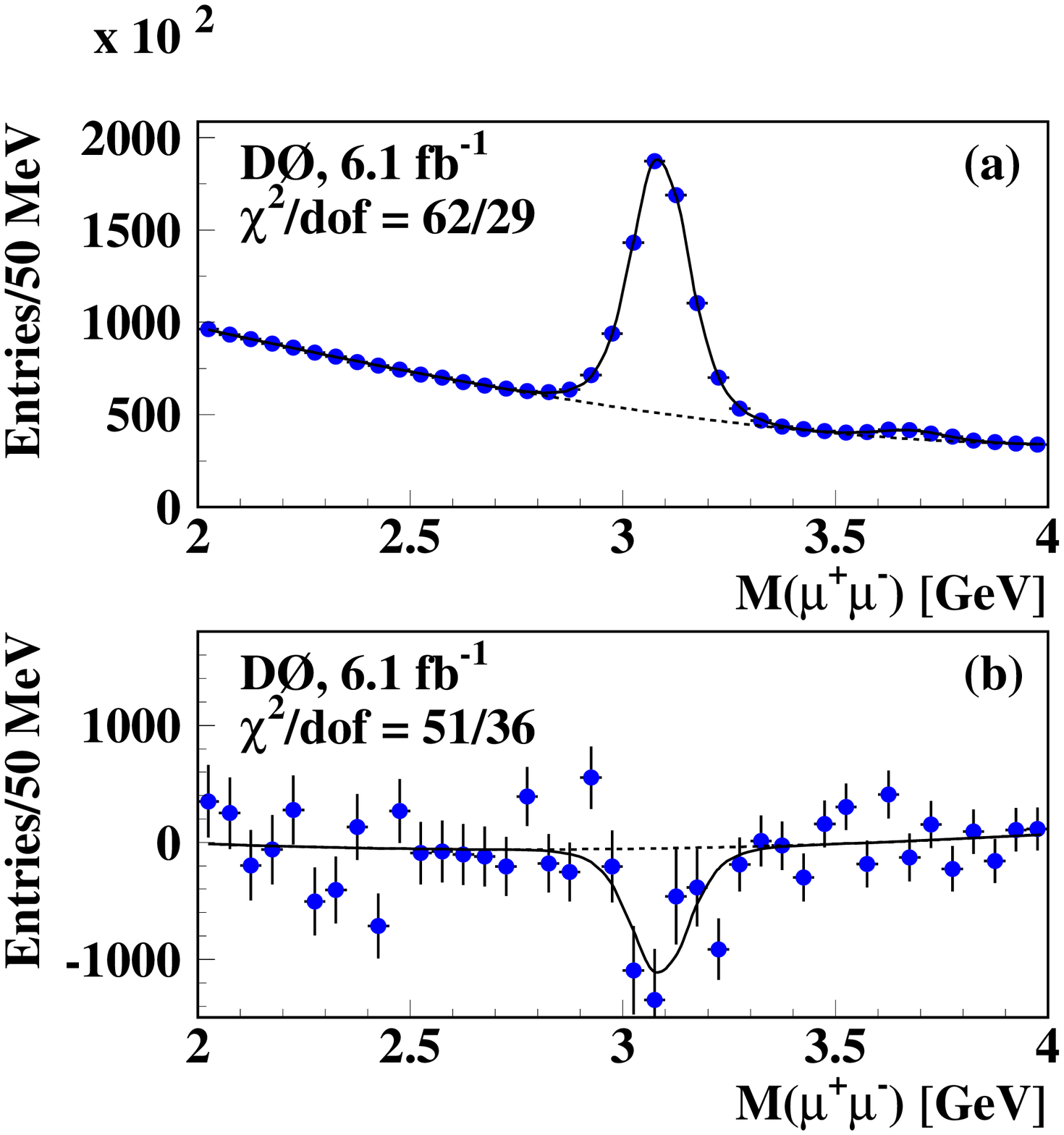}
\caption{The $\mu^+ \mu^-$ invariant mass distributions used to obtain
(a) $\Sigma_{2} = N_{2+}+N_{2-}$ and (b) $\Delta_{2} = N_{2+} - N_{2-}$.
The solid line presents the result
of the fit; the dashed line shows the non-resonant background contribution.}
\label{fig_jpsi_s2}
\end{center}
\end{figure}

\begin{figure}
\begin{center}
\includegraphics[width=0.50\textwidth]{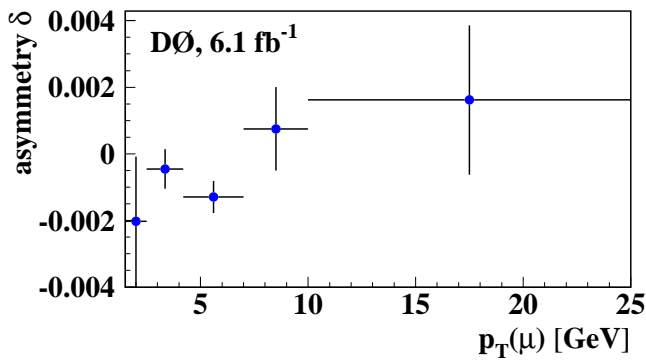}
\caption{Muon reconstruction asymmetry as a function of the muon $p_T$.
}
\label{fig_delta}
\end{center}
\end{figure}

Since the number of measured quantities is greater than the number of unknowns,
$P_i$, $P^i_j$, and $\delta_i$ can be obtained from Eqs.~(\ref{pij}--\ref{array})
by minimizing the $\chi^2$ of the difference between the observed and
expected quantities.
The quantities $N_{ii}$, $\Sigma_{ij}$, $\Delta_{ij}$ $(i < j)$,
$\Sigma_i$, and $\Delta_i$ are obtained from fits to the  $J/\psi$ mass peak in the dimuon
invariant mass distribution $M(\mu^+\mu^-)$ in each of the five $p_T$ bins.
The $J/\psi$ signal is described by the sum of two Gaussians. An additional
Gaussian is included to take into account the contribution from the $\psi'$.
The background is parametrized by a third degree polynomial.
The mean position and R.M.S. of all Gaussian functions in the fit of
$\Delta_{ij}$ and $\Delta_i$ are fixed to the values obtained in the fit of
the corresponding quantities $\Sigma_{ij}$ and $\Sigma_i$. Examples of the fits
are shown in Figs.~\ref{fig_jpsi_s23} and~\ref{fig_jpsi_s2}.
The values of $\delta_i$ obtained as a function of the muon $p_T$
are given in Table~\ref{tab1} and are shown in Fig.~\ref{fig_delta}.
The correlations between values of $\delta_i$ in different bins are given
in Table~\ref{tab1a}.
The weighted average for the residual muon asymmetry in the inclusive
muon and the like-sign dimuon samples, calculated using
weights given by the fraction of muons in a given $p_T$ interval
$f_\mu^i$ ($F_\mu^i$) in the inclusive muon (dimuon) sample, are given respectively by
\begin{eqnarray}
\delta = & \sum_{i=0}^4 f_\mu^i \delta_i = -0.00076 \pm 0.00028, \\
\Delta = & \sum_{i=0}^4 F_\mu^i \delta_i = -0.00068 \pm 0.00023,
\end{eqnarray}
where only the statistical uncertainties are given.
The correlations among different $\delta_i$ are taken into account.
These small values of the residual muon reconstruction asymmetry
are a direct consequence of the regular reversal of the magnets
polarities in the D0 experiment.

\begin{table}
\caption{\label{tab1}
Muon reconstruction asymmetry $\delta_i$ for different muon $p_T$ bins.
The correspondence between the bin number and the $p_T$ range is given
in Table~\ref{tab7}. Only the statistical uncertainties are given.
}
\begin{ruledtabular}
\newcolumntype{A}{D{A}{\pm}{-1}}
\newcolumntype{B}{D{B}{-}{-1}}
\begin{tabular}{cA}
Bin & \multicolumn{1}{c}{$\delta_i$} \\
\hline
0 &  -0.00203\ A \ 0.00194 \\
1 &  -0.00045\ A \ 0.00059 \\
2 &  -0.00130\ A \ 0.00048 \\
3 &  +0.00075\ A \ 0.00125 \\
4 &  +0.00162\ A \ 0.00230
\end{tabular}
\end{ruledtabular}
\end{table}

\begin{table}
\caption{\label{tab1a}
Correlation coefficients among values of $\delta_i$ in different bins.
The correspondence between the bin number and the $p_T$ range is given in
Table~\ref{tab7}.
}
\begin{ruledtabular}
\begin{tabular}{c|ccccc}
Bin & 0 & 1 & 2 & 3 & 4 \\
\hline
0 & +1.000 & $-0.189$ & $-0.155$ &  +0.024 & $-0.051$ \\
1 &$-0.189$ &  +1.000 & $-0.449$ & $-0.117$ & $-0.059$ \\
2 &$-0.155$ & $-0.449$ &  +1.000 & $-0.242$ & $-0.124$ \\
3 & +0.024 & $-0.117$ & $-0.242$ &  +1.000 & $-0.006$ \\
4 &$-0.051$ & $-0.059$ & $-0.124$ & $-0.006$ & +1.000
\end{tabular}
\end{ruledtabular}
\end{table}

\section{Measurement of $a_K$, $a_\pi$, $a_p$}
\label{sec_ak}

The largest detector-related charge asymmetry is produced by \ktomu~ tracks.
It is caused, as discussed in Sec.~\ref{strategy}, by the difference between
the $K^- N$ and $K^+ N$ interaction cross sections \cite{pdg}, resulting in a positive
charge asymmetry of muons coming from kaon decay or punch-through.

The asymmetry $a_K$ of \ktomu~ tracks is measured directly in data using
$K^{*0} \to K^+ \pi^-$ and $\phi \to K^+ K^-$ decays. In both cases we select
candidates with \ktomu~ tracks, in the entire inclusive muon sample.
We calculate separate mass distributions for positive and negative \ktomu~tracks,
and fit the sum and the difference of these distributions to extract the quantity
$\Delta_K$, corresponding to the difference in the number of $K^{*0}$ or $\phi$
meson decays with positive and negative \ktomu~ tracks, and the quantity $\Sigma_K$,
corresponding to their sum. The selection of events and the fitting procedure used
to extract the number of signal decays are described in Appendix~\ref{sec_fits}.
The asymmetry $a_K$ is measured as:
\begin{equation}
a_K = C \Delta_K/\Sigma_K,
\label{asym}
\end{equation}
where the coefficient $C$ is the fraction of correctly reconstructed kaons among all
\ktomu~ tracks as in Eq.~(\ref{coeffC}). In this measurement of $a_K$,
we require that the kaon decays after having been reconstructed in the tracking system,
since its track parameters must be measured in order to reconstruct
the $K^{*0}$ or $\phi$ meson. However, the \ktomu~ tracks in the inclusive sample
also include kaons decaying before being reconstructed in the tracker.
Since the kaon asymmetry is caused by the interactions of kaons with the material of
the detector, and the amount of material near the interaction point is negligible,
the kaons decaying before being reconstructed by the tracker do not produce any
significant asymmetry. They contribute only in the denominator of Eq.~(\ref{asym}).
The factor $C$ takes into account the contribution of these tracks. Its numerical
value is given in Eq.~(\ref{coeffC}). It should be noted that this factor cancels
in the products $f_K a_K$, etc., since both $f_K$ and $a_K$ are measured using
the correctly reconstructed \ktomu~tracks.

\begin{figure}
\begin{center}
\includegraphics[width=0.50\textwidth]{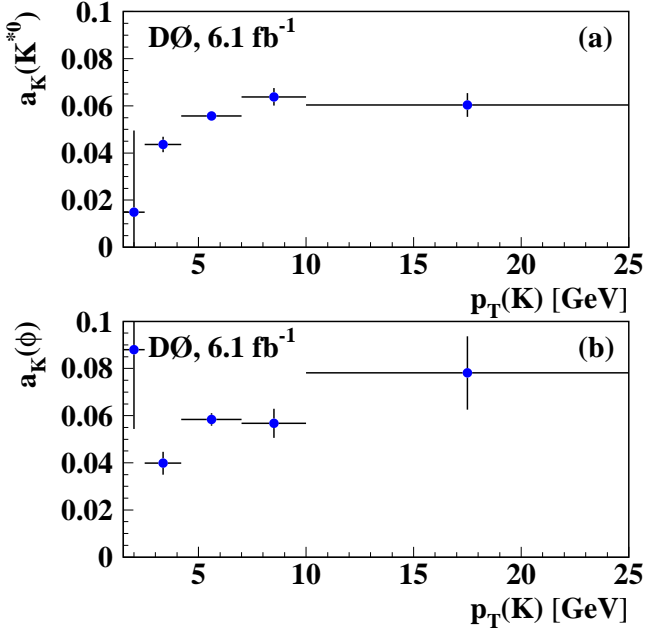}
\caption{The asymmetry $a_K$ measured with (a) $K^{*0} \to K^+ \pi^-$ and
(b) $\phi \to K^+ K^-$ decays as a function of the \ktomu~ $p_T$.
}
\label{fig_ak1}
\end{center}
\end{figure}

Figure~\ref{fig_ak1}(a) shows the value of $a_K(K^{*0})$ measured in
$K^{*0} \to K^+ \pi^-$ decay as a function of the $p_T$ of the \ktomu~ track.
The asymmetry in $\phi \to K^+ K^-$ decays $a_K^{\rm meas}$ has to be corrected
for the charge asymmetry of the second kaon track $a_K^{\rm track}$:
\begin{equation}
a_K(\phi) = a_K^{\rm meas} - a_K^{\rm track}.
\end{equation}
The kaon track reconstruction asymmetry $a_K^{\rm track}$ is discussed and measured
as a function of kaon momentum in~\cite{bjk} using the decay $D^{*+} \to D^0 \pi^+$
with $D^0 \to K^- \mu^+ \nu$, and is taken from that article. It is convoluted with
the $p_T$ distribution of the second kaon for each bin of the \ktomu~$p_T$.
Figure~\ref{fig_ak1}(b) shows the resulting asymmetry $a_K(\phi)$.

\begin{figure}
\begin{center}
\includegraphics[width=0.50\textwidth]{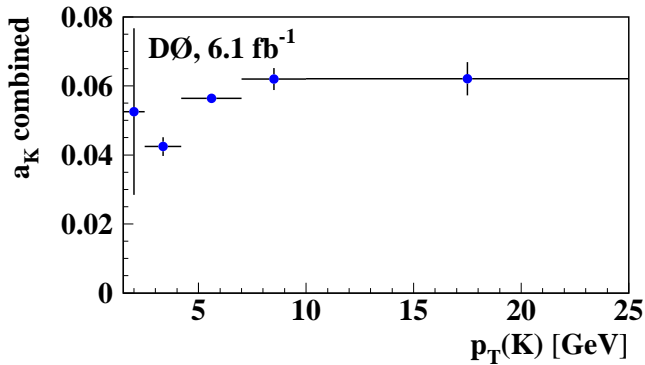}
\caption{The combined asymmetry $a_K$ as a function of $p_T$.
}
\label{fig_ak2}
\end{center}
\end{figure}

The two measurements of $a_K$ are consistent. The $\chi^2/{\rm d.o.f.}$ for their difference
is 5.40/5. Therefore they can be combined and the resulting asymmetry $a_K$ is shown
in Fig.~\ref{fig_ak2} and in Table~\ref{tab2}. Due to the requirement of $p_T > 4.2$ GeV or
$|p_z| > 6.3$ GeV, the first two bins in Fig.~\ref{fig_ak2} correspond to muons that
traverse the forward toroids of the D0 detector. These muons have a larger momentum
and a longer path length before the calorimeter than central muons. As a result,
$a_K$ drops at low $p_T$.
\begin{table}
\caption{\label{tab2}
Asymmetries $a_K$, $a_\pi$, and $a_p$ for different $p_T$ bins.
The correspondence between the bin number and $p_T$ range is given
in Table~\ref{tab7}.
The last line shows the mean asymmetries averaged over the inclusive muon sample.
Only the statistical uncertainties are given.
}
\begin{ruledtabular}
\newcolumntype{A}{D{A}{\pm}{-1}}
\newcolumntype{B}{D{B}{-}{-1}}
\begin{tabular}{cAcc}
Bin &  \multicolumn{1}{c}{$a_K$} & $a_\pi$ & $a_p$  \\
\hline
0     & +0.0526\ A \ 0.0242 & \multirow{2}{*}{$+0.0027 \pm 0.0021$}
                            & \multirow{2}{*}{$-0.104 \pm 0.076$} \\
1     & +0.0424\ A \ 0.0027 &                                       &  \\
2     & +0.0564\ A \ 0.0013 & $+0.0013 \pm 0.0012$ & $+0.028 \pm 0.035$ \\
3     & +0.0620\ A \ 0.0032 & \multirow{2}{*}{$+0.0164 \pm 0.0044$}
                            & \multirow{2}{*}{$+0.077 \pm 0.055$}  \\
4     & +0.0620\ A \ 0.0048 &                                       & \\ \hline
All & +0.0551\ A \ 0.0011 & $+0.0025 \pm 0.0010$ & $+0.023 \pm 0.028$
\end{tabular}
\end{ruledtabular}
\end{table}

The asymmetry $a_\pi$ of \pitomu~ tracks and the asymmetry $a_p$ of \ptomu~
tracks are expected to be much smaller. We measure these asymmetries using
$K_S \to \pi^+ \pi^-$ and $\Lambda \to p \pi^-$ decays, respectively. The
details of the $K_S$ and $\Lambda$ selections are given in Appendix~\ref{sec_fits}.
The technique used to measure the asymmetry is the same as in $a_K$ measurement.
The same factor $C$ is used to measure the asymmetry $a_\pi$ from $K_s \to \pi^+ \pi^-$
decays. The uncertainty in $C$ takes into account its difference for
kaon and pion tracks. Since the proton is stable, this factor is not used
in the computation of the asymmetry of \ptomu~ tracks.

\begin{figure}
\begin{center}
\includegraphics[width=0.50\textwidth]{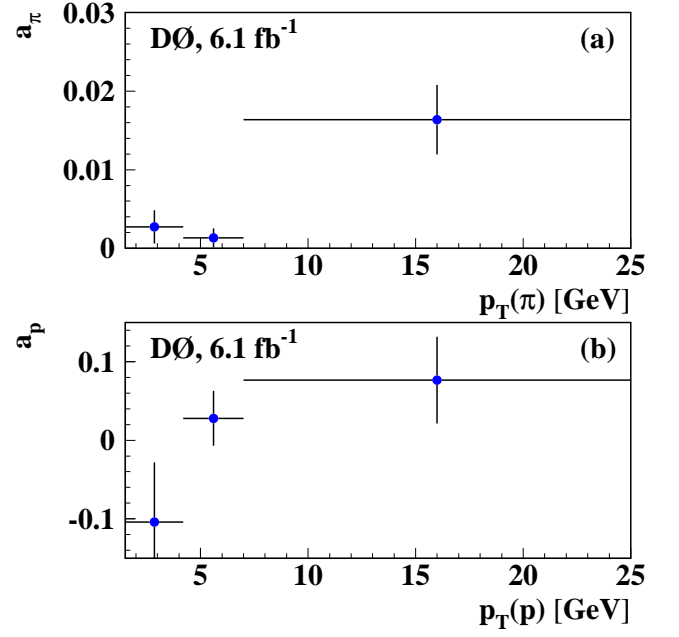}
\caption{The asymmetry (a) $a_\pi$ and (b) $a_p$
as a function of the $p_T$ of the pion and proton, respectively.}
\label{fig_api}
\end{center}
\end{figure}

The asymmetries $a_{\pi}$ and $a_p$ are shown in Fig.~\ref{fig_api} as
a function of the  $p_T$ of the \pitomu~and \ptomu~tracks, respectively.
The values of these asymmetries and their averages are listed in Table~\ref{tab2}
for different $p_T$ bins. We use the mean value
of these quantities in bins 0 and 1 and in bins 3 and 4 since the
statistics available in the first and last bin are not sufficient to perform
separate measurements.

The asymmetries $A_K$, $A_{\pi}$ and $A_p$ are obtained from
$a_K$, $a_{\pi}$ and $a_p$ using Eq.~\ref{FkAk} (and analogous
relations for pions and protons).

\section{Corrections due to background asymmetries}
\label{sec_sumback}
The corrections for the asymmetries of the background, obtained from
Tables~\ref{tab7}, \ref{tab3}, \ref{tab4}
and~\ref{tab2}, are summarized in Tables~\ref{corr_inclmu} and~\ref{corr_lsdimu}.
The values $f_K a_K$, $F_K A_K$, etc., are computed by averaging the
corresponding quantities with weights given by the fraction of muons
in a given $p_T$ interval $f_\mu^i$ ($F_\mu^i$) in the inclusive muon (dimuon) sample,
see Eqs.~(\ref{fkak}) and~(\ref{FkAk}). We use the mean value of
$f_\pi$, $F_\pi$, $f_p$, $F_p$, $a_\pi$, and $a_p$ in bins 0 and 1 and in bins 3 and 4
as the statistics available in the first and last bin are not sufficient to perform
separate measurements.

\begin{table}
\caption{\label{corr_inclmu}
Corrections due to background asymmetries
$f_K a_K$, $f_\pi a_\pi$, and $f_p a_p$ for different $p_T$ bins.
The last line shows the weighted averages obtained using weights
given by the fraction of muons in a given $p_T$
interval $f_\mu^i$ in the inclusive muon sample.
Only the statistical uncertainties are given.
}
\begin{ruledtabular}
\newcolumntype{A}{D{A}{\pm}{-1}}
\newcolumntype{B}{D{B}{-}{-1}}
\begin{tabular}{cAcc}
Bin &  \multicolumn{1}{c}{$f_K a_K \times 10^2$} & $f_\pi a_\pi \times 10^2$ &
$f_p a_p \times 10^2$ \\
\hline
0     & +0.760\ A \ 0.353 & \multirow{2}{*}{$+0.095 \pm 0.076$}
                          & \multirow{2}{*}{$-0.061 \pm 0.060$}  \\
1     & +0.600\ A \ 0.040 &                                     &  \\
2     & +0.889\ A \ 0.023 & $+0.033 \pm 0.030$                  & $+0.020 \pm 0.026$ \\
3     & +0.968\ A \ 0.054 & \multirow{2}{*}{$+0.337 \pm 0.109$}
                          & \multirow{2}{*}{$+0.053 \pm 0.067$} \\
4     & +0.946\ A \ 0.081 &                                     &  \\ \hline
All   & +0.854\ A \ 0.018 & $+0.095 \pm 0.027$                  & $+0.012 \pm 0.022$ \\

\end{tabular}
\end{ruledtabular}
\end{table}

\begin{table}
\caption{\label{corr_lsdimu}
Corrections due to background asymmetries
$F_K A_K$, $F_\pi A_\pi$ and $F_p A_p$ for different $p_T$ bins.
The last line shows the weighted averages obtained using weights
given by the fraction of muons in a given $p_T$
interval $F_\mu^i$ in the dimuon sample.
Only the statistical uncertainties are given.
}
\begin{ruledtabular}
\newcolumntype{A}{D{A}{\pm}{-1}}
\newcolumntype{B}{D{B}{-}{-1}}
\begin{tabular}{cAcc}
Bin &  \multicolumn{1}{c}{$F_K A_K \times 10^2$} & \multicolumn{1}{c}{$F_\pi A_\pi \times 10^2$} &
\multicolumn{1}{c}{$F_p A_p \times 10^2$} \\
\hline
0     & +0.953\ A \ 0.501 & \multirow{2}{*}{$+0.086 \pm 0.069$}
                          & \multirow{2}{*}{$-0.056 \pm 0.054$}  \\
1     & +0.594\ A \ 0.061 &                                     &  \\
2     & +0.910\ A \ 0.048 & $+0.030 \pm 0.027$                  & $+0.019 \pm 0.024$ \\
3     & +0.741\ A \ 0.106 & \multirow{2}{*}{$+0.294 \pm 0.098$}
                          & \multirow{2}{*}{$+0.046 \pm 0.058$} \\
4     & +1.332\ A \ 0.176 &                                     &  \\ \hline
All   & +0.828\ A \ 0.035 & $+0.095 \pm 0.025$                  & $+0.000 \pm 0.021$ \\
\end{tabular}
\end{ruledtabular}
\end{table}

\section{Asymmetries $a_S$ and $A_S$}
\label{sec_Ab}

In the absence of new particles or interactions,
the only non-instrumental source of the asymmetries $a_S$ and $A_S$
is the semileptonic charge asymmetry $\aslb$ given by Eq.~(\ref{asl-w}).
Both $a_S$ and $A_S$ are proportional to \aslb, through the coefficients
\begin{eqnarray}
c_b & \equiv & a_S / \aslb, \label{c_b_def}  \\
C_b & \equiv & A_S / \aslb  \label{c_b_def_2},
\end{eqnarray}
which are determined from simulation.

The decays producing an $S$ muon in the inclusive muon sample,
and their weights relative to the semileptonic decay $b \to \mu X$~\cite{charge},
are listed in Table~\ref{tab8}. All weights are computed using simulated
events. The main process, denoted as $T_1$, is the direct semileptonic decay of
a $b$ quark. It includes the decays $b \to \mu X$ and $b \to \tau X$, with $\tau \to \mu X$.
The weights $w_{1a}$ and $w_{1b}$ for semileptonic decays of $B$ mesons
with and without oscillations are computed using the mean mixing probability
\begin{equation}
\chi_0  =  f'_d \chi_{d0} + f'_s \chi_{s0},
\end{equation}
where $f'_d$ and $f'_s$ are the fractions of \Bd~ and \Bs~ mesons in a sample of semileptonic
$B$-meson decays, and $\chi_{d0}$ and $\chi_{s0}$ are the \Bd~ and \Bs~ mixing probabilities
integrated over time. We use the value $\chi_0 = 0.147 \pm 0.011$ measured at the Tevatron
and given in~\cite{pdg,hfag}.  The second process $T_2$ concerns the sequential decay $b \to c \to \mu X$.
For simplicity we use the same value of $\chi_0$ to compute the weights
$w_{2a}$ and $w_{2b}$ of non-oscillating and oscillating sequential decays $b \to c \to \mu X$.
The process $T_3$ is the decay of a $b$ hadron to a $c \bar{c}$ pair, with
either the $c$ or $\bar c$ quark producing a muon, while $T_4$ includes the decays of short-lived
mesons $\eta$, $\omega$, $\rho^0$, $\phi(1020)$, $J/\psi$, and $\psi'$ to a $\mu^+ \mu^-$ pair.
We take into account both the decays of $b$ hadrons to these particles and their prompt production.
The process $T_5$ represents four-quark production of $b \bar{b} c \bar{c}$ with either the $c$ or
$\bar c$ quark decaying to a muon. The decays of $b$ or $\bar{b}$ quark to a muon in this process 
and the four-quark production of $b\bar{b}b\bar{b}$ are
taken into account through processes $T_1$, $T_2$, and $T_3$. Finally, the process $T_6$ involves $c \bar{c}$ production
followed by $c \to \mu X$ decay. We separate the processes $T_5$ and $T_6$ because only $T_5$
contributes to the like-sign dimuon sample, while both $T_5$ and $T_6$ contribute to the
inclusive muon sample.

The uncertainty in the weights of different processes contains contributions
from the uncertainty in the momentum of the generated $b$ hadrons and from the uncertainties
of branching fractions for $b$-hadron decays. We reweight the simulated $b$-hadron momentum
to get agreement of the of muon momentum spectrum in data and in MC, and the difference
in weights is assigned as the systematic uncertainty on the momentum distribution. The uncertainties
in the inclusive branching fractions $B \to \mu X$, $B \to c X$ and $B \to \bar{c} X$
taken from~\cite{pdg} are propagated into the uncertainties on the corresponding weights.
We assign an additional uncertainty of 10\% to the weights $w_5$ and $w_6$ due to
the uncertainties on the production cross sections of $c \bar{c}$ and $b \bar b c \bar c$ processes.

\begin{table}
\caption{\label{tab8}
Heavy quark decays contributing to the inclusive muon and like-sign dimuon samples.
Abbreviation ``nos" stands for ``non-oscillating," and ``osc" for ``oscillating."
All weights are computed using the MC simulation.
}
\begin{ruledtabular}
\newcolumntype{A}{D{A}{\pm}{-1}}
\newcolumntype{B}{D{B}{-}{-1}}
\begin{tabular}{lll}
   & Process & Weight \\
\hline
$T_1$   & $b \to \mu^-X$ & $w_1 \equiv 1.$ \\
$T_{1a}$ & ~~~$b \to \mu^-X$ (nos) & $w_{1a} = (1-\chi_0) w_1$ \\
$T_{1b}$ & ~~~$\bar{b} \to b \to \mu^-X$ (osc) & $w_{1b} = \chi_0 w_1$ \\
$T_2$ &  $b \to c \to \mu^+X$ & $w_2 = 0.113 \pm 0.010$ \\
$T_{2a}$ & ~~~$b \to c \to \mu^+X$ (nos) & $w_{2a} = (1-\chi_0) w_2$ \\
$T_{2b}$ & ~~~$\bar{b} \to b \to c \to \mu^+X$ (osc) & $w_{2b} = \chi_0 w_2$ \\
$T_3$ & $b \to c \bar c q$ with $c \to \mu^+X$ or $\bar c \to \mu^-X$ & $w_3 = 0.062 \pm 0.006$ \\
$T_4$ & $\eta, \omega, \rho^0, \phi(1020), J/\psi, \psi' \to \mu^+ \mu^-$ & $w_4 = 0.021 \pm 0.001$ \\
$T_5$ & $b \bar b c \bar c$ with $c \to \mu^+X$ or $\bar c \to \mu^-X$ & $w_5 = 0.013 \pm 0.002$ \\
$T_6$ & $c \bar c$ with $c \to \mu^+X$ or $\bar c \to \mu^-X$ & $w_6 = 0.660 \pm 0.077$

\end{tabular}
\end{ruledtabular}
\end{table}

Among all processes listed in Table~\ref{tab8}, the process $T_{1b}$ is directly
related to the semileptonic charge asymmetry \aslb (see Appendix~\ref{theory} for details).
The process $T_{2b}$ produces the flavor-specific charge asymmetry $A_{fs}$. We set
$A_{fs} = - \aslb$, where the negative sign appears because the charge of the muon in
the process $T_{2b}$ is opposite to the charge of the muon in the process $T_{1b}$.
No other process contributes to the charge asymmetry and therefore they just dilute
the value of \aslb. The coefficient $c_b$ is found from:
\begin{equation}
c_b = \frac{w_{1b} - w_{2b}}{w_1 + w_2 + w_3 + w_4 + w_5 + w_6} = 0.070 \pm 0.006.
\label{c_b}
\end{equation}

The computation of the coefficient $C_b$ is more complicated. One of the selections for
the like-sign dimuon sample requires that the invariant mass of the two muons be greater
than 2.8 GeV. This requirement suppresses the contribution from processes in which
both muons arise from the decay of the same quark. The probability that the initial
$b$ quark produces a $\mu^-$ is
\begin{equation}
P_b \propto w_{1a} + w_{2b} + 0.5 (w_3 + w_4 + w_5),
\end{equation}
where we apply the coefficient 0.5 because processes $T_3$, $T_4$, and $T_5$
produce an equal number of positive and negative muons.
The probability that the accompanying $\bar b$ quark also produces a $\mu^-$ is
\begin{equation}
P_{\bar b} \propto w_{1b} + w_{2a} + 0.5(w_3 + w_4 + w_5).
\end{equation}
The total probability of observing like-sign dimuon events from decays of heavy
quarks is
\begin{equation}
P_{\rm tot} \propto P_b P_{\bar b}.
\end{equation}
The probability of processes contributing to the charge asymmetry of dimuon events is
\begin{eqnarray}
P_{\rm as} & \propto & w_{1b}[w_{1a} + 0.5 (w_3 + w_4 + w_5)] - \nonumber \\
       &         & w_{2b}[w_{2a} + 0.5 (w_3 + w_4 + w_5)].
\label{pas}
\end{eqnarray}
The coefficient $C_b$ is obtained from the ratio
\begin{equation}
C_b = P_{\rm as} / P_{\rm tot} = 0.486 \pm 0.032.
\label{C_b}
\end{equation}

This relation assumes that the processes producing the two muons are independent and is
verified by calculating the coefficient $C_b$ in simulated like-sign dimuon events.
We exclude the process $T_6$ with $c \bar{c}$ pair production, since the mixing probability of $D^0$
meson is small and these events do not
contribute significantly to the like-sign dimuon sample.
We count the number of direct-direct $b$-quark decays, $N_{dd}$, of direct-sequential decays, $N_{ds}$,
of sequential-sequential decays, $N_{ss}$, of direct-random events, $N_{dr}$ (``random" includes
processes $T_3$, $T_4$, and $T_5$), of sequential-random decays, $N_{sr}$, 
to obtain
\begin{equation}
C_b = \frac{N_{dd} - N_{ss} + \chi_0 (N_{dr} - N_{sr})}
{N_{\textrm{ls}}}
 = 0.448 \pm  0.071,
\label{AhAbls}
\end{equation}
where $N_{\rm ls}$ is the total number of like-sign dimuon events.
This result agrees well with the value in Eq.~(\ref{C_b}).
The uncertainty of this method is larger because of the small statistics
of simulated like-sign dimuon events.

\section{Asymmetry \aslb}
\label{sec_ah}

The uncorrected asymmetries $a$ and $A$ are obtained by counting the number
of events of each charge in the inclusive muon and the like-sign dimuon samples,
respectively. In total, there are $1.495 \times 10^9$ muons in the inclusive
muon sample, and $3.731 \times 10^6$ events in the like-sign dimuon sample.
We obtain
\begin{eqnarray}
\label{value_a}
a & = & +0.00955 \pm 0.00003, \\
A & = & +0.00564 \pm 0.00053.
\label{value_A}
\end{eqnarray}

The results obtained in Secs.~\ref{sec_fk}--\ref{sec_Ab} are
used to calculate the asymmetries $a_S$ and $A_S$ from these values,
which are then used to evaluate the charge asymmetry for semileptonic $B$ meson decays.

\begin{figure}
\begin{center}
\includegraphics[width=0.50\textwidth]{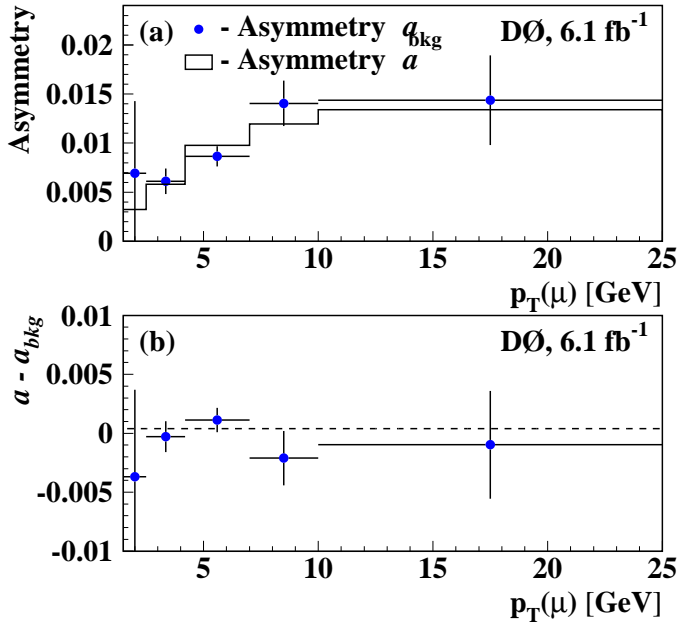}
\caption{(a) 
The asymmetry $a_{\rm bkg}$ (points with error
bars) as expected from our measurements of the fractions and
asymmetries of the background processes is compared to the measured asymmetry $a$ 
of the inclusive muon sample (shown as histogram, since the statistical uncertainties are negligible).
The asymmetry from CP violation is negligible
compared to the background in the inclusive muon sample;
(b)~the difference $a - a_{\rm bkg}$. The horizontal dashed line shows the mean value of 
this difference.
}
\label{fig_asym}
\end{center}
\end{figure}

The asymmetry $\aslb$, extracted from the asymmetry $a$ of the inclusive muon
sample using Eqs.~(\ref{inclusive_mu_a}) and~(\ref{c_b_def}), is 
\begin{equation}
\aslb = +0.0094 \pm 0.0112~({\rm stat}) \pm 0.0214~({\rm syst}).
\label{ah1}
\end{equation}
The contributions to the uncertainty on this value are given in Table~\ref{tab5}.
Figure~\ref{fig_asym}(a) shows a comparison of the asymmetry $a$ and
the background asymmetry $a_{\rm bkg} = f_S \delta + f_K a_K + f_\pi a_\pi + f_p a_p$,
as a function of the muon $p_T$. There is excellent agreement between these
two quantities, with the $\chi^2/{\rm d.o.f.}$ for their difference being 2.4/5.
Figure~\ref{fig_asym}(b) shows the value of $f_S a_S = a - a_{\rm bkg}$,
which is consistent with zero. The values $a$ and $a_{\rm bkg}$ are given
in Table~\ref{tab6}. This result agrees with expectations, since
the value of the asymmetry $a$ should be determined mainly by the
background, and the contribution from $\aslb$ should be strongly
suppressed by the small factor of $c_b = 0.070 \pm 0.006$.

\begin{table}
\caption{\label{tab5}
Sources of uncertainty on $\aslb$ in Eqs.~(\ref{ah1}),
(\ref{ah2}), and~(\ref{ah3}). The first eight rows contain statistical uncertainties, 
the next three rows contain systematic uncertainties.
}
\begin{ruledtabular}
\newcolumntype{A}{D{A}{\pm}{-1}}
\newcolumntype{B}{D{B}{-}{-1}}
\begin{tabular}{cccc}
Source & $\delta\sigma(\aslb)(\ref{ah1})$ & $\delta\sigma(\aslb)$(\ref{ah2}) & $\delta\sigma(\aslb)$(\ref{ah3}) \\
\hline
$A$ or $a$ (stat)       & 0.00066 & 0.00159 & 0.00179 \\
$f_K$ or $F_K$ (stat)   & 0.00222 & 0.00123 & 0.00140 \\
$P(\pitomu)/P(\ktomu)$  & 0.00234 & 0.00038 & 0.00010 \\
$P(\ptomu)/P(\ktomu)$   & 0.00301 & 0.00044 & 0.00011 \\
$A_K$                   & 0.00410 & 0.00076 & 0.00061 \\
$A_\pi$                 & 0.00699 & 0.00086 & 0.00035 \\
$A_p$                   & 0.00478 & 0.00054 & 0.00001 \\
$\delta$ or $\Delta$    & 0.00405 & 0.00105 & 0.00077 \\
\hline
$f_K$ or $F_K$ (syst)   & 0.02137 & 0.00300 & 0.00128 \\
$\pi$, $K$, $p$
multiplicity            & 0.00098 & 0.00025 & 0.00018 \\
$c_b$ or $C_b$          & 0.00080 & 0.00046 & 0.00068 \\
\hline
Total statistical       & 0.01118 & 0.00266 & 0.00251 \\
Total systematic        & 0.02140 & 0.00305 & 0.00146 \\
Total                   & 0.02415 & 0.00405 & 0.00290
\end{tabular}
\end{ruledtabular}
\end{table}

\begin{table}
\caption{\label{tab6}
The measured asymmetry $a$ and the expected background asymmetry $a_{\rm bkg}$ in the inclusive muon sample
for different $p_T$ bins. For the background asymmetry,
the first uncertainty is statistical, the second is systematic.
}
\begin{ruledtabular}
\newcolumntype{A}{D{A}{\pm}{-1}}
\newcolumntype{B}{D{B}{-}{-1}}
\begin{tabular}{cAA}
bin &  \multicolumn{1}{c}{$a \times 10^2$} &
\multicolumn{1}{c}{$a_{\rm bkg} \times 10^2$} \\
\hline
0     & 0.324\ A \ 0.036 & 0.693\ A \ 0.379 \pm 0.632 \\
1     & 0.582\ A \ 0.007 & 0.611\ A \ 0.109 \pm 0.072 \\
2     & 0.978\ A \ 0.003 & 0.865\ A \ 0.054 \pm 0.088 \\
3     & 1.193\ A \ 0.008 & 1.405\ A \ 0.159 \pm 0.168 \\
4     & 1.339\ A \ 0.011 & 1.438\ A \ 0.206 \pm 0.408
\end{tabular}
\end{ruledtabular}
\end{table}

The consistency of $\aslb$  with zero in Eq.~(\ref{ah1})
and the good description of the charge asymmetry $a$ for
different values of the muon $p_T$ shown in Fig.~\ref{fig_asym}
constitute an important tests of the validity of the background
model and of the analysis method discussed in this article.

The second measurement of the asymmetry $\aslb$, obtained from the uncorrected asymmetry $A$ of
the like-sign dimuon sample using Eqs.~(\ref{dimuon_A}) and~(\ref{c_b_def_2}), is
\begin{equation}
\aslb = -0.00736 \pm 0.00266~({\rm stat}) \pm 0.00305~({\rm syst}).
\label{ah2}
\end{equation}
The contributions to the uncertainty on $\aslb$ for this measurement
are also listed in Table~\ref{tab5}.

The results~(\ref{ah1}) and~(\ref{ah2}) represent two different measurements
of \aslb. The uncertainties in Eq.~(\ref{ah1}) are much larger because the asymmetry
$a_s$ is divided by the small coefficient $c_b$. Since the same background processes
contribute to the uncorrected asymmetries $a$ and $A$, their uncertainties in
Eqs.~(\ref{ah1}) and~(\ref{ah2}) are strongly correlated. We take advantage of
this correlation to obtain a single optimized value of $\aslb$, with higher precision,
using a linear combination of the uncorrected asymmetries
\begin{equation}
A' \equiv A - \alpha a,
\label{combination}
\end{equation}
and choosing the coefficient $\alpha$ in order to minimize the total
uncertainty on the value of \aslb.

\begin{figure}
\begin{center}
\includegraphics[width=0.50\textwidth]{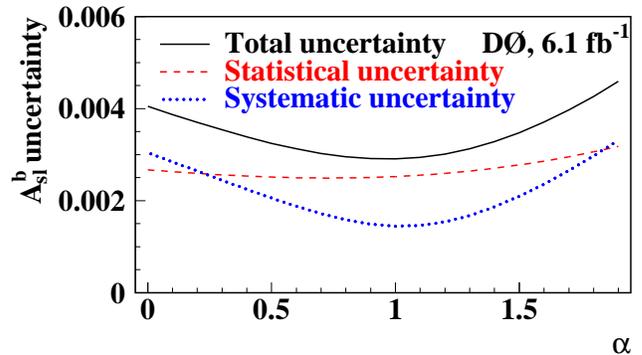}
\caption{Statistical (dashed line), systematic (doted line), and
total (full line) uncertainties on $\aslb$ as a function of the parameter $\alpha$
of Eq.\ (\ref{combination}).
}
\label{fig_scan}
\end{center}
\end{figure}

It is shown in Secs.~\ref{sec_fk}--\ref{sec_sumback} that the contributions from background sources
in Eqs.~(\ref{inclusive_mu_a}) and~(\ref{dimuon_A}) are of the same order of magnitude.
On the other hand, the dependence of $A$ and $a$ on the asymmetry $\aslb$,
according to Eqs.~(\ref{c_b}) and~(\ref{C_b}), is significantly
different, with $C_b \gg c_b$. As a result,
we can expect a reduction of background uncertainties in~(\ref{combination}) for $\alpha \approx 1$
with a limited reduction of the statistical sensitivity on $\aslb$.
Figure~\ref{fig_scan} shows the statistical, systematic, and total
uncertainties on $\aslb$ as a function of the parameter $\alpha$.
The total uncertainty on $\aslb$ has a minimum for $\alpha = 0.959$,
and the corresponding value of the asymmetry $\aslb$ is
\begin{equation}
\label{ah3}
\aslb = -0.00957 \pm 0.00251~({\rm stat}) \pm 0.00146~({\rm syst}).
\end{equation}
This value is our final result for \aslb. It
differs by 3.2 standard deviations from the standard model prediction of $\aslb$ given in
Eq.~(\ref{in_aslbsm}). The different
contributions to the total uncertainty of $\aslb$ in Eq.~(\ref{ah3})
are listed in Table~\ref{tab5}.

\section{Consistency checks}
\label{sec_consistency}

To check the stability of the result, we repeat this measurement with modified
selections, or with subsets of the available data sample. Changes
are implemented in a variety of tests:
\begin{itemize}
\item{Test \textsf{A}:} Using only the part of the data sample corresponding to the
first 2.8 fb$^{-1}$.
\item{Test \textsf{B}:} In addition to the reference selections, requiring at least
three hits in muon wire chamber layers B or C, and the $\chi^2$ for a fit to a track
segment reconstructed in  the muon detector to be less than 8.
\item{Test \textsf{C}:} Since the background muons are produced by decays of kaons
and pions, their track parameters measured by the central tracker and by the muon
system are different. Therefore, the fraction of background strongly depends on
the $\chi^2$ of the difference between these two measurements.
The requirement on this $\chi^2$ is changed from 40 to 4 in this study.
\item{Test \textsf{D}:} The maximum value of the transverse impact parameter
is changed from 0.3 to 0.05 cm,
and the requirement on the longitudinal distance between the point of closest approach
to the beam and the associated interaction vertex is changed from 0.5 to 0.05 cm. This test serves
also as a cross-check against the possible contamination from muons from cosmic rays in the
selected sample.
\item{Test \textsf{E}:} Using only low-luminosity events with fewer than three
interaction vertices.
\item{Test \textsf{F}:} 
Using only events corresponding to two of the four possible
configurations of the magnets, for which the solenoid and
toroid polarities are identical.
\item{Test \textsf{G}:} Changing the requirement on the invariant mass of the two muons from 2.8 GeV to 12 GeV.
\item{Test \textsf{H}:} Using the same muon $p_T$ requirement, $p_T > 4.2$~GeV, over the full detector acceptance.
\item{Test \textsf{I}:} Requiring the muon $p_T$ to be $<7.0$~GeV.
\item{Test \textsf{J}:} Requiring the azimuthal angle $\phi$ of the muon track to be in the range
$0 < \phi < 4$ or $5.7 < \phi < 2 \pi$. This selection excludes muons directed to the region
of poor muon identification efficiency in the support structure of the detector.
\item{Test \textsf{K}:} Requiring the muon $\eta$ to be in the range $ |\eta| < 1.6$ (this test
serves also as a cross-check against the possible contamination from muons associated with the beam halo).
\item{Test \textsf{L}:} Requiring the muon $\eta$ to be in the range $ |\eta| < 1.2$
or $1.6 < |\eta| < 2.2$.
\item{Test \textsf{M}:} Requiring the muon $\eta$ to be in the range $ |\eta| < 0.7$
or $1.2 < |\eta| < 2.2$.
\item{Test \textsf{N}:} Requiring the muon $\eta$ to be in the range
$0.7 < |\eta| < 2.2$.
\item{Test \textsf{O}:} Using like-sign dimuon events passing at least one single muon trigger,
while ignoring the requirement of a dimuon trigger for these events.
\item{Test \textsf{P}:} Using like-sign dimuon events passing both single muon and dimuon triggers.
\end{itemize}

A summary of the results from these studies is presented in Tables~\ref{tab10} and~\ref{tab11}.
The last line, denoted as ``significance",
gives the difference between the reference result (column \textsf{Ref})
and each modification, divided by its uncertainty, and taking
into account the overlap between the samples.
The statistical uncertainties are used in the calculation of the
significance of the difference between two results.
These tests demonstrate an impressive stability of the $\aslb$ result,
and provide a strong confirmation of the validity of the method.
As a result of the variations of the selection criteria,
all input quantities are changed over a wide range, while
the asymmetry $\aslb$ remains well within the assigned uncertainties.
For example, the uncorrected asymmetry $A$ changes by a factor $\approx 1.5$ in test \textsf{C}, while
the asymmetry $\aslb$ changes by less than 7\%. It should also be noted that reducing the kaon background
in test \textsf{C} yields a negative asymmetry $A$.

\begin{table*}
\caption{\label{tab10}
Measured asymmetry $\aslb$ with reference selections (column \textsf{Ref}) and variations
\textsf{A -- H}.
}
\begin{ruledtabular}
\newcolumntype{A}{D{A}{\pm}{-1}}
\newcolumntype{B}{D{B}{.}{-1}}
\begin{tabular}{lrrrrrrrrr}
& \textsf{Ref} & \textsf{A} & \textsf{B} & \textsf{C} & \textsf{D} & \textsf{E} & \textsf{F}
& \textsf{G} & \textsf{H}
\\ \hline
$N(\mu \mu) \times 10^{-6}$  &  3.731   &  1.809 &  2.733 &  1.809 &  1.785 &  2.121 &  1.932 &  1.736 &  1.783 \\
$a \times 10^2$              & +0.955   & +0.988 & +0.791 & +0.336 & +1.057 & +0.950 & +1.029 & +0.955 & +1.032 \\
$A \times 10^2$              & +0.564   & +0.531 & +0.276 & -0.229 & +0.845 & +0.543 & +0.581 & +0.821 & +0.632 \\
$\alpha$                     &  0.959   &  0.901 &  0.942 &  1.089 &  1.083 &  0.902 &  0.915 &  1.029 &  0.877 \\
$[(2-F_{\rm bkg}) \Delta - \alpha f_S \delta]
\times 10^2$                 & $-0.065$ & $-0.072$ & $-0.143$ & $-0.200$ & $-0.074$ & $-0.075$ & $-0.069$ & $-0.023$ & $-0.061$ \\
$F_{\rm bkg}$                &  0.409   &  0.372 &  0.401 &  0.303 &  0.384 &  0.385 &  0.426 &  0.449 &  0.343 \\
\hline
$\aslb \times 10^2$          & $-0.957$ & $-0.976$ & $-1.084$ & $-0.892$ & $-1.107$ & $-0.888$ & $-1.096$ & $-0.873$ & $-0.769$ \\
$\sigma(\aslb) \times 10^2$ (stat)
                             &  0.251   & 0.330  &  0.293 &  0.315 &  0.402 &  0.328 &  0.375 &  0.388 &  0.336 \\
Significance                 &          & 0.090  &  0.846 & 0.324  &  0.478 &  0.326 &  0.498 &  0.281 &  0.779
\end{tabular}
\end{ruledtabular}
\end{table*}

\begin{table*}
\caption{\label{tab11}
Measured asymmetry $\aslb$ with reference selections (column \textsf{Ref}) and variations
\textsf{I -- P}.
}
\begin{ruledtabular}
\newcolumntype{A}{D{A}{\pm}{-1}}
\newcolumntype{B}{D{B}{.}{-1}}
\begin{tabular}{lrrrrrrrrr}
& \textsf{Ref} & \textsf{I} & \textsf{J} & \textsf{K} & \textsf{L} & \textsf{M}
& \textsf{N} & \textsf{O} & \textsf{P}
\\ \hline
$N(\mu \mu) \times 10^{-6}$  &  3.731 &  2.569 &  2.208 &  1.884 &  1.909 &  2.534 &  2.122 &  2.002 &  1.772 \\
$a \times 10^2$              & +0.955 & +0.896 & +1.002 & +0.984 & +1.098 & +0.679 & +1.097 & +0.968 & +0.968 \\
$A \times 10^2$              & +0.564 & +0.407 & +0.648 & +0.576 & +0.630 & +0.353 & +0.748 & +0.722 & +0.692 \\
$\alpha$                     &  0.959 &  0.975 &  0.913 &  0.895 &  0.877 &  0.940 &  0.949 &  0.983 &  0.934 \\
$[(2-F_{\rm bkg}) \Delta - \alpha f_S \delta]
\times 10^2$                 & $-0.065$ & $-0.101$ & $-0.079$ & $-0.125$ & $-0.142$ & $-0.081$ & $-0.019$ & $-0.046$ & $-0.044$ \\
$F_{\rm bkg}$                &  0.409 &  0.439 &  0.412 &  0.363 &  0.365 &  0.412 &  0.452 &  0.419 &  0.398 \\
\hline
$\aslb \times 10^2$          & $-0.957$ & $-1.295$ & $-0.710$ & $-0.851$ & $-0.801$ & $-0.759$ & $-1.102$ & $-0.897$ & $-0.833$ \\
$\sigma(\aslb) \times 10^2$ (stat) 
                             &  0.251 &  0.314 &  0.320 &  0.320 &  0.383 &  0.275 &  0.344 &  0.346 &  0.349 \\
Significance                 &        &  1.798 &  1.241 &  0.482 &  0.539 &  1.317 &  0.622 &  0.240 &  0.485
\end{tabular}
\end{ruledtabular}
\end{table*}

\begin{figure}
\begin{center}
\includegraphics[width=0.50\textwidth]{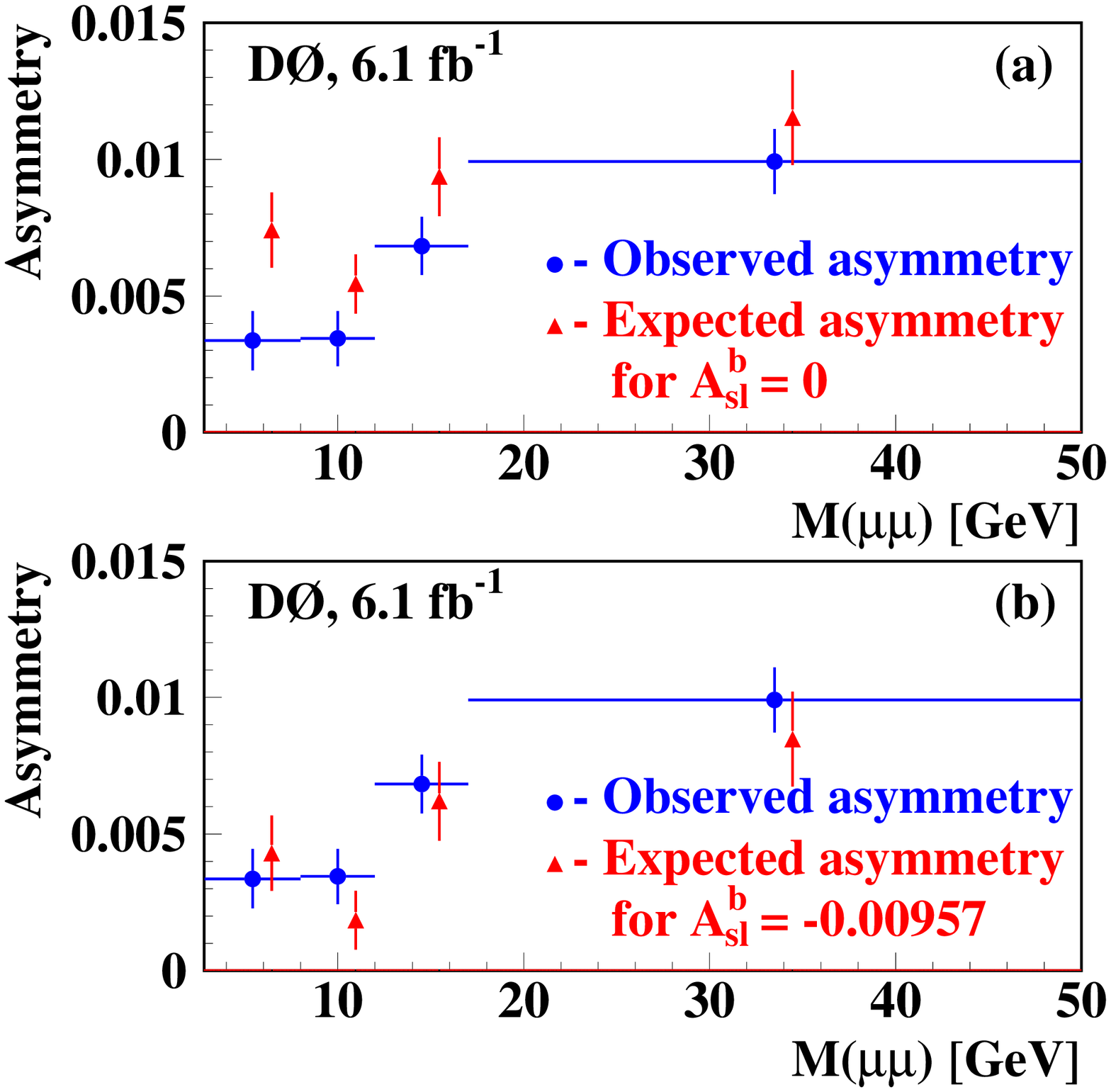}
\caption{The observed and expected like-sign dimuon charge asymmetries
in bins of dimuon invariant mass. The expected asymmetry
is shown for (a) $\aslb = 0.0$ and (b) $\aslb = -0.00957$.
}
\label{a-mmm}
\end{center}
\end{figure}

Figure~\ref{a-mmm} shows the observed and expected uncorrected like-sign dimuon charge
asymmetry as a function of the dimuon invariant mass. The expected asymmetry is computed
using Eq.~(\ref{dimuon_A}) and all the measurements of the sample composition and of
the asymmetries. We compare the expected uncorrected asymmetry using two different
assumptions for $\aslb$. In Fig.~\ref{a-mmm}(a) the observed asymmetry is compared
to the expectation for $\aslb = 0$, while Fig.~\ref{a-mmm}(b) shows the expected asymmetry
for $\aslb = -0.00957$. A possible systematic discrepancy between the observed and expected
asymmetries can be observed for $\aslb = 0$, while it essentially disappears for the
measured $\aslb$ value corresponding to Eq.~(\ref{ah3}). It can also be seen that the
observed asymmetry changes as a function of the dimuon invariant mass, and that the
expected asymmetry reproduces this effect when $\aslb = -0.00957$. 
This dependence of the asymmetry on the invariant
mass of the muon pair is a complex function of the production mechanism, of
the mass of the particles being produced and of their decays. The agreement between
the observed and expected asymmetries indicates that the physics
leading to the observed asymmetry is well described by the contributions from the
backgrounds and from decaying $b$ hadrons.

We conclude that our method of analysis provides a consistent description
of the dimuon charge asymmetry for a wide range of input parameters, even
for significantly modified selection criteria.

In addition to the described consistency checks, we perform other studies to verify the validity
of the analysis method. These tests are described in Appendices~\ref{sec_tras},
\ref{sec_trig}, and~\ref{sec_alt}. We determine 
the asymmetry of track reconstruction and
the asymmetry of trigger selection. We also measure the ratio $F_K/f_K$ 
using an alternative fitting procedure.
These studies do not show any bias in the extracted value of \aslb.

\section{Comparison with existing measurements}
\label{sec_comp}

\begin{figure}
\begin{center}
\includegraphics[width=0.45\textwidth]{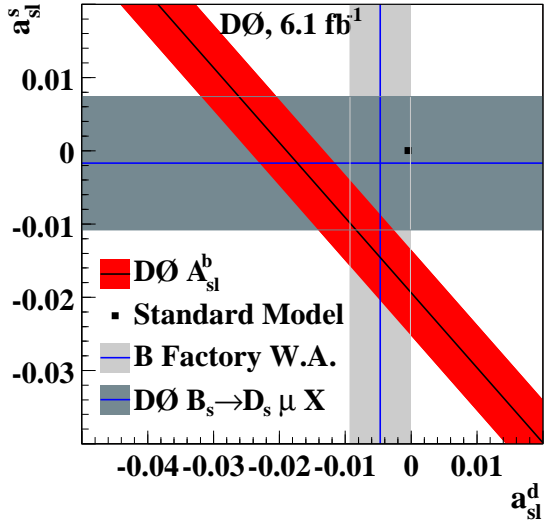}
\caption{(Color online) Comparison of $\aslb$ in data with the standard model prediction
for $\asld$ and $\asls$. Also shown are the existing measurements of $\asld$~\cite{hfag} and
$\asls$~\cite{asl-d0}. The error bands represent the $\pm 1$ standard deviation uncertainties on
each individual measurement.}
\label{asl_svsd}
\end{center}
\end{figure}

The measured value of $\aslb$ places
a constraint on the charge asymmetries of semileptonic
decays of $\Bd$ and $\Bs$ mesons, and the $CP$-violating phases
of the $\Bd$ and $\Bs$ mass mixing matrices.
Calculating the coefficients in Eq.~(\ref{Ab1}) assuming the current PDG
values~\cite{pdg} for all parameters (details are given in
Appendix~\ref{theory}), we obtain
\begin{equation}
\aslb = (0.506 \pm 0.043) \asld + (0.494 \pm 0.043) \asls.
\label{Ab_7}
\end{equation}
Figure~\ref{asl_svsd} presents this measurement in the $\asld$--$\asls$ plane, together with
the existing direct measurements of $\asld$ from the B-Factories~\cite{hfag} and of
our independent measurement of 
$\asls$ in $\Bs \to \Ds \mu X$ decays~\cite{asl-d0}.
Using Eqs.~(\ref{ah3}) and~(\ref{Ab_7})
and the current experimental value of $\asld = -0.0047 \pm 0.0046$~\cite{hfag}, we obtain
\begin{equation}
\asls = -0.0146 \pm 0.0075.
\end{equation}
This agrees with our direct measurement of
$\asls = -0.0017 \pm 0.0091$ ~\cite{asl-d0}.

An independent method for measuring $\phi_s$ is through
$B^0_s \rightarrow J/\psi \phi$ decays.
Such measurements have been performed by the
D0~\cite{D0-phi} and CDF~\cite{CDF-phi} Collaborations.
All measurements are consistent
and the combined value of $\phi_s$ differs from the standard model prediction
by about two standard deviations~\cite{5928}. 

\begin{figure}
\begin{center}
\includegraphics[width=0.45\textwidth]{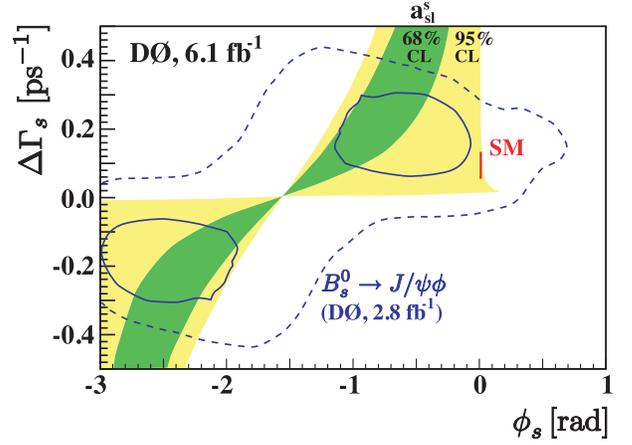}
\caption{(Color online) The 68\% and 95\% C.L. regions of probability for
$\Delta \Gamma_s$ and $\phi_s$ values obtained from this measurement,
considering the experimental constraints on $\asld$~\cite{hfag}.
The solid and dashed curves show respectively the 68\% and 95\% C.L.
contours from the $\Bs \to J/\psi \phi$ measurement~\cite{D0-phi}.
Also shown is the standard model (SM) prediction
for $\phi_s$ and $\Delta \Gamma_s$.
}
\label{dg-phi}
\end{center}
\end{figure}

\begin{figure}
\begin{center}
\includegraphics[width=0.45\textwidth]{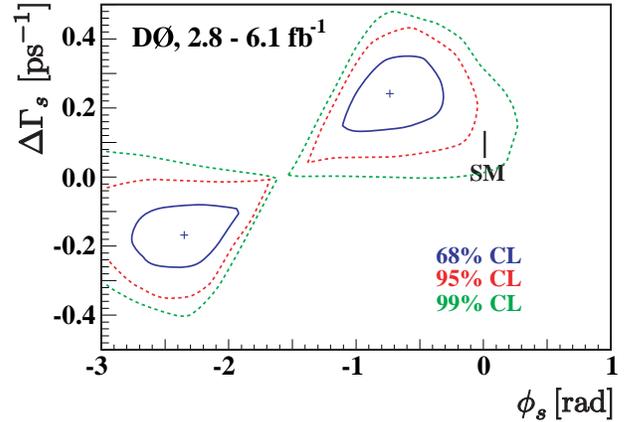}
\caption{(Color online) Probability contours in the
($\phi_s$,$\Delta \Gamma_s$) plane for the combination of this measurement
with the result of Ref.~\cite{D0-phi}, using the experimental constraints on $\asld$~\cite{hfag}.
}
\label{dg-comb}
\end{center}
\end{figure}

Taking into account the experimental constraints on $\asld$~\cite{hfag}, Fig.~\ref{dg-phi}
shows the 68\% and 95\% C.L. regions of $\Delta \Gamma_s$ and $\phi_s$ obtained from our measurement.
The 68\% and 95\% C.L. regions from the D0 measurement using the
$\Bs \to J/\psi \phi$ decay~\cite{D0-phi} are also included in this figure.
Since the sign of $\Delta \Gamma_s$ is not known,
there is also a mirror solution with $\phi_s \rightarrow -\pi - \phi_s$, corresponding
to the change $\Delta \Gamma_s \to - \Delta \Gamma_s$. It can be seen that the D0 results
are consistent. Figure~\ref{dg-comb} shows the probability contours in the
($\phi_s$,$\Delta \Gamma_s$) plane for the combination of our measurement with
the result of Ref.~\cite{D0-phi}.

\section{Conclusions}
\label{conclusions}

We have measured the like-sign dimuon charge asymmetry
$\aslb$ of semileptonic $b$-hadron decays:
\begin{equation}
\aslb = -0.00957 \pm 0.00251~({\rm stat}) \pm 0.00146~({\rm syst}).
\label{Ab_6}
\end{equation}
This measurement is obtained from a data set corresponding to 6.1 fb$^{-1}$ of
integrated luminosity collected with the D0 detector at
Fermilab Tevatron collider. It is consistent with our previous measurement
\cite{D01} obtained with 1 fb$^{-1}$ and supersedes it.
This asymmetry is in disagreement with the prediction of the
standard model by 3.2 standard deviations.
This is the first evidence for anomalous CP-violation in the mixing of neutral
$B$-mesons.

\begin{acknowledgments}

%
We thank the staffs at Fermilab and collaborating institutions,
and acknowledge support from the
DOE and NSF (USA);
CEA and CNRS/IN2P3 (France);
FASI, Rosatom and RFBR (Russia);
CNPq, FAPERJ, FAPESP and FUNDUNESP (Brazil);
DAE and DST (India);
Colciencias (Colombia);
CONACyT (Mexico);
KRF and KOSEF (Korea);
CONICET and UBACyT (Argentina);
FOM (The Netherlands);
STFC and the Royal Society (United Kingdom);
MSMT and GACR (Czech Republic);
CRC Program and NSERC (Canada);
BMBF and DFG (Germany);
SFI (Ireland);
The Swedish Research Council (Sweden);
and
CAS and CNSF (China).

\end{acknowledgments}

\appendix

\section{Theory}
\label{theory}

This Appendix is included for completeness and to
define the notations. Assuming $CPT$ symmetry, the
mixing and decay of the $\Bq, \barBq$ pair ($q=s,d$) is
described~\cite{CPV} by
{\small
\begin{eqnarray}
i \frac{d}{dt}
\left(
\begin{array}{c}
\Bq(t) \\
\barBq(t)
\end{array}
\right) & = &
\left( \left[ \begin{array}{cc}
M_q & M_q^{12} \\
(M_{q}^{12})^* & M_q
\end{array}
\right] - \frac{i}{2}
\left[ \begin{array}{cc}
\Gamma_q & \Gamma_q^{12} \\
(\Gamma_{q}^{12})^* & \Gamma_q
\end{array}
\right] \right) \nonumber \\
& & \cdot \left(
\begin{array}{c}
\Bq(t) \\
\barBq(t)
\end{array}
\right),
\label{B_Bbar}
\end{eqnarray}
}
where $M_q$, $M_q^{12}$, $\Gamma_q$, and $\Gamma_q^{12}$
are the elements of the mass matrix of the $\Bq \barBq$ system.
The matrix element $M_q^{12}$ is due to
box diagrams~\cite{pdg}. New particles foreseen
in extensions of the standard model can contribute to these
box diagrams, and physics beyond the standard model can therefore modify
the phase and amplitude of $M_q^{12}$.

The eigenvalues of the mass matrix in Eq.~(\ref{B_Bbar}) are
\begin{eqnarray}
M_q + \frac{1}{2} \Delta M_q - \frac{i}{2}(\Gamma_q - \frac{1}{2} \Delta \Gamma_q), \\
M_q - \frac{1}{2} \Delta M_q - \frac{i}{2}(\Gamma_q + \frac{1}{2} \Delta \Gamma_q),
\end{eqnarray}
where, by definition, $\Delta M_q > 0$.
Notice the sign conventions for $\Delta M_q$ and $\Delta \Gamma_q$.
With this convention, $\Delta \Gamma_q$
is positive in the standard model. A violation of the $CP$ symmetry is caused by a non-zero
value of the phase
\begin{equation}
\phi_q \equiv \arg \left( - \frac {M_q^{12}}{\Gamma_q^{12}} \right).
\end{equation}

The observable quantities are $M_q$, $\Gamma_q$,
$\Delta M_q$, $\Delta \Gamma_q$ and $\phi_q$, with
\begin{eqnarray}
\Delta M_q = 2 \left| M_q^{12} \right|, \qquad
\Delta \Gamma_q = 2 \left| \Gamma_q^{12} \right| \cos \phi_q.
\end{eqnarray}

The charge asymmetry \aslq~ for ``wrong-charge"
semileptonic $B^0_q$-meson decay induced by oscillations is defined as
\begin{equation}
\aslq = \frac{\Gamma(\bar{B}^0_q(t)\rightarrow \mu^+ X) -
              \Gamma(    {B}^0_q(t)\rightarrow \mu^- X)}
             {\Gamma(\bar{B}^0_q(t)\rightarrow \mu^+ X) +
              \Gamma(    {B}^0_q(t)\rightarrow \mu^- X)}.
              \label{aslq}
\end{equation}
This quantity is independent of the lifetime $t$, and can be expressed as
\begin{equation}
\aslq = \frac{\left| \Gamma_q^{12} \right|}{\left| M_q^{12} \right|} \sin \phi_q =
\frac{\Delta \Gamma_q}{\Delta M_q} \tan \phi_q.
\label{phiq}
\end{equation}

The like-sign dimuon charge asymmetry
$\aslb$ for semileptonic decays of $b$ hadrons produced in
proton-antiproton ($p \bar{p}$) collisions is defined as
\begin{equation}
\aslb \equiv \frac{N^{++}_{b} - N^{--}_{b}}{N^{++}_{b} + N^{--}_{b}},
\end{equation}
where $N^{++}_{b}$ and $N^{--}_{b}$ are the numbers of events
containing two $b$ hadrons that decay semileptonically, producing two positive or
two negative muons, respectively, with only the direct semileptonic decays $b \rightarrow \mu X$
considered in the definition of $N^{++}_{b}$ and $N^{--}_{b}$.
The asymmetry $\aslb$ can be expressed~\cite{Grossman} as
\begin{equation}
\aslb = \frac{f_d Z_d \asld + f_s Z_s \asls}{f_d Z_d + f_s Z_s},
\label{Ab1}
\end{equation}
where
\begin{eqnarray}
Z_q & \equiv & \frac{1}{1 - y_q^2} - \frac{1}{1 + x_q^2}, \\
y_q & \equiv & \frac{\Delta \Gamma_q}{2 \Gamma_q}, \\
x_q & \equiv & \frac{\Delta M_q}{\Gamma_q}.
\end{eqnarray}
with $q = d, s$. The quantities $f_d$ and $f_s$ are the
production fractions for $\bar{b} \rightarrow \Bd$ and
$\bar{b} \rightarrow \Bs$ respectively.
These fractions have been measured for $p \bar p$ collisions at the Tevatron~\cite{pdg}:
  \begin{eqnarray}
f_d & = & 0.323 \pm 0.037, \nonumber \\
f_s & = & 0.118 \pm 0.015.
\end{eqnarray}
All other parameters in~(\ref{Ab1}) are also taken from Ref.~\cite{pdg}:
\begin{eqnarray}
x_d & = & 0.774 \pm 0.008, \nonumber \\
y_d & = & 0 \nonumber, \\
x_s & = & 26.2 \pm 0.5, \nonumber \\
y_s & = & 0.046 \pm 0.027.
\end{eqnarray}
Substituting these values in Eq.~(\ref{Ab1}), we obtain
\begin{equation}
\aslb = (0.506 \pm 0.043) \asld + (0.494 \pm 0.043) \asls.
\end{equation}
Using the values of \asld, \asls~ from Ref.~\cite{Nierste},
\begin{eqnarray}
\asld({\rm SM}) & = & (-4.8 ^{+1.0}_{-1.2}) \times 10^{-4} \nonumber \\
\asls({\rm SM}) & = & (2.1 \pm 0.6) \times 10^{-5},
\end{eqnarray}
the predicted value of $\aslb$ in the standard model is
\begin{equation}
\aslb({\rm SM}) = (-2.3^{+0.5}_{-0.6}) \times 10^{-4}.
\label{aslbsm}
\end{equation}
The current experimental values of the two semileptonic asymmetries are
$\asld = -0.0047 \pm 0.0046$~\cite{hfag} and
$\asls = -0.0017 \pm 0.0091$~\cite{asl-d0}.

It can be concluded from Eq.~(\ref{aslbsm}) that the standard model predicts a small negative value
of $\aslb$ with rather small uncertainty. Any significant deviation of
$\aslb$ from the SM prediction on a scale larger than
that of the uncertainty on $\aslb$ , would
be an unambiguous signal of new physics.

The asymmetry $\aslb$ is also equivalent to the charge
asymmetry of semileptonic decays of $b$ hadrons to ``wrong charge" muons
that are induced by oscillations~\cite{Grossman}, i.e.,
\begin{equation}
a^b_{\rm sl} \equiv \frac{\Gamma(\bar B \to \mu^+ X) - \Gamma(B \to \mu^- X)}
             {\Gamma(\bar B \to \mu^+ X) + \Gamma(B \to \mu^- X)} = \aslb.
\end{equation}
Without initial flavor tagging it is impossible to correctly select the decays
producing a muon of ``wrong charge" from a sample of semileptonic decays of
$b$ quarks.  The charge asymmetry of semileptonic $b$ hadron decays, contrary
to the charge asymmetry of like-sign dimuons, is therefore reduced by the
contribution of decays producing a muon with ``correct" charge, and is
consequently less sensitive to the asymmetry \aslb.

New physical phenomena can change the phase and magnitude
of the standard model $M_{s}^{12,SM}$ to
\begin{equation}
M_s^{12} \equiv M_{s}^{12,SM} \cdot \Delta_s
= M_{s}^{12,SM} \cdot \left| \Delta_s \right| e^{i \phi^\Delta_s},
\end{equation}
where
\begin{eqnarray}
\phi_s & = & \phi^{SM}_s + \phi^\Delta_s, \nonumber \\
\phi^{SM}_s & = & 0.0042 \pm 0.0014.
\label{phi_np}
\end{eqnarray}
Other changes expected as a result of new sources of $CP$ violation~\cite{Nierste} are
\begin{equation}
\Delta M_s = \Delta M^{SM}_s \cdot \left| \Delta_s \right| = (19.30 \pm 6.74) \textrm{ ps}^{-1}
\cdot \left| \Delta_s \right|,
\end{equation}
\begin{equation}
\Delta \Gamma_s = 2 \left| \Gamma_s^{12} \right| \cos \phi_s = (0.096 \pm 0.039) \textrm{ ps}^{-1}
\cdot \cos \phi_s,
\end{equation}
\begin{equation}
\frac{\Delta \Gamma_s}{\Delta M_s} = \frac{\left| \Gamma_s^{12} \right|}{\left| M_{s}^{12,SM} \right|}
\cdot \frac{\cos \phi_s}{\left| \Delta_s \right|} =
(4.97 \pm 0.94) \cdot 10^{-3} \cdot \frac{\cos \phi_s}{\left| \Delta_s \right|},
\end{equation}
\begin{equation}
\asls = \frac{\left| \Gamma_s^{12} \right|}{\left| M_{s}^{12,SM}\right|}
\cdot \frac{\sin \phi_s}{\left| \Delta_s \right|} =
(4.97 \pm 0.94) \cdot 10^{-3} \cdot \frac{\sin \phi_s}{\left| \Delta_s \right|}.
\end{equation}

The $\Bs \rightarrow J/\psi\phi$ decay can also be
used to investigate $CP$ violation. In that case, the $CP$-violating phase
obtained from fits to the $\Bs \rightarrow J/\psi \phi$ angular distributions
is modified as follows~\cite{Nierste}: \\
\begin{equation}
\phi_s^{J/\psi \phi} = -2 \beta^{SM}_s + \phi^\Delta_s,
\end{equation}
where
\begin{equation}
\beta^{SM}_s = \arg [ -V_{ts} V^*_{tb} / (V_{cs} V^*_{cb})] = 0.019 \pm 0.001
\end{equation}
and the quantities $V_{ts}$, $V_{tb}$, $V_{cs}$, and $V_{cb}$ are the parameters of the CKM matrix.
The contribution of new physics to $\phi_s$ and $\phi_s^{J/\psi \phi}$ are identical.

\section{Reconstruction of exclusive decays}
\label{sec_fits}

\subsection{Reconstruction of $K_S$ mesons}
The $\ks$ meson is used to reconstruct the $\kstp$ meson~\cite{charge}
and to measure the fraction and asymmetry of \pitomu~ tracks.
The $\ks \to \pi^+ \pi^-$ decay is reconstructed by requiring two
tracks with opposite charge. Each track must have an impact parameter significance
with respect to the interaction vertex $>3$, where the significance
is defined as $\sqrt{[\epsilon_T/\sigma(\epsilon_T)]^2+
[\epsilon_L/\sigma(\epsilon_L)]^2}$, and $\epsilon_T$ ($\epsilon_L$) is
the projection of the track impact parameter on
the plane transverse to the beam direction (along the
beam direction), and $\sigma(\epsilon_T ) [\sigma(\epsilon_L)]$ is its uncertainty.
At least one of the tracks must have an impact parameter significance $>4$,
and at least one of the particles must have $p_T> 1.5$~GeV. The two tracks must share a
common vertex that is separated from the primary interaction point by more than 4 mm in
the transverse plane. The significance of the reconstructed impact parameter for the $\ks$
must be $<4$.  All $\ks$ candidates satisfying these selection criteria are used to reconstruct the
$\kstp \to \ks \pi^+$ decay. In addition, for the measurement of the fraction and asymmetry of
\pitomu~ tracks, we require that one of the pions from $\ks$ decay
pass the muon selection given in Sec.~\ref{selection}.

\begin{figure}
\begin{center}
\includegraphics[width=0.50\textwidth]{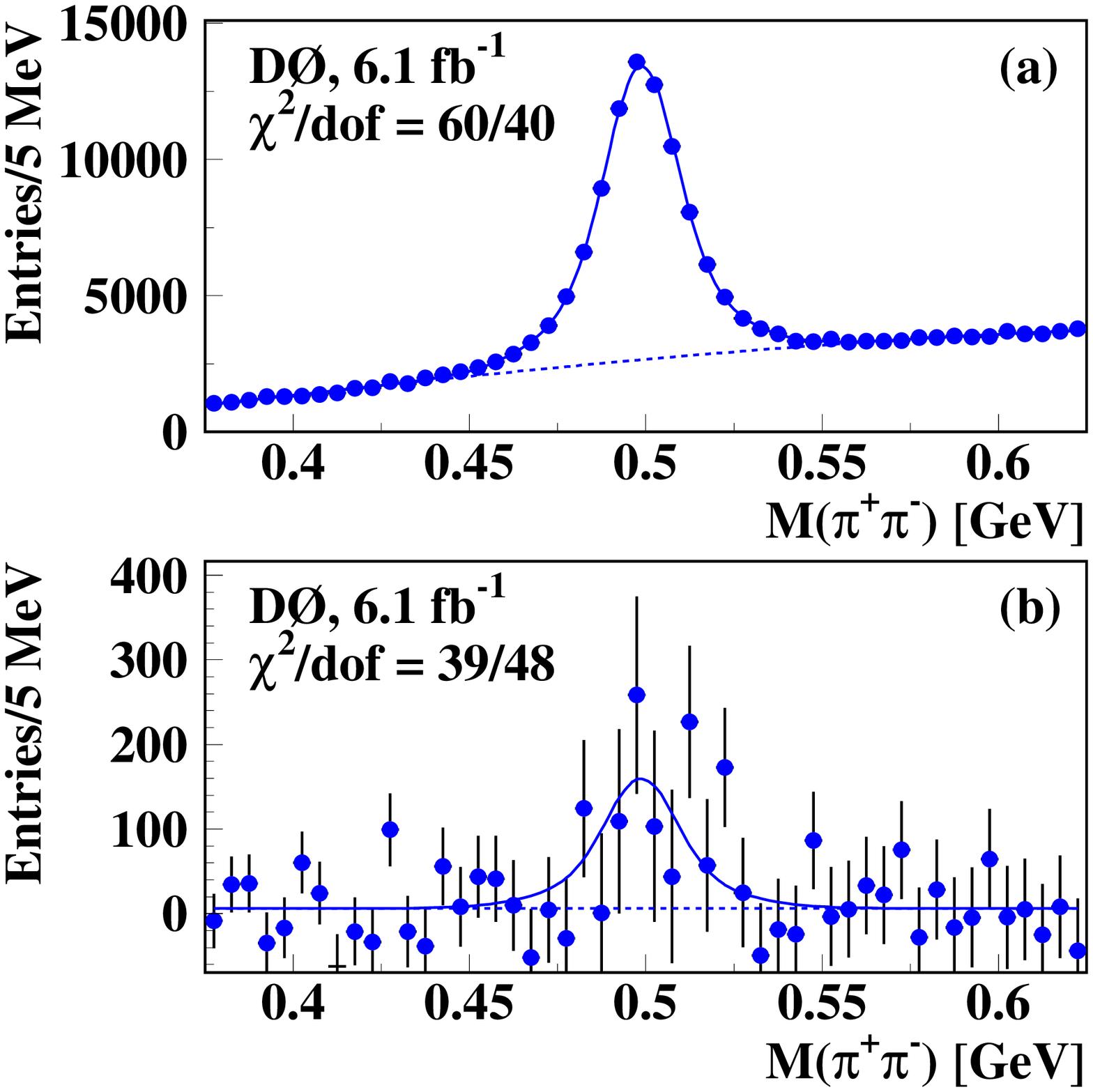}
\caption{The $\pi^+ \pi^-$ invariant mass distribution of $\ks$ candidates in the inclusive muon sample
for \pitomu~ with $7.0 < p_T < 10.0$ GeV. (a) The sum of the distributions
for positive and negative \pitomu~ tracks and (b) their difference.
The solid lines present the result of the fit; the dashed lines show the background contribution. }
\label{fig_ks}
\end{center}
\end{figure}

Figure~\ref{fig_ks} displays the $\pi^+ \pi^-$ invariant mass distribution of
$\ks \to \pi^+ \pi^-$ candidates in the inclusive muon sample for all \pitomu~ with
$7.0 < p_T < 10.0$ GeV. We show separately the sum and the difference of the distributions
for the samples with positive and negative $\pitomu$ tracks, which are used to measure
the asymmetry of $\pitomu$ tracks. The $\ks$ signal is fitted with a double Gaussian, and
the background is parameterized by a third degree polynomial for the sum
of the two distributions and a straight line for their difference. While fitting
the difference of the distributions all the parameters describing the
$\ks$ signal, except its normalization, are fixed to the values obtained from
the fit to the sum of the distributions.

\subsection{Reconstruction of $\kstp$ mesons}

The $\kstp$~\cite{charge} signal is obtained by combining the reconstructed $\ks$ meson with an additional
track which is assigned the mass of the charged pion. The $\ks$ candidate must satisfy the
track selection criteria given in Sec.~\ref{selection}, except for the requirements on the number
of hits in the tracking detectors and the $\chi^2$ of the track fit. The invariant mass
of the $\pi^+ \pi^-$ system must be $480 < M(\pi^+ \pi^-) < 515$ MeV. The additional track
must have at least 2 axial and 1 stereo hits in the silicon microstrip
detector, at least 3 axial and 3 stereo hits in the fiber tracker, and a track impact
parameter significance $<3$ relative to the interaction vertex. The cosine of the angle
between the direction of the $\ks$ meson and the additional track must be greater than 0.3.
The $\ks$ and the additional track must be consistent with sharing the same interaction vertex.

\begin{figure}
\begin{center}
\includegraphics[width=0.50\textwidth]{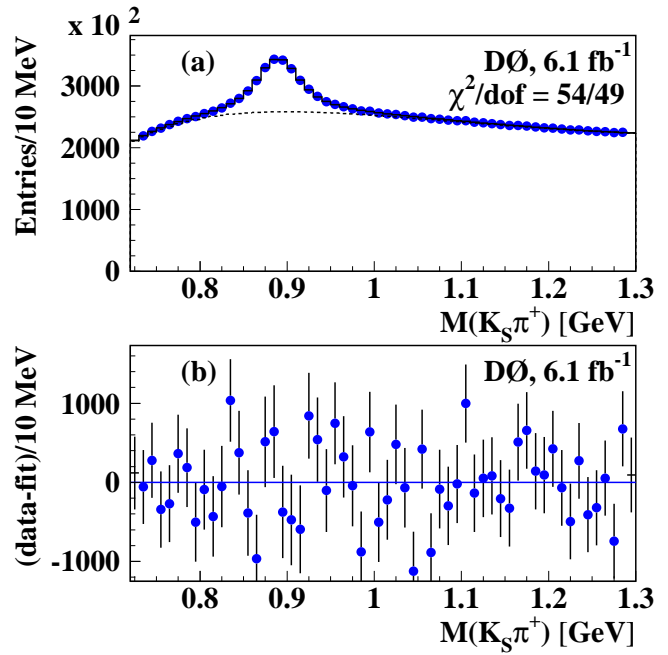}
\caption{(a) The $K_S \pi^+$ invariant mass distribution of $K^{*+}$ candidates
in the inclusive muon sample.
The solid line presents the result of the fit; the dashed line shows the background contribution.
(b) The difference between data and the fit result.
}
\label{fig_kstp}
\end{center}
\end{figure}

Figure~\ref{fig_kstp} shows the $\ks \pi^+$ invariant mass distribution.
The $\kstp$ signal is fitted with a relativistic Breit-Wigner function convoluted with a
Gaussian resolution, and the background is parameterized by the function
\begin{eqnarray}
f_{\rm bck}(M) & = & (M-M_K-M_\pi)^{p_0} \nonumber \\
& \times & \exp(p_1 M + p_2 M^2 + p_3 M^3),
\label{bckg1}
\end{eqnarray}
which includes a threshold factor. Here $M$ is the $\ks \pi^+$ invariant mass, and
$p_0$, $p_1$, $p_2$ and $p_3$ are free parameters. Figure~\ref{fig_kstp}(b), which shows the
difference between data points and the result of the fit, demonstrates the good quality of the
fit with $\chi^2/{\rm d.o.f.} = 54/49$. The measured width of the $\kstp$ meson is
$\Gamma(K^{*+}) = 47.9 \pm 1.4$(stat) MeV, which is consistent with the
current PDG value~\cite{pdg}.

\subsection{Reconstruction of $\kstneu$ mesons}

The $\kstneu$ meson is reconstructed by selecting two tracks of opposite charge
and assigning one of them the mass of the charged kaon. This particle is required to be identified as
a muon and to pass the muon selection criteria given in Sec.~\ref{selection}.
The second track is assigned the mass of a pion and required to satisfy the
criteria used to select the pion in the $\kstpm$ reconstruction.

\begin{figure}
\begin{center}
\includegraphics[width=0.50\textwidth]{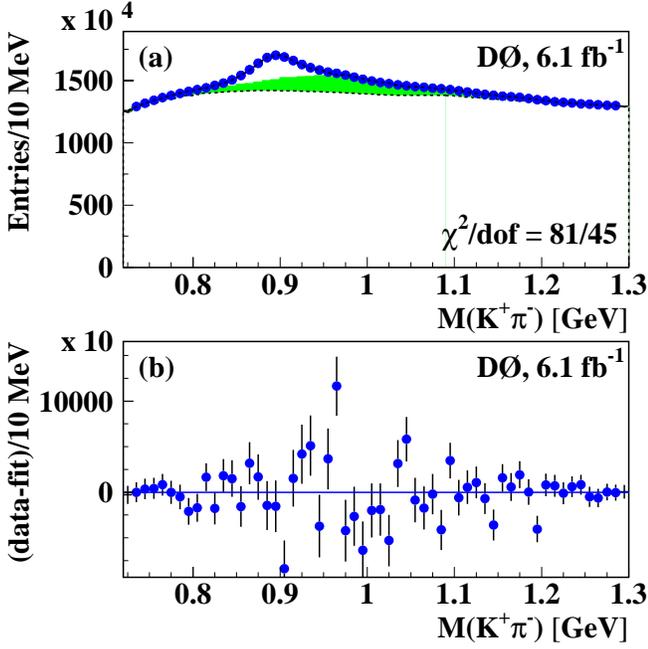}
\caption{(a) The $K^+ \pi^-$ invariant mass distribution of $K^{*0}$ candidates
in the inclusive muon sample. The solid line corresponds to the result of the fit and the
dashed line shows the contribution from the combinatorial background. The shaded
histogram is the contribution of $\rho^0 \to \pi^+ \pi^-$ events. (b) The difference between data
and the result of the fit.}
\label{fig_kst0_incl}
\end{center}
\end{figure}

\begin{figure}
\begin{center}
\includegraphics[width=0.50\textwidth]{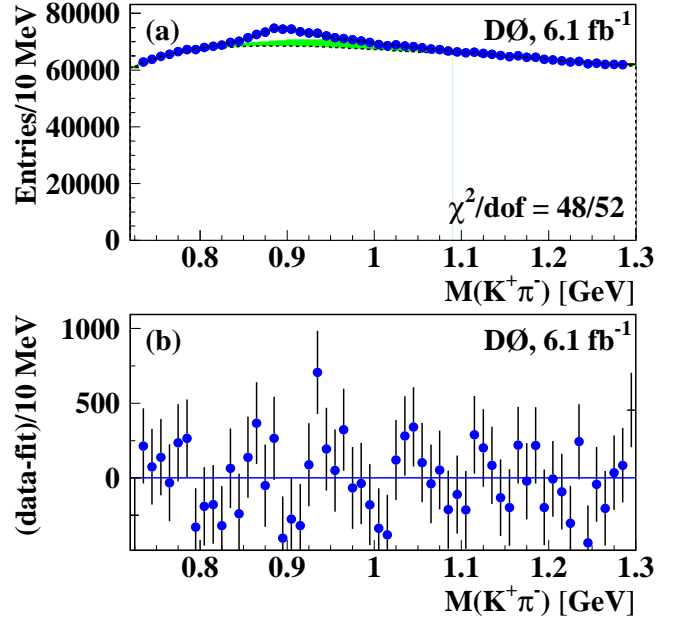}
\caption{(a) The $K^+ \pi^-$ invariant mass distribution of $K^{*0}$ candidates
in the like-sign dimuon sample. The solid line corresponds to the result of the fit and the
dashed line shows the contribution from the combinatorial background. The shaded
histogram is the contribution of $\rho^0 \to \pi^+ \pi^-$ events. (b) The difference between data
and the result of the fit.}
\label{fig_kst0_mu2}
\end{center}
\end{figure}

Figure~\ref{fig_kst0_incl} shows the $K^+ \pi^-$~\cite{charge}
invariant mass distribution of the $K^{*0}$ candidates with \ktomu~ in the inclusive muon sample,
while Fig.~\ref{fig_kst0_mu2} shows the corresponding mass distribution in the like-sign dimuon sample.

\begin{figure}
\begin{center}
\includegraphics[width=0.50\textwidth]{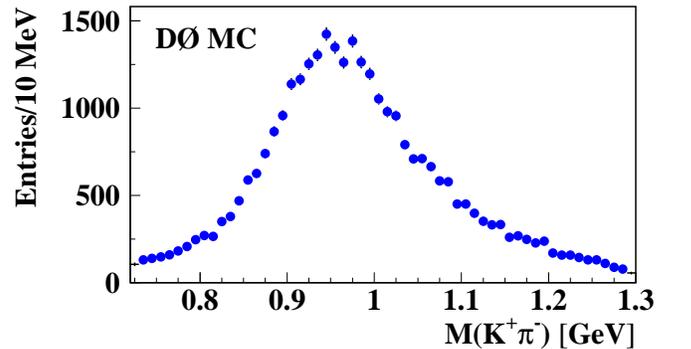}
\caption{The ``$K^+ \pi^-$" invariant mass distribution of simulated
$\rho^0 \to \pi^+ \pi^-$ decays, where the mass of the charged kaon is assigned
to one of the two reconstructed tracks.}
\label{fig_rho}
\end{center}
\end{figure}

The measurement of the number of $K^{*0} \to K^+ \pi^-$ decays with \ktomu~ is
complicated because of the large combinatorial background under the $K^{*0}$ signal,
and because of the contribution of light meson resonances decaying to $\pi^+ \pi^-$.
The most important contribution comes from the $\rho^0 \to \pi^+ \pi^-$ decay with \pitomu.
It produces a peak in the mass region close to the $K^{*0}$ mass.  Figure~\ref{fig_rho}
shows the mass distribution of simulated $\rho^0 \to \pi^+ \pi^-$ decays with one pion assigned
the kaon mass. This pion is also required to satisfy the track selections.

To overcome these misidentification difficulties, the fit to the $K^+ \pi^-$ mass distribution
is performed in several steps, assuming for the width of the $K^{*0}$ meson the value obtained
in the previous section for the $K^{*+}$ meson. Contrary to the $K^+ \pi^-$ system, the decays
of light resonances do not contribute into the $K_S \pi^+$ mass distribution because the $K_S$
meson is identified unambiguously, and the $K^{*+}$ signal is clean and unbiased.

\begin{figure}
\begin{center}
\includegraphics[width=0.50\textwidth]{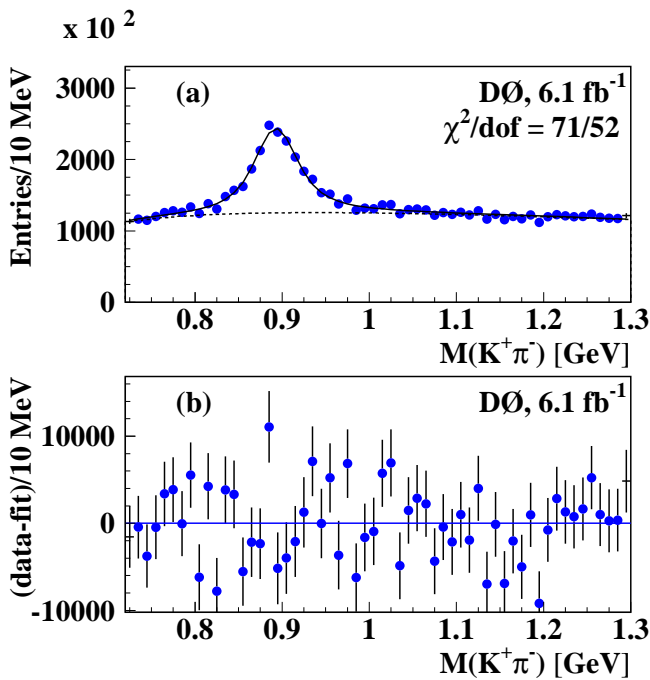}
\caption{(a) The difference of the $K^+ \pi^-$ and $K^- \pi^+$ mass distributions of $K^{*0}$
candidates in the inclusive muon sample. The solid line represents the result of the fit, while the dashed line
shows the background contribution. (b) The difference between data and the result of the fit.}
\label{fig_kst0_diff}
\end{center}
\end{figure}

The $K^{*0}$ mass and detector resolution for $K^{*0} \to K^+ \pi^-$ with \ktomu~are obtained
from a fit to the difference of the $K^+ \pi^-$ and $K^- \pi^+$ mass distributions.
The reconstruction efficiency of \ktomu~ tracks demonstrates a charge asymmetry $\approx$6\%,
(see Sec.~\ref{sec_ak}). Such an asymmetry is significantly smaller for \pitomu~ tracks, and
the contribution of $\rho \to \pi^+ \pi^-$ and other light resonances is therefore suppressed
in the difference of the $K^+ \pi^-$ and $K^- \pi^+$ mass distributions. In addition, the
contribution of the combinatorial background is significantly reduced, as can be seen in
Fig.~\ref{fig_kst0_diff}(a). The $K^{*0}$ signal is fitted with a relativistic Breit-Wigner
function convoluted with a Gaussian resolution, and the background is parameterized by the
function~(\ref{bckg1}). Figure~\ref{fig_kst0_diff}(b), which shows the difference between data
and the result of the fit, indicates a moderate quality for the fit, with $\chi^2/{\rm d.o.f.} = 71/52$.
The fit gives $\sigma(M) = 12.2 \pm 1.5$(stat) MeV for the $K^{*0}$ mass resolution
of the detector. The mass difference between $K^{*0}$ and $K^{*+}$ is
\begin{equation}
M(K^{*0}) - M(K^{*+}) = 3.50 \pm 0.66 \mbox{(stat) MeV},
\end{equation}
which is consistent with the PDG value of $4.34 \pm 0.36$ MeV~\cite{pdg}.

The number of $K^{*0}$ events in the inclusive muon and in the like-sign dimuon samples
is determined from the fit of the mass distributions shown in Figs.~\ref{fig_kst0_incl}
and~\ref{fig_kst0_mu2}. The signal is parameterized with the convolution of a
relativistic Breit-Wigner function and a Gaussian resolution.
The $K^{*0}$ mass and width
and the detector resolution are fixed in the fit to the values obtained from the fit 
to the distribution in Fig.\ \ref{fig_kst0_diff}a.
The mass distribution of the $\rho^0 \to \pi^+ \pi^-$ background is taken from the MC simulation.
To improve the quality of the fit, the parameterization of the background is modified by adding
two additional Gaussian terms:
\begin{eqnarray}
f'_{\rm bck}(M) & = & f_{\rm bck}(M) \nonumber \\
                & + & p_4 \exp(-\frac{(M-M_1)^2}{2 \sigma_1^2}) \nonumber \\
                & + & p_5 \exp(-\frac{(M-M_2)^2}{2 \sigma_2^2}).
\end{eqnarray}
Here $f_{\rm bck}(M)$ is given in~(\ref{bckg1}).
The additional terms are needed to describe the distortion of the smooth behavior of the combinatorial
background at large masses of $M\approx1.15$ GeV, due to the contributions of other
light resonances, and should be considered as a parameterization of the observed mass
distribution rather than the contribution from specific sources. These terms are significant
only because of the large statistics of the inclusive muon sample, which contains about
 $10^7$ entries per bin in Fig.~\ref{fig_kst0_incl}. The fit of the $\kstneu$ signal in the
like-sign dimuon sample is of the same quality without these additional terms. The results do not
change significantly if these terms are omitted. The impact of these terms on the final measurement
is included in the systematic uncertainties of the fractions $f_K$ and $F_K$ discussed
in Sec.~\ref{sec_syst}.

In the fit to the inclusive muon distribution, the parameters $M_1$, $M_2$, $\sigma_1$, and
$\sigma_2$ and the contribution of $\rho^0 \to \pi^+ \pi^-$, are allowed to vary.
The ratio of the fractions of $\rho^0$ and $\kstneu$ mesons is constrained within 10\%
of the value obtained in the simulation.
The fit yields $M_1 = 1.095 \pm 0.005$ GeV and $M_2 = 1.170 \pm  0.007$ GeV, which is far
from the region of the $K^{*0}$ mass, and does not influence the fitted number of
$K^{*0}$ mesons. The $\chi^2/{\rm d.o.f.}$ of the fit is 81/45. The results of the
fit and corresponding residuals are shown in Fig.~\ref{fig_kst0_incl}.

In the fit to the like-sign dimuon distribution the parameters $M_1$, $M_2$, $\sigma_1$, and
$\sigma_2$ are fixed to the values obtained in the fit of the inclusive muon sample.
The $\chi^2/{\rm d.o.f.}$ of the fit is 48/52. The results of the
fit and corresponding residuals are shown in Fig.~\ref{fig_kst0_mu2}.

\subsection{Reconstruction of $\phin$ mesons}

The $\phin$ meson is reconstructed by selecting two tracks with opposite charge,
and assigning both of them the mass of the charged kaon. One track is required to pass the
track selections of Sec.~\ref{selection}. The second one satisfies the same selection
criteria as the pion in the $\kstpm$ and $\kstneu$ reconstructions.

\begin{figure}
\begin{center}
\includegraphics[width=0.50\textwidth]{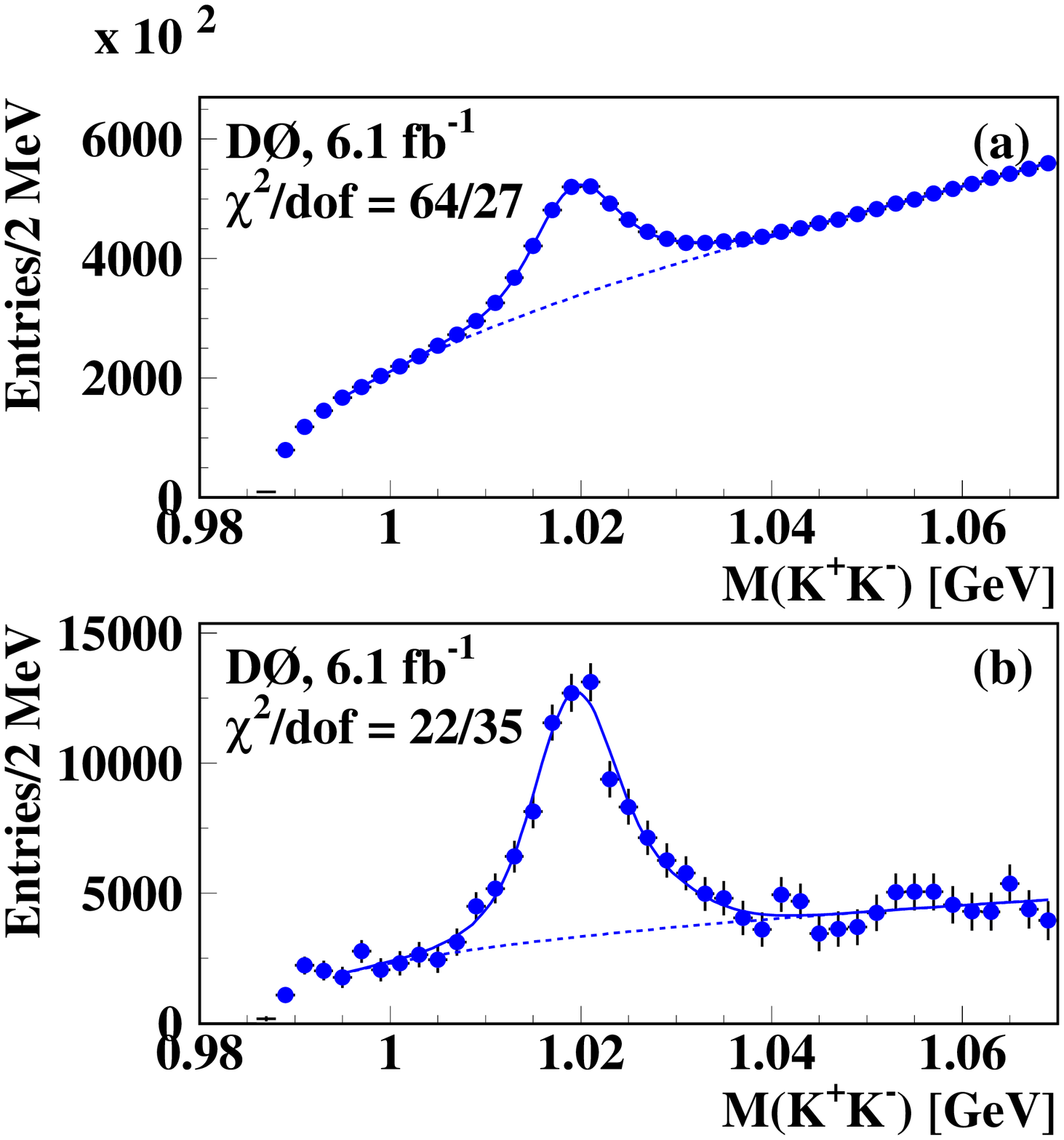}
\caption{The $K^+ K^-$ invariant mass distribution of $\phin$ candidates in the inclusive muon sample
for \ktomu~ with $4.2 < p_T < 7.0$ GeV. (a) The sum of distributions
of positive and negative \ktomu~ tracks and (b) their difference.
The solid lines present the result of the fit and the dashed lines show the background contribution.
}
\label{fig_phi}
\end{center}
\end{figure}

Figure~\ref{fig_phi} shows the $K^+ K^-$ invariant mass distribution of the $\phin\to K^+ K^-$
candidates in the inclusive muon sample, with an additional requirement on the
transverse momentum of the kaon misidentified as a muon, $4.2 < p_T < 7.0$ GeV.
We display separately the sum and the difference of the distributions for
the \ktomu~ tracks with positive and negative charges, as done in the case of $\ks$
candidates, and use these distributions to measure the asymmetry for \ktomu~ tracks.
The $\phin$ signal is fitted with a double Gaussian and the background is parameterized
by the threshold function
\begin{eqnarray}
f_{\rm bck}(M) & = & (M-2\, M_K)^{p_0} \nonumber \\
& \times & \exp(p_1 M + p_2 M^2 + p_3 M^3).
\label{bckg2}
\end{eqnarray}
All the parameters describing the signal, except its normalization, are fixed in
the fit of the difference of the invariant mass distributions to the values obtained
from the fit to the sum of the distributions.

\subsection{Reconstruction of $\Lambda$ baryons}

\begin{figure}
\begin{center}
\includegraphics[width=0.50\textwidth]{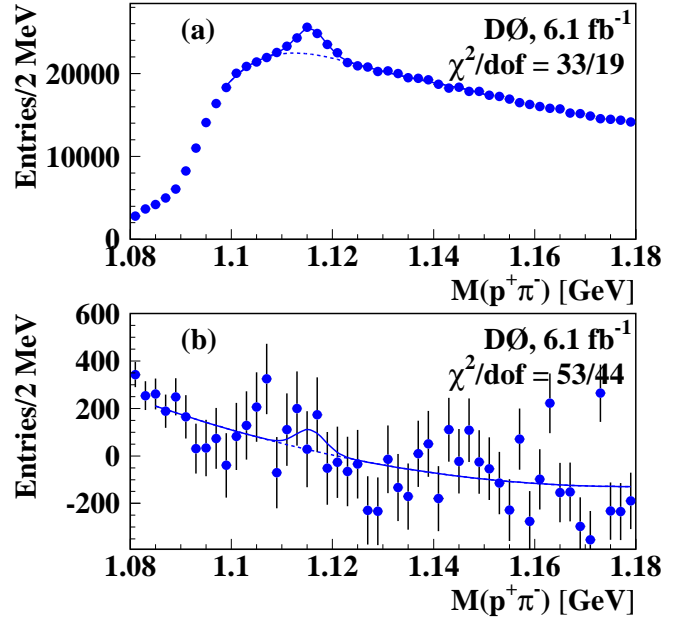}
\caption{The $p \pi^-$ invariant mass distribution of $\Lambda$ candidates in the inclusive muon sample
for \ptomu~ with $4.2 < p_T < 7.0$ GeV. (a) The sum of distributions
of positive and negative \ptomu~ tracks
and (b) their difference.
The solid lines present the result of the fit and the dashed lines show the background contribution.
}
\label{fig_lam}
\end{center}
\end{figure}

The selection of $\Lambda \to p \pi^-$ decays~\cite{charge} follows that of $\ks \to \pi^+ \pi^-$,
except that one of the tracks is assigned the mass of the proton. Figure~\ref{fig_lam}
shows the $p \pi^-$ invariant mass distribution of $\Lambda \to p \pi^-$ candidates
in the inclusive muon sample, with an additional requirement on the
transverse momentum of the proton misidentified as a muon, for $4.2 < p_T < 7.0$ GeV.
Also in this case we display separately the distributions for the sum and the difference
of the distributions for $\Lambda$ and $\bar{\Lambda}$ decays, and use them to determine
the asymmetry for $p\to\mu$ tracks. The $\Lambda$ baryon signal is fitted with a Gaussian, while
the background is parameterized by a fourth (second) degree polynomial for the sum (difference)
of the invariant mass distributions. All parameters describing the signal, except its
normalization, are fixed in the fit of the difference of the invariant mass distributions
to the values obtained from the fit to the sum of the distributions.

\section{Track reconstruction asymmetry}
\label{sec_tras}

In this measurement of $\aslb$, we assume that the charge asymmetry of track reconstruction
cancels as a result of the regular reversal of the magnets polarity.
The method developed in Secs.~\ref{sec_fk} and~\ref{sec_ak} provides a way of evaluating
this assumption in data by comparing the track charge asymmetry $a_{\rm track}$
with that expected from already analysed sources.
We select events with one reconstructed muon and at least one additional track
satisfying the selection criteria of Sec.~\ref{selection}. The muon is not used in this
study, but is required since the events are collected with single muon triggers.
Nevertheless, the charges of the muon and the additional track can be correlated, and
the asymmetry of the inclusive muon events can bias the observed track asymmetry. To eliminate this bias,
we consider separately events in which the muon and the track have equal and opposite charges.
The asymmetries in these samples can be expressed as
\begin{eqnarray}
a_{\rm opp} & = & a_{\rm track} - a, \nonumber \\
a_{\rm equal} & = & a_{\rm track} + a.
\label{atrk}
\end{eqnarray}
The asymmetry $a$ is defined in Eq.~(\ref{asym_a}), and the asymmetries
$a_{\rm opp}$ and $a_{\rm equal}$ are computed as
\begin{eqnarray}
a_{\rm opp} & = & \frac{n^{+-} - n^{-+}}{n^{+-} + n^{-+}}, \nonumber \\
a_{\rm equal} & = & \frac{n^{++} - n^{--}}{n^{++} + n^{--}},
\end{eqnarray}
where $n^{ij}$ is the number of events containing a track with charge $i$ and a
muon with charge $j$. The track asymmetry $a_{\rm track}$ is computed as
\begin{equation}
a_{\rm track} = \frac{1}{2}(a_{\rm opp} + a_{\rm equal}).
\label{asym_track1}
\end{equation}

As discussed in Sec.~\ref{sec_ak}, we expect the value of $a_{\rm track}$ to contain
a contribution from the asymmetry of kaon reconstruction, even after averaging over
the different magnet polarities.  The expected value of $a_{\rm track}$ is therefore given by:
\begin{equation}
a_{\rm track} = a_K^{\rm track} f_K^{\rm track},
\label{asym_track}
\end{equation}
where $f_K^{\rm track}$ is the fraction of reconstructed tracks that are kaons, and
$a_K^{\rm track}$ is the charge asymmetry of kaon reconstruction.

The fraction $f_K^{\rm track}$ is measured using $K^{*0} \to K^+ \pi^-$ decays~\cite{charge}.
The selected charged particle is assigned the kaon mass and combined with an additional
track to produce the $K^{*0}$ candidate. The $K^{*0}$ selection and fitting
procedure is described in Appendix~\ref{sec_fits}. The measured number of $K^{*0}$ mesons is
converted into the number of kaons using a method similar to that presented in Sec.~\ref{sec_fk}.
The measured $f_K^{\rm track}$ fraction is assigned a systematic uncertainty as
in Sec.~\ref{sec_syst}.

The same $K^{*0} \to K^+ \pi^-$ decay is used to measure
the kaon reconstruction asymmetry. This measurement can be biased by
the asymmetry of the muon, because the charge of the kaon and the muon can be correlated.
The kaon asymmetry is therefore measured separately in the samples in which
the muon and the kaon have equal or opposite charges. The asymmetry $a_K^{\rm track}$ is computed
in a way similar to Eq.~(\ref{asym_track1}):
\begin{equation}
a_K^{\rm track} = \frac{1}{2}(a_K^{\rm opp} + a_K^{\rm equal}).
\end{equation}
The $K^{*0}$ mass distribution is plotted separately
for positive and negative kaons in each sample,
and the sum and the difference of these distributions
is fitted to extract the quantity $\Delta$, corresponding to the difference in the number
of $K^{*0}$ decays with positive and negative kaons,
and the quantity $\Sigma$, corresponding to their sum. The asymmetry $a_K^{\rm opp}$ is measured as
$a_K^{\rm opp} = \Delta_{\rm opp}/\Sigma_{\rm opp}$, and a similar relation is used to obtain
the asymmetry $a_K^{\rm equal}$.

\begin{figure}
\begin{center}
\includegraphics[width=0.50\textwidth]{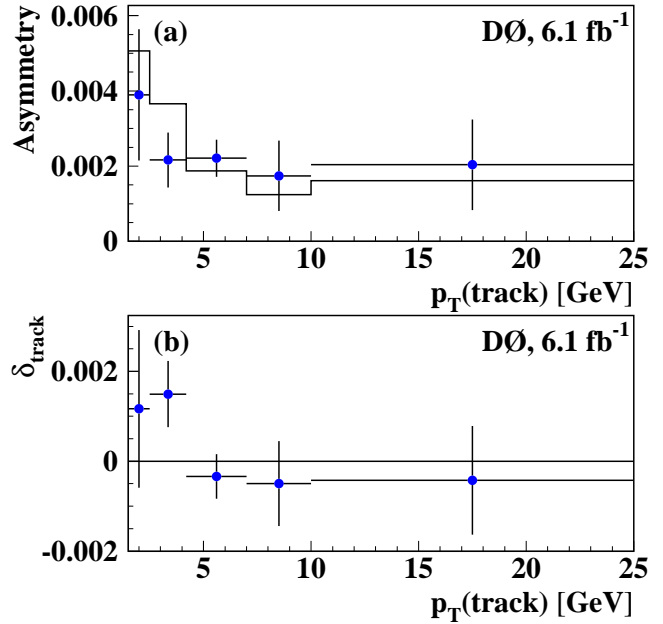}
\caption{(a) The expected (points with errors) and measured
(histogram with negligible uncertainties) track reconstruction asymmetry.
(b) The quantity $\delta_{\rm track} = a_{\rm track} - a_K^{\rm track} f_K^{\rm track}$.
}
\label{fig_tras}
\end{center}
\end{figure}

The expected and observed track reconstruction
asymmetry, i.e., the right and left side of Eq.~(\ref{asym_track}),
are compared in Fig.~\ref{fig_tras} as a function of track $p_T$.
There is excellent agreement between these
two quantities. The $\chi^2/{\rm d.o.f.}$ for their difference is 5.4/5.
The fit of $\delta_{\rm track} = a_{\rm track} - a_K^{\rm track} f_K^{\rm track}$
to a constant yields the following estimate for a residual track asymmetry of
\begin{equation}
\delta_{\rm track} = +0.00011 \pm 0.00035.
\label{delta_tr}
\end{equation}
We conclude that the residual track asymmetry is consistent with zero as expected. The
uncertainty on this value is about a factor of 16 times smaller than the observed charge asymmetry
in the like-sign dimuon events. This study provides an additional confirmation of the validity
of the method used in this analysis.

\section{Trigger asymmetry}
\label{sec_trig}

We determine the kaon, pion and proton charge asymmetries using events passing
at least one single muon trigger, and we apply the same asymmetries to the like-sign dimuon events
collected with dimuon triggers. Similarly, we measure the muon reconstruction asymmetry
using events passing the dimuon triggers, and we use the same quantity for events collected
with the single muon triggers. If the trigger selection is charge asymmetric, and this asymmetry is
different for single muon and for dimuon triggers, the obtained value of $\aslb$ can be biased.
Since we extract the asymmetry $\aslb$ from the difference $A - \alpha a$, and
because the value of $\alpha$ is very close to unity, our measurement is especially sensitive
to a difference between the charge asymmetry of dimuon and single muon triggers.

To examine the impact of this difference on our result,
we repeat the measurement of $\aslb$ using dimuon events passing
any single muon trigger without requiring dimuon triggers.
The result of this test, given in column \textsf{O} of Table~\ref{tab11}, does not
indicate a bias from the trigger selection.
In addition, we measure $\aslb$ using dimuon events passing
both single muon and dimuon triggers. The result of this test is given
in column \textsf{P} of Table~\ref{tab11}. These two tests provide a residual
difference $\delta_T$ between the asymmetry of dimuon triggers and single muon triggers of
\begin{equation}
\delta_T = +0.00010 \pm 0.00029.
\end{equation}

From this result, we conclude that the trigger selections do not produce any significant bias to the value of \aslb.

\section{Alternative measurement of $F_K/f_K$}
\label{sec_alt}

%
%

We consider two methods to obtain the parameters $\alpha_i \equiv F^i_K / f^i_K$ 
in the $p_T$ interval $i$ from
\begin{equation}
\alpha_i  = 2 \frac{N_i(K^{*0} \rightarrow K \rightarrow \mu)}{N_i}
\frac{n_i}{n_i(K^{*0} \rightarrow K \rightarrow \mu)},
\label{alpha_i_1}
\end{equation}
where $N_i$ refers to the like-sign dimuon sample and
$n_i$ refers to the inclusive muon sample.
$N_i(K^{*0} \rightarrow K \rightarrow \mu)$ and
$n_i(K^{*0} \rightarrow K \rightarrow \mu)$ are obtained
by fitting the invariant mass histograms of
$K^{*0} \rightarrow \pi K \rightarrow \mu$ in the
like-sign dimuon and inclusive muon samples, respectively.
These fits require precise modeling of the $\rho^0$ resonance
and of other backgrounds. In the first method we use the results listed 
in Table~\ref{tab3}
and obtain the values of $\alpha_i$ listed in Table~\ref{alphas}.

\begin{table}
\caption{\label{alphas}
Values of $\alpha_i \equiv F^i_K/f^i_K$ obtained through two methods,
with their statistical uncertainties.
}
\begin{ruledtabular}
\newcolumntype{A}{D{A}{\pm}{-1}}
\newcolumntype{B}{D{B}{-}{-1}}
\begin{tabular}{cAA}
Bin &  \multicolumn{1}{c}{$\alpha_i$ from Table~\ref{tab3}} & \multicolumn{1}{c}{$\alpha_i$
from null fit} \\
\hline
0     & 1.309 \ A \ 0.340 & 0.954 \ A \ 0.217 \\
1     & 0.987 \ A \ 0.082 & 0.942 \ A \ 0.069 \\
2     & 1.022 \ A \ 0.050 & 1.031 \ A \ 0.027 \\
3     & 0.758 \ A \ 0.101 & 0.806 \ A \ 0.055 \\
4     & 1.406 \ A \ 0.159 & 1.292 \ A \ 0.079 \\ \hline
All   & 0.998 \ A \ 0.038 & 0.990 \ A \ 0.022  
\end{tabular}
\end{ruledtabular}
\end{table}

A second set of values for the $\alpha_i$ parameters can be
obtained by finding a scale factor which minimizes
the differences between invariant mass distributions for the $K^{*0}$ candidates
in the inclusive muon and dimuon samples (null fit method). Invariant mass distributions for
$K^{*0}$ candidates analogous to the ones shown in Fig.~\ref{fig_kst0_incl}
and~\ref{fig_kst0_mu2} are built for each bin of \ktomu~transverse momentum.
We scale the invariant mass distribution in the inclusive muon sample by a factor
$\alpha_i N_i/(2 n_i)$ and subtract it from the invariant mass
distribution obtained from the dimuon sample. The contributions from the
$\rho$ resonance and from other backgrounds cancel to first order
in the difference of the two invariant mass distributions,
simplifying the convergence of mass fits. We then vary the factor $\alpha_i$
and find the value that yields a null normalization factor for the
residual $K^{*0}$ signal. The statistical uncertainty on $\alpha_i$ is
obtained by choosing the value of the scale factor which yields
a positive (negative) value of this normalization factor which
is equal to its uncertainty. The values of $\alpha_i$ obtained
with this method are also listed in Table~\ref{alphas}, with their
statistical uncertainties. There is remarkable agreement between
the values of $\alpha_i$ obtained using the two methods.

%


\begin{thebibliography}{99}
\bibitem{charge}
Charge conjugation invariance is implied throughout this article.
\bibitem{rapidity}
Pseudorapidity $\eta \equiv -\ln{ \left[ \tan(\theta/2) \right] }$, where
$\theta$ is the angle of the track relative to the direction of the
proton beam; $\phi$ is the polar angle with respect to the proton beam,
and $\phi = 90^0$ defined as the vertical axis.
\end{thebibliography}
\end{document}